\documentclass[10pt,a4paper,twoside]{article}
\usepackage{a4}
\usepackage{amsfonts}
\usepackage{amssymb}
\usepackage{epsfig}
\usepackage{psfig}

\def\epsfig{\psfig}
\newcommand{\beqn}{\begin{eqnarray}}
\newcommand{\eeqn}{\end{eqnarray}}
\newcommand{\be}{\begin{equation}}
\newcommand{\ee}{\end{equation}}
\newcommand{\non}{\nonumber \\}
\newcommand{\tr}{\mbox{tr}} 
\newcommand{\Tr}{\mbox{Tr}}
\newcommand{\dt}{{\mbox{det}}}
\newcommand{\Dt}{\mbox{Det}}
\newcommand{\al}{\alpha^\prime}
\newcommand{\eps}{\epsilon}

\newcommand{\host}{\ ^*\hspace{-0.05cm}} 
\newcommand{\sz}{\scriptsize}

\input{epsf.sty}
\def\axowidth{0.5 }
\def\axoscale{1.0 }
\def\axoxoff{0 }
\def\axoyoff{0 }

\def\Gluon(#1,#2)(#3,#4)#5#6{
\put(\axoxoff,\axoyoff){
}
}

\def\Photon(#1,#2)(#3,#4)#5#6{
\put(\axoxoff,\axoyoff){
}
}

\def\ZigZag(#1,#2)(#3,#4)#5#6{
\put(\axoxoff,\axoyoff){
}
}

\def\PhotonArc(#1,#2)(#3,#4,#5)#6#7{
\put(\axoxoff,\axoyoff){
}
}

\def\GlueArc(#1,#2)(#3,#4,#5)#6#7{
\put(\axoxoff,\axoyoff){
}
}

\def\ArrowArc(#1,#2)(#3,#4,#5){
\put(\axoxoff,\axoyoff){
}
}

\def\LongArrowArc(#1,#2)(#3,#4,#5){
\put(\axoxoff,\axoyoff){
}
}

\def\DashArrowArc(#1,#2)(#3,#4,#5)#6{
\put(\axoxoff,\axoyoff){
}
}

\def\ArrowArcn(#1,#2)(#3,#4,#5){
\put(\axoxoff,\axoyoff){
}
}

\def\LongArrowArcn(#1,#2)(#3,#4,#5){
\put(\axoxoff,\axoyoff){
}
}

\def\DashArrowArcn(#1,#2)(#3,#4,#5)#6{
\put(\axoxoff,\axoyoff){
}
}

\def\ArrowLine(#1,#2)(#3,#4){
\put(\axoxoff,\axoyoff){
}
}

\def\LongArrow(#1,#2)(#3,#4){
\put(\axoxoff,\axoyoff){
}
}

\def\DashArrowLine(#1,#2)(#3,#4)#5{
\put(\axoxoff,\axoyoff){
}
}

\def\Line(#1,#2)(#3,#4){
\put(\axoxoff,\axoyoff){
}
}

\def\DashLine(#1,#2)(#3,#4)#5{
\put(\axoxoff,\axoyoff){
}
}

\def\CArc(#1,#2)(#3,#4,#5){
\put(\axoxoff,\axoyoff){
}
}

\def\DashCArc(#1,#2)(#3,#4,#5)#6{
\put(\axoxoff,\axoyoff){
}
}

\def\Vertex(#1,#2)#3{
\put(\axoxoff,\axoyoff){
}
}

\def\Text(#1,#2)[#3]#4{
\dimen0=\axoxoff \unitlength
\dimen1=\axoyoff \unitlength
\advance\dimen0 by #1 \unitlength
\advance\dimen1 by #2 \unitlength
\makeatletter
\@killglue\raise\dimen1\hbox to\z@{\kern\dimen0 \makebox(0,0)[#3]{#4}\hss}
\ignorespaces
\makeatother
}

\def\BCirc(#1,#2)#3{
\put(\axoxoff,\axoyoff){
}
}

\def\GCirc(#1,#2)#3#4{
\put(\axoxoff,\axoyoff){
}
}

\def\EBox(#1,#2)(#3,#4){
\put(\axoxoff,\axoyoff){
}
}

\def\BBox(#1,#2)(#3,#4){
\put(\axoxoff,\axoyoff){
}
}

\def\GBox(#1,#2)(#3,#4)#5{
\put(\axoxoff,\axoyoff){
}
}

\def\Boxc(#1,#2)(#3,#4){
\put(\axoxoff,\axoyoff){
}
}

\def\BBoxc(#1,#2)(#3,#4){
\put(\axoxoff,\axoyoff){
}
}

\def\GBoxc(#1,#2)(#3,#4)#5{
\put(\axoxoff,\axoyoff){
}
}

\def\SetOffset(#1,#2){\def\axoxoff{#1 } \def\axoyoff{#2 }}

\def\fsize{10 }

\def\PText(#1,#2)(#3)[#4]#5{
\ifx#4 lt{\def\fmode{0 }}\else{
\ifx#4 tl{\def\fmode{0 }}\else{
\ifx#4 lb{\def\fmode{2 }}\else{
\ifx#4 bl{\def\fmode{2 }}\else{
\ifx#4 l{\def\fmode{1 }}\else{
\ifx#4 rt{\def\fmode{6 }}\else{
\ifx#4 tr{\def\fmode{6 }}\else{
\ifx#4 rb{\def\fmode{8 }}\else{
\ifx#4 br{\def\fmode{8 }}\else{
\ifx#4 r{\def\fmode{7 }}\else{
\ifx#4 t{\def\fmode{3 }}\else{
\ifx#4 b{\def\fmode{5 }}\else{ \def\fmode{4 } }\fi
}\fi}\fi}\fi}\fi}\fi}\fi}\fi}\fi}\fi}\fi}\fi
\put(#1,#2){\makebox(0,0)[]{\special{"/pfont findfont /fsize scale setfont
\axoxoff \axoyoff #3 \fmode \fsize (#5) ptext }}}  }

\def\GOval(#1,#2)(#3,#4)(#5)#6{
\put(\axoxoff,\axoyoff){
}
}

\def\Oval(#1,#2)(#3,#4)(#5){
\put(\axoxoff,\axoyoff){
}
}

\let\eind=]

\def\kromme(#1,#2)#3{#1 #2 \ifx #3\eind\else\expandafter\kromme\fi#3}

\def\LogAxis(#1,#2)(#3,#4)(#5,#6,#7,#8){
\put(\axoxoff,\axoyoff){
}
}

\def\LinAxis(#1,#2)(#3,#4)(#5,#6,#7,#8,#9){
\put(\axoxoff,\axoyoff){
}
}

\begin{document} 

\title{{\bf Recent Developments in String Theory:
From Perturbative Dualities to M-Theory}} 
\author{}
\date{Lectures given by D. L\"ust at the Saalburg Summer School in September 1998}

\maketitle

\begin{center}

{\bf Michael Haack}\footnote{Email: 
Michael@Hera1.Physik.Uni-Halle.De}\hspace{-0.03cm}$^{,*}$, 
{\bf Boris K\"ors}\footnote{Email: Koers@Physik.HU-Berlin.De}
\hspace{-0.11cm}$^{,\dagger}$, {\bf Dieter L\"ust}\footnote{Email: 
Luest@Physik.HU-Berlin.De}\hspace{-0.01cm}$^{,\dagger}$ \\

\vspace{0.5cm}

$^*$ Martin-Luther-Universit\"at Halle-Wittenberg \\
{\small{Institut f\"ur Physik, Friedemann Bach Platz 6, 06108 Halle/Saale, Germany}}

\vspace{0.3cm}

$^\dagger$ Humboldt Universit\"at zu Berlin \\
{\small{Institut f\"ur Physik, Invalidenstr. 110, 10115 Berlin, Germany}} 

\vspace{2cm}

\end{center}

\begin{abstract}
These lectures intend to give a pedagogical introduction into some of the developments
in string theory during the last years. They include perturbative
T-duality and non perturbative S- and U-dualities, their unavoidable demand
for D-branes, an example of enhanced gauge symmetry at fixed points of the
T-duality group, a review of classical solitonic solutions in general
relativity, gauge theories and tendimensional supergravity, a discussion of
their BPS nature, Polchinski's observations that allow to view D-branes as RR
charged states in the non perturbative string spectrum, the application of all
this to the computation of the black hole entropy and Hawking radiation and
finally a brief survey of how everything fits together in M-theory.
\end{abstract}

\thispagestyle{empty}

\clearpage

\tableofcontents
\clearpage

\section{Introduction}
\label{intro}

These notes are a summary and a substantial extension of the 
material that D. L\"ust presented in his lectures at the summer 
school at Saalburg in 1998. They are intended to give a basic overview over 
non perturbative effects
and duality symmetries in string theory including recent developments. 
After a
short review of the status of perturbative string theory, as it presented
itself before the (second) string revolution in 1995, and a brief summary of
the recent progress especially concerning non perturbative aspects, the main
text falls into two pieces. In chapter 2 we will go into some details of
T-duality. Afterwards chapter 3 and chapter 4 will focus on some
non perturbative phenomena. The text is however not meant as an introduction
to string theory but rather relies on some basic knowledge (see e.g. the
lectures given at this school by O. Lechtenfeld or \cite{LT,GSW,JP}). The references we give are never intended to be exhaustive but only to display the material that is essentially needed to justify our arguments and calculations.

\subsection{Perturbative string theory}
\label{pertubative}

\begin{figure}[h]
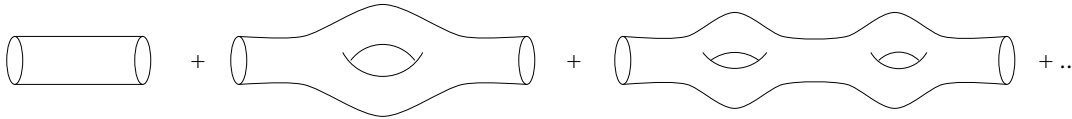

\begin{center}
\input pert.pstex_t
\end{center}
\caption{String perturbation expansion}
\label{figspe}
\end{figure}

\begin{figure}[h]
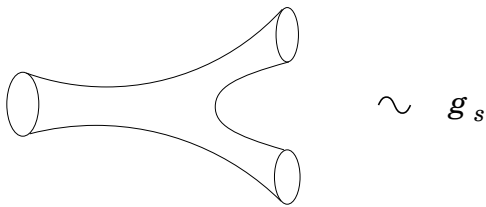

\begin{center}
\vspace{-1cm}
\input vertex.pstex_t
\end{center}
\caption{Fundamental string interaction vertex}
\label{figfsiv}
\end{figure}

Before 1995 string theory was only defined via its perturbative expansion. As the string moves in time, it sweeps out a two dimensional worldsheet $\Sigma$ which is embedded via its coordinates in a Minkowski target space $\cal{M}$:
\be
X^{\mu}(\sigma,\tau) : \Sigma \rightarrow \cal{M}.  \label{eq1}
\ee
This worldsheet describes (after a Wick rotation in the time
variable $\tau$) a Riemann surface (possibly with boundary). Propagators or general Green's functions of scattering processes can be expanded in the different topologies of Riemannian surfaces, which corresponds to an expansion in the string coupling constant $g_{\mbox{\sz S}}$ (see fig.\ref{figspe}).
The reason for this is, that all string diagrams can be built out of the fundamental splitting respectively joining vertex (fig.\ref{figfsiv}). This vertex comes along with a factor $g_{\mbox{\sz S}}$, which is given by the vacuum expectation value (VEV) of a scalar field $\Phi$, the so called dilaton:
\be
g_{\mbox{\sz S}} \sim e^{\langle \Phi \rangle}.  \label{eq2}
\ee
As there is no potential for the dilaton in string perturbation theory, its VEV is an arbitrary parameter, which can be freely chosen. Only if it is small, the above expansion in Riemann surfaces makes sense. Statements about the strong coupling regime on the other hand require some knowledge about non perturbative characteristics of string theory such as duality relations combining weakly coupled string theories with strongly coupled ones. 
First quantizing the string amounts to quantizing the embedding coordinates, regarded as fields of a two dimensional (conformal) field theory living on the world sheet. This is a two dimensional analog of point particle quantum mechanics. A sensible second quantized string field theory is very difficult to achieve and will not be discussed here any further. 
\begin{table}[h]
\begin{center}
\begin{tabular}{l c c c}
Type & Gauge group & \# of supercharges & N \\ [2mm]
Heterotic & $E_{8} \times E_{8}$ & $16$ & $1$ \\ [2mm]
Heterotic$^\prime$ & $SO(32)$ & $16$ & $1$ \\ [2mm]
I (includes open strings) & $SO(32)$ & $16$ & $1$ \\ [2mm]
IIA (nonchiral) & - & $16 + \overline{16}$ & $2$ \\ [2mm]
IIB (chiral) & - & $16 + 16$ & $2$\\
\end{tabular}
\end{center}
\caption{The five consistent superstring theories in d=10}
\label{tab2}
\end{table}
In the perturbative regime there exist five consistent ten
dimensional superstring theories (see table \ref{tab2}). Type IIA and IIB at first sight do not contain any open strings. In fact they do however appear if one introduces the non perturbative objects called {\em D-branes}, which are hyperplanes on which open strings can end. Thus from the world sheet viewpoint a D-brane manifests itself by cutting a hole into the surface and imposing Dirichlet boundary conditions. These objects will be studied in more detail below.
To get a string theory in lower dimensional space-time, such as in the
phenomenologically most interesting case $d=4$, one has to compactify the
additional space dimensions. There are several methods to construct
fourdimensional string theories and in fact there are many
different ways to get rid of the extra dimensions. A priori each compactification gives
rise to a different string vacuum with different particle content, gauge group
and couplings. This huge vacuum degeneracy in four dimensions is known as the
{\em vacuum problem}. But despite of the large number of different known vacua
it has not yet been possible to find a compactification yielding in its low
energy approximation precisely the standard model of particle physics.

\subsection{T-Duality}
\label{tduality}

T-duality (or target space duality) \cite{GPR} denotes the equivalence of two string
theories compactified on different background spaces. 'Both' theories can in fact be considered as one and the same string
theory as they contain exactly the same physics. The equivalence
transformation can thus be considered as some kind of transformation of
variables, in which the theory is described. Nevertheless we will always use
the usual terminology, speaking of different theories when we actually mean
different equivalent formulations of the same physical theory. 
T-duality is a perturbative symmetry in the sense, that the T-duality transformation maps the weak coupling region of one theory to the weak coupling regime of another theory. Thus it can be tested in perturbation theory, e.g. by comparing the perturbative string spectra. Examples of T-dualities are:
\begin{table}[h]
\begin{center}
\begin{tabular}{c c c}
Het on $S^{1}$ with radius $\frac{R}{\sqrt{\alpha'}}$ & $\stackrel{\mbox{\sz T-dual}}{\leftrightarrow}$ & Het on $S^{1}$ with radius $R_{D} = \frac{\sqrt{\alpha'}}{R}$ \\ [2mm]
IIA on $S^{1}$ with radius $\frac{R}{\sqrt{\alpha'}}$ & $\stackrel{\mbox{\sz T-dual}}{\leftrightarrow}$ & IIB on $S^{1}$ with radius $R_{D} = \frac{\sqrt{\alpha'}}{R}$.
\end{tabular}
\end{center}
\end{table}
\vspace{-4.5mm}

\noindent These are special cases of the so called {\em mirror symmetry}. As we will see in chapter \ref{Tduality}, T-duality transformations for closed strings exchange the winding number around some circle with the corresponding (discrete) momentum quantum number. Thus it is clear, that this symmetry relation has no counterpart in ordinary point particle field theory as the ability of closed strings to wind around the compactified dimension is essential.

\subsection{Non-perturbative dualities}
\label{strongcoupling}

At strong coupling the higher topologies of the expansion (fig. \ref{figspe}) become large and the series expansion does not make sense anymore. Non perturbative effects dominantly contribute to the scattering processes. Their contributions behave like:
\be
{\cal A} \sim e^{-1/g_{\mbox{\sz S}}} \quad \mbox{or} \quad {\cal A} \sim e^{-1/g_{\mbox{\sz S}}^{2}}.   \label{eq3}
\ee
The second exponential (with $g_{\mbox{\sz S}} \leftrightarrow g_{\mbox{\sz YM}}$) is the typical non perturbative suppression factor in gauge field theoretic amplitudes involving solitons like magnetic monopoles or instanton effects. Solitons also play a role in general relativity in the form of black holes. 
In general solitons are non trivial solutions of the field equations which have a finite action integral. Their energy is localized in space and they have properties similar to point particles. 
Clearly it is of some interest to ask, what kind of solitonic objects appear
in string theory giving rise to the behavior of eq. (\ref{eq3}). The answer is
that the string solitons are extended $p$-(spatial)dimensional flat objects called
{\em p-branes} (i.e. $(p+1)$-dimensional hypersurfaces in space-time). The
special values $p=0,1,2$ therefore give point particles, strings and membranes
respectively. Such objects can indeed be found as classical solutions of the
effective low energy field theories derived from the various superstring
theories (see section \ref{stringsolitons}). It was however Polchinski's
achievement to realize, that some of them (namely those which do not arise in
the universal sector) have an alternative description as hyperplanes on which
open strings can end \cite{POL3}. As these objects are necessarily contained in type IIA
and IIB string theory, it is apparent, that these theories have to contain
open strings. Unlike in type I theory the open strings just have to start and
end on the p-branes and are not allowed to move freely in the whole of
space-time. It is obvious that the boundary conditions of the open strings
have changed from Neumann to Dirichlet ones in the space dimensions transvers
to the branes. That is why these p-branes are called {\em D-branes} (in
contrast to the p-branes from the universal sector which are sometimes also called {\em NS-branes}). Note that precisely the D-branes are responsible for contributions to scattering amplitudes that are suppressed by the first type of suppression factor. \\

This new insight into the nature of the non perturbative degrees of freedom in string theory is a fundamental ingredient of the recently conjectured non perturbative duality symmetries. Like T-duality, these dualities are supposed to establish an equivalence of two (seemingly different) full string theories, but in their case the duality transformations map the weak coupling regions of one theory to the strong coupling regions of the other one and vice versa. Thus they e.g. exchange elementary excitations and the solitonic p-branes. Several different kinds of such non perturbative dualities have to be distinguished:

\subsubsection{S-duality}
\label{sduality}

By S-duality we mean a selfduality, which maps the weak coupling regime 
of one string
theory to the strong coupling region of the same theory. 
The existence of such a strong-weak coupling duality in string 
theory was first conjectured in \cite{LU} in the context of the 
compactification of the heterotic string to four dimensions. 
After some time accumulating
evidence for the S-duality of the heterotic string compactified on $T^6$
was found found \cite{REY}-\cite{SENSCHW}.
More recently it was realized that also the type IIB superstring 
in ten dimensions is S-dual to itself \cite{SCH} and another  
In both cases the transformation acts via an element of $SL(2, \mathbb{Z})$ 
on
a complex scalar $\lambda$, whose imaginary part's VEV is related to the
coupling constant of the string theory (see (\ref{eq2})). 
In the heterotic theory it is given by $\lambda = \tilde{a} + i e^{-\Phi}$, 
where $\Phi$ is the dilaton and $\tilde{a}$ the scalar which is equivalent 
to the antisymmetric tensor $B_{\mu \nu}$ in four dimensions. 
For the type IIB theory we have instead 
$\lambda = \tilde{a} + i e^{-\Phi/2}$, where now $\tilde{a}$ is the 
second scalar present in the ten dimensional spectrum coming from the 
RR sector. Both theories are invariant under the S-duality transformation 
\be
\lambda \rightarrow \frac{a \lambda + b}{c \lambda +d} \quad \mbox{with} 
\quad \left( \begin{array}{cc}
                 a & b \\
                 c & d 
                 \end{array} 
          \right)  \in SL(2,\mathbb{Z})  \label{eq3A}
\ee
combined with a transformation of either the four dimensional gauge 
bosons, mixing $F$ and its dual,
\beqn
\left( \begin{array}{c} 
F \\
\tilde F
\end{array}
\right) \rightarrow \left( \begin{array}{cc}
                 a & b \\
                 c & d 
                 \end{array} 
          \right)
\left( \begin{array}{c}
       F \\
       \tilde F
       \end{array}
\right) ,
\eeqn
or the two antisymmetric tensors 
\beqn
\left( \begin{array}{c} 
B_{\mu \nu} \\
B_{\mu \nu}^{'}
\end{array}
\right) \rightarrow \left( \begin{array}{cc}
                 a & b \\
                 c & d 
                 \end{array} 
          \right)
\left( \begin{array}{c}
       B_{\mu \nu} \\
       B_{\mu \nu}^{'}
       \end{array}
\right)
\eeqn
in the heterotic or type IIB case respectively. The case
$\tilde{a}=0$, $a=d=0$,
$b=-c=1$ shows, that the S-duality transformations comprise the inversion of
the coupling constant. In the first example the theory is 
additionally invariant under a T-duality group (see below (\ref{eq2k2})).

\subsubsection{U-duality}
\label{uduality}

The U-duality group (see e.g. \cite{obers} for a review)
of a given string theory is the group which comprises T- and 
S-duality and embeds them into a generally larger group with new symmetry generators. The main example is the
system  type IIA/IIB in $d \leq 8$ on $T^{10-d}$. It will be seen in section
\ref{twoatwob} that for $d \leq 9$ type IIA and type IIB have the same
moduli space and are T-dual to each other in the sense that type IIA at large
compactification radius is equivalent to type IIB at small radius and vice
versa. The two tendimensional theories are different limits of a single space of
compactified theories, which will in the following sometimes be called the moduli space of type II theory (meaning all compactifications of type IIA and IIB). As indicated above, the theory is invariant under T-duality, which relates compactifications of type IIA (IIB) to
those of type IIB (IIA) for a T-duality transformation in an odd number of
directions and to those of type IIA (IIB) for a transformation in an even
number. Furthermore it inherits the S-duality of the type IIB in ten
dimensions. All these transformations act however
only on the scalars of the NSNS sector, i.e. the Kaluza-Klein scalars coming
from the metric and antisymmetric tensor (including the scalars coming from
the ten dimensional RR scalar and RR antisymmetric tensor of IIB). It has
however been conjectured \cite{HT} that there is a much larger symmetry group called U-duality group, which contains the S- and T-duality group\footnote{The product $\boxtimes$ is non-commutative, as the S-duality transformation also acts non trivially on the antisymmetric tensor.} $SL(2, \mathbb{Z}) \boxtimes SO(D,D,\mathbb{Z})$ as its subgroup but transforms all scalars into each other, including those coming from the RR sector. In particular the U-duality groups of type II string theory on a $(10-d)$-dimensional torus are:
\begin{table}[h]
\begin{center}
\begin{tabular}{c c c c c c c c}
d & 8 & 7 & 6 & 5 & 4 & 3 & 2 \\ [2mm]
U-duality & $SL(2,\mathbb{Z})$ & $SL(5,\mathbb{Z})$ & $SO(5,5,\mathbb{Z})$ & $E_{6(6)}(\mathbb{Z})$ & $E_{7(7)}(\mathbb{Z})$ & $E_{8(8)}(\mathbb{Z})$ & $\hat{E}_{8(8)}(\mathbb{Z})$  \\
group & $\times SL(3,\mathbb{Z})$ & & & & & & 
\end{tabular}
\end{center}
\end{table}

\noindent where $E_{n(n)}$ denotes a noncompact version of the exceptional group $E_{n}$ for $n = 6,7,8$ (see e.g. \cite{LIE}) and $\hat{G}$ for any group $G$ means the loop group of $G$, i.e. the group of mappings from the circle $S^{1}$ into $G$.

\subsubsection{String-string-duality}
\label{stringstringduality}

This duality 
(some aspects of the
string-string duality are reviewed in \cite{ntwo})
relates two different string theories in a way that the 
perturbative expansions get mixed up. The perturbative regime of the one theory is equivalent to the non perturbative regime of the other one. The elementary excitations on one side are mapped to the solitonic objects on the other side and vice versa. Examples are:
\begin{table}[h]
\begin{center}
\begin{tabular}{ccc}
Het on $T^{4}$ & $\leftrightarrow$ & IIA on K3 \\
Het with gauge group $SO(32)$ in d=10 & $\leftrightarrow$ & I in d=10
\end{tabular} 
\end{center}
\end{table}

\subsubsection{Duality to elevendimensional supergravity and M-theory}

Let us now consider the type IIA superstring. 
At weak coupling it is the known tendimensional theory. 
However if we increase the coupling, a 
new eleventh dimension opens up \cite{WITTENAA}. 
Or more precisely stated, the effective Lagrangian of the tendimensional type IIA supergravity perfectly agrees with that of the elevendimensional supergravity compactifed on a circle of radius $R_{11}$, if the following identification of the type IIA string coupling and $R_{11}$ is made:
\beqn
g^{2/3}_{\mbox{\sz S}}=R_{11}.
\eeqn
The Kaluza-Klein states of the elevendimensional theory get masses proportional to $1/R_{11}$ and they are mapped by the duality transformation to the D0-branes of the type IIA superstring theory. Something analogous happens for the heterotic string, where the strong coupling limit is dual to elevendimensional theory compactified on an interval $S_1/Z_2$ \cite{HOW1,HOW2}. \\

\begin{figure}[h]
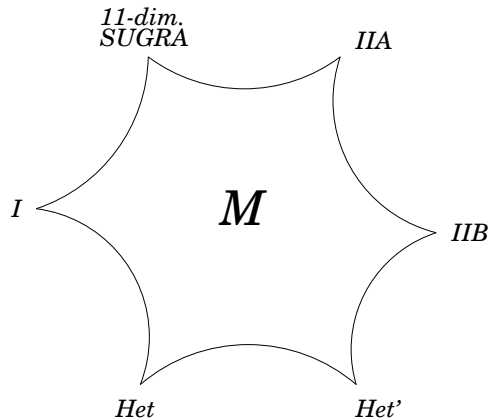

\begin{center}
\vspace{1cm}
\input M.pstex_t
\end{center}
\caption{M-theory}
\label{figm}
\end{figure}

All these different dualities have now led to the conjecture that all superstring theories are connected to each other via duality transformations in different dimensions. This suggests, that there is only one underlying unique fundamental theory and the different string vacua are just different weak coupling regions in the moduli space of this fundamental theory called {\em M-theory} (see fig. \ref{figm}). From the type IIA string theory we have learned that this M-theory is supposed to be an elevendimensional theory whose low energy effective Lagrangian should coincide with that of elvendimensional supergravity. However a fundamental formulation of M-theory is still lacking. We will come back to M-theory in chapter \ref{mtheory} and give more convincing arguments in favour of the above claims.

\section{T-duality}
\label{Tduality}

We are now going to have a closer look at the perturbative dualities, namely T-duality. We will first consider the most simple case, the bosonic string on a circle respectively D-dimensional torus, and afterwards generalize to the superstring. During the investigation of the type I string we will see that it is unavoidable to introduce D-branes, i.e. hyperplanes on which the open string ends. That is because T-duality changes the boundary conditions of open strings from Neumann to Dirichlet.

\subsection{Closed bosonic string theory}
\label{closedbosonic}

The principal effects of T-duality in closed bosonic string theory can be studied within the context of

\subsubsection{Compactification on a circle}
\label{circle}

The string action of the bosonic string moving in a flat background is given in the conformal gauge:
\be
S = \frac{1}{4 \pi \alpha'} \int d\tau d\sigma \, \partial_{\alpha} X^{\mu} (\sigma, \tau) \partial^{\alpha} X_{\mu} (\sigma, \tau).   \label{eq2a}
\ee
The resulting equation of motion is simply the two dimensional wave equation:
\be
\left( \frac{\partial^{2}}{\partial \sigma^{2}} - \frac{\partial^{2}}{\partial \tau^{2}} \right) \, X^{\mu} (\sigma, \tau) = 0    \label{eq2b}
\ee
leading to the usual decomposition:
\be
X^{\mu} (\sigma, \tau) = X^{\mu}_{R} (\sigma^{-}) + X^{\mu}_{L}(\sigma^{+})   \label{eq2c}
\ee
where $X^{\mu}_{R}$ and $X^{\mu}_{L}$ are arbitrary functions of their arguments $\sigma^{-} = \tau - \sigma$ respectively $\sigma^{+} = \tau + \sigma$, just constrained to obey certain boundary conditions, which depend on the background the string is moving in. Besides the tendimensional Minkowski space also a space-time with one (or several) dimension(s) compactified on a circle (or higher dimensional torus) has the property of being Ricci flat, which is required for the background of any consistent string theory. We first assume, that only one coordinate is compact, namely 
\be
X^{25} \simeq X^{25} + 2 \pi R \label{eq2d}
\ee
and therefore have to implement in our solution (\ref{eq2c}) the periodicity condition
\be
X^{25}(\sigma + 2 \pi, \tau) = X^{25}(\sigma, \tau) + 2 \pi m R   \label{eq2e}
\ee
in the compact direction. The general solution is given by 
\beqn
X^{25}_{R} (\sigma^{-}) & = & x^{25}_{R} + \sqrt{\frac{\alpha'}{2}} p^{25}_{R} (\tau - \sigma) + i \sqrt{\frac{\alpha'}{2}} \sum_{l \neq 0} \frac{1}{l} \alpha^{25}_{R,l} e^{-i l (\tau - \sigma)}  , \non  
X^{25}_{L} (\sigma^{+}) & = & x^{25}_{L} + \sqrt{\frac{\alpha'}{2}} p^{25}_{L} (\tau + \sigma) + i \sqrt{\frac{\alpha'}{2}} \sum_{l \neq 0} \frac{1}{l} \alpha^{25}_{L,l} e^{-i l (\tau + \sigma)}   \label{eq2f} ,
\eeqn
where we have used 
\beqn
p^{25}_{R} & = & \frac{1}{\sqrt{2}} \left( \frac{\sqrt{\alpha'}}{R} n - \frac{R}{\sqrt{\alpha'}} m \right)   , \non 
p^{25}_{L} & = & \frac{1}{\sqrt{2}} \left( \frac{\sqrt{\alpha'}}{R} n + \frac{R}{\sqrt{\alpha'}} m \right)  \label{eq2g}
\eeqn
for $m,n \in \mathbb{Z}$. The canonical momentum is $p^{25} =
(p^{25}_{L} + p^{25}_{R})/ \sqrt{2 \alpha'} = n/R$. The solution in
(\ref{eq2f}) has to be supplemented with the usual solution of the wave
equation in the non compact directions (i.e. replace $p^{25}_{L}$ and
$p^{25}_{R}$ in (\ref{eq2f}) by the same continous $p^{\mu}$ and both
$x^{25}_{L}$ and $x^{25}_{R}$ by $\frac{1}{2} x^{\mu}$). A string state in 26
uncompactified dimensions is characterized by specifying its momentum and oscillations. Analogously the states of the compactified string theory depend on the two quantum numbers $m$ and $n$, denoting the discrete momentum and winding of the string in the 25$^{\mbox{\tiny th}}$ dimension, its momentum in the non compact dimensions and the oscillations (internal and external ones). The winding is obviously a typical string effect with no analog in field theory. 
The mass of the (perturbative) states is given by $M^{2} = M^{2}_{L} + M^{2}_{R}$ with 
\beqn
M^{2}_{L} & = & -\frac{1}{2} p^{\mu}p_{\mu} \hspace{2mm} = \hspace{2mm}
\frac{1}{2} (p^{25}_{L})^{2} + \frac{2}{\alpha'} (N_{L} - 1) ,  \non
M^{2}_{R} & = & -\frac{1}{2} p^{\mu}p_{\mu} \hspace{2mm} = \hspace{2mm}
\frac{1}{2} (p^{25}_{R})^{2} + \frac{2}{\alpha'} (N_{R} - 1)  , \label{eq2h}  
\eeqn 
where $N_{L}$ and $N_{R}$ denote both the internal and the external oscillations. T-duality in this case refers to the symmetry of the mass spectrum under the $\mathbb{Z}_{2}$-transformation:
\beqn
\frac{R}{\sqrt{\alpha'}} & \leftrightarrow & \frac{\sqrt{\alpha'}}{R} , \non
m & \leftrightarrow & n.   \label{eq2i}
\eeqn
As the transformation just maps the perturbative mass spectrum into (and onto) itself, the T-duality is from the target space point of view a perturbative symmetry. Concerning the two dimensional world sheet point of view however there is an exchange of the elementary excitations (momentum states) with the solitonic ones (winding states). \\

It is obvious from (\ref{eq2i}) and (\ref{eq2g}) that the transformation maps $p^{25}_{L}$ to itself and $p^{25}_{R}$ to minus itself. If the whole theory is supposed to be invariant under T-duality this should be especially the case for all interactions (i.e. the interactions of states in one theory should be the same as those of the dual states in the `other' theory).  Therefore the vertex operators should also be invariant. They contain however phase factors like $\exp \left(i p^{25}_{L} X^{25}_{L} \right)$ and $\exp \left( i p^{25}_{R} X^{25}_{R} \right)$  which are only invariant if we demand:
\beqn
X^{25}_{L} & \rightarrow & X^{25}_{L}  ,\non
X^{25}_{R} & \rightarrow & - X^{25}_{R}, \quad \mbox{i.e.} \hspace{.2cm} \alpha^{25}_{R,i} \rightarrow -\alpha^{25}_{R,i} \ \mbox{and}\ x^{25}_{R} \rightarrow -x^{25}_{R}.   \label{eq2j}
\eeqn
Now it is possible to show that this change of the signs of the right-moving $\alpha^{25}_{R,i}$ leaves all the correlation functions invariant and therefore is a symmetry of perturbative closed string theory. It should be emphasized that T-duality thus is a space-time parity operation on the right moving degrees of freedom only, which will become important in the context of type II string theory. \\

The moduli space of a theory which depends on one (or more) parameter(s) is defined as the range of the parameter(s) leading to distinct physics. In our case the relevant parameter is the radius of the compactification circle. But whereas in field theory every radius $R \in \mathbb{R}_{+}$ leads to different physics the situation in string theory is different. Here T-duality relates small with large radii and there is a `smallest' (resp. biggest) radius, namely the fixed point of the transformation (\ref{eq2i}), i.e. $R_{\mbox{\sz fix}}=\sqrt{\alpha '}$. Thus the moduli space is ${\cal M}=\left\{ R  \leq R_{\mbox{\sz fix}} \right\}$ resp. ${\cal M}= \left\{R \geq R_{\mbox{\sz fix}} \right\}$, which can be expressed in a more formal way as 
\be
{\cal M}=\left\{R \in \mathbb{R}_{+}/\mathbb{Z}_{2} \right\}.    \label{eq2k}
\ee
This is a general feature of T-duality, that the moduli space in string theory can be obtained from the one in field theory by modding out a discrete symmetry group, namely the T-duality group. 
In string theory these parameters or moduli describing different vacuum configurations are typically given by vacuum expectation values of massless scalar fields. In the case of circle compactification the relevant scalar is 
\be 
|\Phi\rangle = \alpha^{25}_{L,-1} \alpha^{25}_{R,-1} |0\rangle,   \label{eq2l}
\ee
whose VEV corresponds to the radius of the circle (to be more precise, it
really corresponds to the difference of the radius and $R_{\mbox{\sz fix}}$). The state $|0\rangle$ denotes the vacuum state without winding or internal momentum.

\subsubsection{Compactification on a torus $T^{D}$}
\label{torus}

We now want to generalize the results of the previous discussion to the case
of higher dimensional torus compactifications on $T^{D}$ (for convenience we
set $\alpha'=2$ throughout this section). A torus can be defined by identifying points of $\mathbb{R}^{D}$, which lie in a $D$-dimensional lattice $\Lambda^{D}$,
\be
T^{D} = \mathbb{R}^{D} / 2 \pi \Lambda^{D},   \label{eq2m}
\ee
where the lattice $\Lambda^{D}$ can be specified by giving $D$ linear independent vectors in $\mathbb{R}^{D}$, namely $\left( R_{i}/ \sqrt{2}\right) \vec{e}_{i}$, $i = 1,\ldots,D$ with $(\vec{e}_{i})^{2} = 2$. I.e. 
\be
\Lambda^{D} = \left\{\vec{L} =  \sum_{i=1}^{D} \sqrt{\frac{1}{2}} m_{i} R_{i} \vec{e}_{i} \, , \, \vec{m} \in \mathbb{Z}^{D} \right\}.   \label{eq2s}
\ee
We also need the notion of the dual lattice $\Lambda^{D \ast}$, given by the $D$ vectors $\left( \sqrt{2} /R_{i} \right) \vec{e}^{\ast i}$, which is defined as the lattice of vectors which have integer scalar products with all the elements of $\Lambda^{D}$. In particular the basis vectors satisfy
\be
\sum_{a=1}^{D} e_{i}^{a} e^{\ast j}_{a} = \delta^{j}_{i}, \quad \sum_{i=1}^{D} e_{i}^{a} e^{\ast i}_{b} = \delta^{a}_{b}  . \label{eq2n}
\ee
The importance of the dual lattice lies in the fact, that the canonical momenta\footnote{Remember that the canonical momenta are in general not the  same as the kinematical momenta, denoted by $p^{i}$.} $\pi^{i}$ in the compactified dimensions have to lie in $\Lambda^{D \ast}$ for compactifications on $\Lambda^{D}$ in order to ensure the single valuedness of $\exp \left( i X^{i} \pi^{i} \right)$, which is the generator of translations in the internal directions.
Furthermore the metric $G_{ij}$ of $\Lambda^{D}$ is given by
\be 
\sum_{a=1}^{D} \sqrt{\frac{1}{2}} R_{i} e^{a}_{i} \sqrt{\frac{1}{2}} R_{j} e^{a}_{j} = G_{ij} ,\quad  \sum_{a=1}^{D} \frac{\sqrt{2}}{R_{i}} e^{\ast i}_{a} \frac{\sqrt{2}}{R_{j}} e^{\ast j}_{a} = (G^{-1})^{ij}.   \label{eq2o}
\ee
In generalization of (\ref{eq2l}) we now have $D^{2}$ massless scalars 
\be
|\Phi^{ij}\rangle = \alpha^{i}_{L,-1} \alpha^{j}_{R,-1}|0\rangle, \quad i,j = 1,\ldots,D.  \label{eq2p}
\ee
These fields correspond to the moduli of the $D$-dimensional torus compactification. Their VEVs can be regarded as the internal components of the metric, i.e. (\ref{eq2o}), and the antisymmetric tensor field $B_{ij}$, yielding $D(D+1)/2$ respectively $D(D-1)/2$ degrees of freedom. Therefore we have to generalize (\ref{eq2a}) to contain the antisymmetric tensor. The action for the internal degrees of freedom of a string moving in a background specified by $G_{ij}$ and $B_{ij}$ then takes the form\footnote{The second term did not show up in the previous section, because there is no antisymmetric tensor in one dimension. The third term does not play any role for our purposes. $\Phi$ is the dilaton and $R^{(2)}$ the two dimensional world sheet curvature scalar. We shall return to this background field action (the linear $\sigma$-model) later on.} 
\be
S = \int d\sigma d\tau \, L = \frac{1}{8 \pi} \int d\sigma d\tau \, \left( G_{ij} \partial_{\alpha} X^{i} \partial^{\alpha} X^{j} + \epsilon^{\alpha \beta} B_{ij} \partial_{\alpha}X^{i} \partial_{\beta} X^{j}  -2 \Phi R^{(2)} \right).    \label{eq2q}
\ee
The fields $X^{i}$, $G_{ij}$ and $B_{ij}$ are all given via their components referring to the basis of $\Lambda^{D}$, namely referring to $\{e_{i}\}$. One could as well take all components referring to the standard basis of $\mathbb{R}$. In this case we will take the indices to be $a,b,\ldots$ in contrast to $i,j,\ldots$. From (\ref{eq2q}) (with $(i,j) \leftrightarrow (a,b)$) we can deduce the canonical momentum:
\be
\pi^{a} = \int_{0}^{2 \pi} d\sigma \, \frac{\partial L}{\partial (\partial_{\tau} X_{a})} = p^{a} + \frac{1}{4 \pi} B^{ab} \left( X_{b}(\sigma = 2\pi) - X_{b}(\sigma = 0) \right) = p^{a} + \frac{1}{2} B^{ab} L_{b}.    \label{eq2r}
\ee
As already mentioned $\vec{\pi}$ has to be an element of $\Lambda^{D \ast}$. We also have 
\beqn
p^{a} = \frac{1}{4 \pi} \int_{0}^{2 \pi} d\sigma \ \partial_{\tau} X^{a} = \frac{1}{2} (p^{a}_{L} + p^{a}_{R})
\eeqn 
and $p^{a}_{L} - p^{a}_{R} = L^{a}$ (compare e.g. (\ref{eq2g})) and therefore we get
\beqn
p^{a}_{L,R} & = & \pi^{a} - \frac{1}{2} B^{ab} L_{b} \pm \frac{1}{2} L^{a} \non
 & = & \sum_{k=1}^{D} \sqrt{2} \frac{n_{k}}{R_{k}} (e^{\ast k})^{a} + \frac{1}{2} \sum_{j,k=1}^{D} \left( \tilde{B}_{kj} \pm G_{kj} \right) \sqrt{2} \frac{m_{k}}{R_{j}} (e^{\ast j})^{a},   \label{eq2t} 
\eeqn
where we have expressed the antisymmetric tensor via its components referring to the basis of the dual lattice, i.e. 
\beqn
B^{ab} = \frac{\sqrt{2}}{R_{k}} (e^{\ast k})^{a} \tilde{B}_{kj} (e^{\ast j})^{b} \frac{\sqrt{2}}{R_{j}} ,
\eeqn
and we have used (\ref{eq2n}) and (\ref{eq2o}) to rewrite 
\beqn
L^{a} = \sum_{k=1}^{D} \frac{1}{\sqrt{2}} m_{k} R_{k} e_{k}^{a} = \sum_{j,k=1}^{D} m_{k} G_{kj} \frac{\sqrt{2}}{R_{j}} (e^{\ast j})^{a} .
\eeqn
The $p^{a}_{L,R}$ are in general no elements of $\Lambda^{D \ast}$, but they are the momenta which enter the analog of (\ref{eq2h}). That is why the spectrum depends on both, the shape of the torus and the VEV of the antisymmetric tensor field:
\be
M^{2} = N_{L} + N_{R} - 2 + \frac{1}{2} \left( (\vec{p}_{L})^{2} + (\vec{p}_{R})^{2} \right).   \label{eq2u}
\ee
It is obvious from this equation, that the spectrum is invariant under seperate rotations $SO(D)_{L}$ and $SO(D)_{R}$ of the vectors $\vec{p}_{L}$ and $\vec{p}_{R}$. 
In order to determine the classical moduli space it is convenient to look at the lattice $\Gamma_{D,D}$ which consists of all vectors $(\vec{p}_{L},\vec{p}_{R})$ and for which we choose a Lorentzian scalar product, i.e. $(\vec{p}_{L},\vec{p}_{R})\cdot(\vec{p}_{L}\hspace{0.01cm}',\vec{p}_{R}\hspace{0.01cm}') = \vec{p}_{L} \cdot \vec{p}_{L}\hspace{0.01cm}' - \vec{p}_{R} \cdot \vec{p}_{R}\hspace{0.01cm}'$. Using (\ref{eq2o}) and the antisymmetry of $\tilde{B}_{ij}$ it is easy to verify that we have
\be
(\vec{p}_{L},\vec{p}_{R})\cdot({\vec{p}_{L}}\hspace{0.01cm}',{\vec{p}_{R}}\hspace{0.01cm}') = \sum_{i=1}^{D} (m_{i} n_{i}' + n_{i} m_{i}') \in  \mathbb{Z}.   \label{eq2v}
\ee
That means the inner product is independent of the background fields $G_{ij}$ and $\tilde{B}_{ij}$. It can be calculated by taking for example $G_{ij}=\delta_{ij} \, (\mbox{i.e.}\, \, e_{i}^{a} = \sqrt{2} \delta^{a}_{i}, \, R_{i}=1)$ and $\tilde{B}_{ij}=0$. In this case it is obvious, that $\Gamma_{D,D}$ is even and self-dual, i.e. the length of any element is even (clear from (\ref{eq2v})) and the lattice is equal to its dual.
But as the scalar product (\ref{eq2v}) is independent of the background fields
the same holds for the self-duality and the eveness of the lattice. Different
values of the $D^{2}$ parameters (\ref{eq2p}) lead to different such
Lorentzian lattices $\Gamma_{D,D}$ and actually all of them can be obtained by
choosing the correct values for the background fields. It is known, that it is
possible to generate all different even and self-dual  Lorentzian lattices via
an $SO(D,D)$ rotation of a reference lattice, e.g. the special one considered
above ($e_{i}^{a} = \sqrt{2} \delta^{a}_{i}, \, R_{i}=1$ and
$\tilde{B}_{ij}=0$). We have seen however in (\ref{eq2u}) a hint that not all
of them yield a different string theory. In fact one can identify the classical moduli space to be
\be
{\cal M}_{\mbox{\sz class}} = \frac{SO(D,D)}{SO(D) \times SO(D)},   \label{eq2x}
\ee
which is a $D^{2}$-dimensional manifold. \\

Like in the case of the circle compactification, special points in the moduli
space are equivalent because of stringy effects, while they were not equivalent classically. Again the equivalent points can be mapped to each other via the
operation of a discrete T-duality group and it is obvious from the
representation $T^{D} = S^{1} \times \ldots \times S^{1}$ that this group
embraces $\mathbb{Z}_{2}^{D}$ coming from the T-duality groups of the $S^{1}$
factors which make up the torus. Actually the whole T-duality group of $T^{D}$
is bigger \cite{GPR}, namely 
\be
\Gamma_{\mbox{\sz T-duality}} = SO(D,D,\mathbb{Z}),   \label{eq2y}
\ee
which consists of all $SO(D,D)$ matrices with integer entries. Roughly speaking the T-duality group is generated by the T-duality transformations of the different circles, basis changes for the defining lattice of the torus and integer shifts in $\tilde{B}_{ij}$. Instead of showing this in general we include an extensive treatment of the example $D=2$ in the appendix. We also demonstrate the feature of gauge symmetry enhancement at fixed points of the duality group on the moduli space there.

\subsection{Heterotic string theory}
\label{heterotic}

The internal bosonic part of the heterotic string action on a D-dimensional torus including the coupling to a background gauge field $V_{a A}$ ($a=1, \ldots, D$, $A=1, \ldots , 16$) and a background antisymmetric tensor is 
\beqn
S & = & \frac{1}{4 \pi \alpha'} \int d\sigma d\tau \left( (G_{ab} \delta_{\alpha \beta} + B_{ab} \epsilon_{\alpha \beta}) \partial^{\alpha} X^{a} \partial^{\beta} X^{b} +(G_{AC} \delta_{\alpha \beta} + B_{AC} \epsilon_{\alpha \beta}) \partial^{\alpha} X^{A}_{L} \partial^{\beta} X^{C}_{L} \right. \non
& & \hspace{6.5cm} \mbox{} \left. + V_{aA} \partial_{\alpha} X^{a}_{L} \partial^{\alpha} X^{A}_{L} + \ldots \right).   \label{eq2i2} 
\eeqn 
The background gauge field is called a {\em Wilson line} and is a pure gauge
configuration in the Cartan subalgebra of the gauge group, which yields
however a non trivial parallel transport around non trivial paths in space
time\footnote{The expression Wilson line refers to both, the gauge
  configuration and (the trace of the pathordered product of) the line integral of the vector potential along
  a closed line: $tr \left( {\cal P} \exp \left( \oint d\vec{X} \cdot \vec{A}
    \right) \right)$.  \label{fn1}}. The potential for the gauge bosons not in
the Cartan subalgebra has for torus compactifications no flat directions so
that it is not possible for them to obtain a VEV. If all the Wilson line
moduli are zero the gauge group is unbroken (i.e. $E_{8} \times E_{8}$ or
$SO(32)$) but for non zero  values of the background gauge field only the
subgroup commuting with the Wilson line remains a gauge symmetry, i.e. it is
broken to $U(1)^{16}$ for generic values. Thus the Wilson line parameters are further $16 D$ moduli specifying the vacuum configuration, which is now characterized by $D(D+16)$ background fields.  \\

It is obvious from (\ref{eq2i2}) that only the left moving part of the internal action has changed compared to the bosonic case. That is because the heterotic string is a hybrid construction of a left moving bosonic and a right moving fermionic string which are compactified on two different tori. The compactification torus for the left moving sector is the product of that one for the right moving sector times the one defined through the dual of the root lattice of $E_{8} \times E_{8}$ or $SO(32)$. Thus it is clear that the right momenta $p_{R}$ from (\ref{eq2t}) do not change, but the left ones do. Nevertheless it is again possible to show, that the resulting vectors $(p_{L}, p_{R})$ form an even self dual Lorentzian lattice with signature $(D+16, D)$. As before all of those can be generated via $SO(D+16, D)$ transformations of a reference lattice and again the spectrum is invariant under individual $SO(D+16) \times SO(D)$ rotations of the left and right momenta. Thus the classical moduli space is
\be
{\cal M}_{\mbox{\sz class}} = \frac{SO(D+16,D)}{SO(D+16) \times SO(D)}   \label{eq2j2},
\ee
which is a $D(D+16)$-dimensional manifold as one would have guessed. Like in the bosonic case the T-duality group is the maximal discrete subgroup of the numerator of the classical moduli space (\ref{eq2j2}), namely:
\be
\Gamma_{\mbox{\sz T-duality}} = SO(D+16,D,\mathbb{Z})   \label{eq2k2}
\ee
and the quantum moduli space is the coset space obtained from the classical one by modding out this T-duality group.

\subsection{Type IIA and IIB superstring theory}
\label{twoatwob}

The world sheet action in a flat space time background for type IIA and IIB is given by
\be
S = \frac{1}{4 \pi \alpha'} \int d\tau d\sigma  \, \left( \partial_{\alpha} X^{\mu} \partial^{\alpha} X_{\mu} + \psi^{\mu}_{R} \partial_{+} \psi_{R \mu} + \psi^{\mu}_{L} \partial_{-} \psi_{L \mu} \right)   \label{eq2l2}
\ee
where we have introduced the notation $\partial_{\pm} = \frac{1}{2} \left( \partial_{\tau} \pm \partial_{\sigma} \right)$. The world sheet fermions $\psi_{R}$ and $\psi_{L}$ are right respectively left moving. Both of them can either obey periodic Ramond (R) or antiperiodic Neveu-Schwarz (NS) boundary conditions. The zero modes of $\psi_{L/R}$ in the R sector lead to a vacuum degeneracy such that the ground states transform according to one of the irreducible spinor representations of $SO(8)$, ${\bf 8_{\mbox{\sz s}}}$ or ${\bf 8_{\mbox{\sz c}}}$ distinguished by their chirality. The lowest lying state in the NS sector is a tachyon and the massless ground state transforms as a vector ${\bf 8_{\mbox{\sz v}}}$ under $SO(8)$. In both, the left and right moving spectrum, the GSO projection keeps only one of the irreducible spinor representations in the massless R sector and skips the tachyon in the NS sector. As the spectrum of the closed string is derived by tensoring the left and the right spectra and as the two choices ${\bf 8_{\mbox{\sz s}}}$ or ${\bf 8_{\mbox{\sz c}}}$ for each individual sector are physically equivalent, differing only by a space time parity transformation, there are two distinguished string theories, depending on whether the chirality of the fermions are the same in both sectors or not. In the first case we end up with the chiral type IIB theory, whose massless particle content is equal to the one of $N=2$ type IIB supergravity  in ten dimensions (namely $({\bf 8_{\mbox{\sz v}}} + {\bf 8_{\mbox{\sz s}}}) \otimes ({\bf 8_{\mbox{\sz v}}} + {\bf 8_{\mbox{\sz s}}})$). The other alternative leads to the nonchiral type IIA theory with the same massless spectrum as $N=2$ type IIA supergravity in ten dimensions ($({\bf 8_{\mbox{\sz v}}} + {\bf 8_{\mbox{\sz s}}}) \otimes ({\bf 8_{\mbox{\sz v}}} + {\bf 8_{\mbox{\sz c}}})$). Space time fermions are made by tensoring states from the left moving R sector with those of the right moving NS sector or vice versa, leading for type IIA to two gravitinos and dilatinos of opposite chirality and for type IIB to two gravitinos and dilatinos of the same chirality. 
\begin{table}[h]
\begin{center}
\begin{tabular}{l c}
IIA & $({\bf 8_{\mbox{\sz v}}} \otimes {\bf 8_{\mbox{\sz c}}}) + ({\bf 8_{\mbox{\sz s}}} \otimes {\bf 8_{\mbox{\sz v}}}) = ({\bf 8_{\mbox{\sz c}}} + {\bf 56_{\mbox{\sz c}}}) + ({\bf 8_{\mbox{\sz s}}} + {\bf 56_{\mbox{\sz s}}})$ \\ 
IIB & $({\bf 8_{\mbox{\sz v}}} \otimes {\bf 8_{\mbox{\sz s}}}) + ({\bf 8_{\mbox{\sz s}}} \otimes {\bf 8_{\mbox{\sz v}}}) = ({\bf 8_{\mbox{\sz s}}} + {\bf 56_{\mbox{\sz s}}}) + ({\bf 8_{\mbox{\sz s}}} + {\bf 56_{\mbox{\sz s}}})$
\end{tabular}
\end{center}
\caption{Fermionic massless spectrum} 
\end{table}
Bosons are obtained in the NSNS and the RR sector. The NSNS sector is universal and leads for both theories to the same states: ${\bf 8_{\mbox{\sz v}}} \otimes {\bf 8_{\mbox{\sz v}}} = {\bf 1} + {\bf 28} + {\bf 35}$, where the occuring states are in turn the dilaton, the antisymmetric tensor and the graviton, common to all string theories. The RR sector is different for both theories. It is shown in table \ref{tab5} (the superscript '$+$' at the 4-form in the IIB spectrum reminds at the fact that its field strength is self dual). 
\begin{table}[h]
\begin{center}
\begin{tabular}{l c}
IIA & $({\bf 8_{\mbox{\sz s}}} \otimes {\bf 8_{\mbox{\sz c}}}) = ({\bf 8_{\mbox{\sz v}}} + {\bf 56})$ \\ [2mm]
 &  \hspace{2.5cm} $A_{M}$  $A^{(3)}_{[MNP]}$  \\ [2mm]
IIB & $({\bf 8_{\mbox{\sz s}}} \otimes {\bf 8_{\mbox{\sz s}}}) = {\bf 1} + {\bf 28} + {\bf 35}$ \\ [2mm]
 & \hspace{2.8cm} $\Phi^{'}$ $A^{(2)}_{[MN]}$ $A^{+}_{[MNPQ]}$
\end{tabular}
\end{center}
\caption{Massless RR states}
\label{tab5} 
\end{table}  

\noindent 
Let us now have a look at the new features coming up in the context of
T-duality in type II superstrings \cite{DHS}. 
Consider first the case of one compact dimension. We have seen in the bosonic case that
T-duality reverses the sign of the compactified coordinate in the right moving
sector (\ref{eq2j}). This remains true in the type II string theory. Conformal invariance requires then also (as the
supercurrent $-\frac{1}{2} \psi_{R \mu} \partial_{+} X^{\mu}_{R}$ should be invariant)
\be
\psi^{9}_{R} \rightarrow -\psi^{9}_{R}.   \label{eq2m2}
\ee   
This however switches the chirality on the right moving side, i.e. ${\bf 8_{\mbox{\sz s}}} \leftrightarrow {\bf 8_{\mbox{\sz c}}}$, which had to be expected as we have already seen in the bosonic case that T-duality is a space-time parity operation on just one side of the worldsheet. Thus the relative chirality between the left and right moving sectors is changed and a T-duality transformation (in one direction) switches between type IIA and IIB:
\be
\mbox{IIA} \hspace{.3cm} \mbox{on $S^{1}\left( \frac{R}{\sqrt{\alpha'}} \right) $} \hspace{.3cm} \cong \hspace{.3cm}  \mbox{IIB} \hspace{.3cm} \mbox{on $S^{1}\left( \frac{\sqrt{\alpha'}}{R} \right) $}.   \label{eq2n2}
\ee
This remains true in higher dimensional torus compactifications if we T-dualize in an odd number of directions. On the other hand, if we T-dualize in an even number of directions, the relative chirality of the left and the right moving sectors is not changed and in this case T-duality is a selfduality (compare the discussion in section \ref{uduality}).

\subsection{Open strings}
\label{typeone}

\begin{figure}[h]
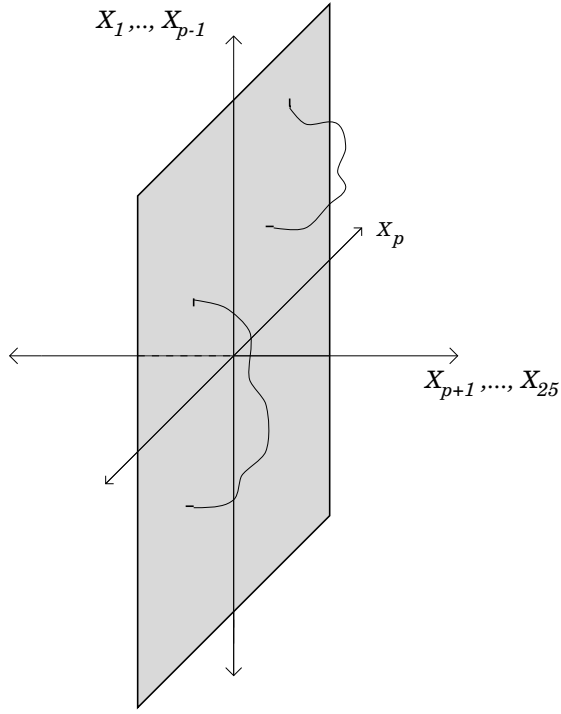

\begin{center}
\input db.pstex_t
\end{center}
\caption{Open strings moving on a Dp-brane}
\label{figdb}
\end{figure}

While type I string theory contains open superstrings most of the generic features we shall be interested in already appear with purely bosonic open strings (for which we
choose a parametrization $0 \leq \sigma \leq \pi$). For them there are two
kinds of boundary conditions possible: The Poincar\'e invariant Neumann
boundary conditions mean that there is no momentum flowing off the edges of
the string: 
\beqn
\left. \partial_{\sigma} X^{\mu} \right|_{\sigma = 0,\pi} = 0.
\eeqn
The ends of the string can however move freely in space time. Dirichlet
boundary conditions on the other hand break 26-dimensional Poincar\'e
invariance. They are given by 
\beqn
\left. \partial_{\tau} X^{\mu} \right|_{\sigma = 0,\pi} = 0 ,\quad \mbox{i.e.}\quad  \left. X^{\mu} \right|_{\sigma = 0, \pi} = c .
\eeqn
Now the
ends of the open string are fixed at position $c$. That both endpoints are
fixed at the same position becomes clear in (\ref{eq2q2}), but in fact the
endpoints of all open strings (not charged under a non abelian gauge group) with Dirichlet boundary conditions in $\mu$
direction are fixed to the same value, as can be deduced by considering a
graviton exchange of two open strings \cite{POL3}. It is of course possible to
choose different boundary conditions in different directions. In case the open
string obeys Neumann boundary conditions in the $(p+1)$ directions
$X^{\mu}, \mu = 0,\ldots,p$  and Dirichlet ones in the remaining $X^{i}, i = p+1, \ldots, 25$ its end points can just move within the $p$-dimensional plane in space transversal to the directions in which Dirichlet boundary conditions are valid (see fig. \ref{figdb}). This plane is called {\em Dp-brane}. 
Introduced in this manner, D-branes are just rigid objects in space time defined via the boundary conditions of open strings. It will be seen in the next chapter that they are in fact dynamical degrees of freedom with a tension $T_{p} \sim 1/g_{\mbox{\sz S}}$. Thus they only seem to be rigid at weak coupling but become dynamical in the strong coupling regime, hinting at the fact that they have to be identified with the semi solitons already announced in the introduction (section \ref{strongcoupling}). \\

Let us now investigate the role of D-branes for T-duality of open strings. Suppose we start with open strings obeying Neumann boundary conditions. Now we compactify one coordinate on a circle of radius $R$, say $X^{25}$, but keeping Neumann boundary conditions. The center of mass momentum in this direction takes only the discrete values $p^{25} = n/R$ like for the closed strings. In contrast to the closed string case however there is no analog of a winding state for the open string as its winding is topologically always trivial. The solutions of the equations of motion for the compactified left and right moving coordinates are\footnote{It is clear from (\ref{eq2g}) with $m=0$ that $p_{R}^{25} = p_{L}^{25} = \sqrt{\frac{\alpha'}{2}} p^{25}$; the additional factors of $2$ compared to (\ref{eq2f}) stem from the different parametrization of the open string world sheet.}:
\beqn
X^{25}_{R}(\sigma,\tau) & = & \frac{x^{25}}{2} - \frac{c}{2} + \alpha' p^{25} (\tau - \sigma) + i \sqrt{\frac{\alpha'}{2}} \sum_{l \neq 0} \frac{1}{l} \alpha^{25}_{l} e^{-2 i l (\tau - \sigma)} , \non
X^{25}_{L}(\sigma,\tau) & = & \frac{x^{25}}{2} + \frac{c}{2} + \alpha' p^{25} (\tau + \sigma) + i \sqrt{\frac{\alpha'}{2}} \sum_{l \neq 0} \frac{1}{l} \alpha^{25}_{l} e^{-2 i l (\tau + \sigma)}.  \label{eq2o2}
\eeqn
We see that the compactified coordinate moves with momentum $p^{25}=\frac{n}{R}$ on $S^{1}_{R}$:
\be
X^{25}(\sigma,\tau) = X^{25}_{L}(\sigma,\tau) + X^{25}_{R}(\sigma,\tau) = x^{25} + 2 \alpha' p^{25} \tau + \mbox{osc}.  \label{eq2p2}
\ee
As the radius is taken to zero only the $n=0$ mode survives and the open string seems to move only in 25 space-time dimensions but nevertheless still vibrates in 26 (or rather in the 24 transverse ones). This is similar to an open string whose endpoints are fixed at a hyperplane with 25 dimensions. \\

This fact can be better understood if one performs a T-duality transformation in the $X^{25}$ direction and introduces the T-dual coordinate $\hat{X}^{25}(\sigma,\tau) = X^{25}_{L}(\sigma,\tau) - X^{25}_{R}(\sigma,\tau)$. This choice is motivated by the fact that T-duality is a one sided space-time parity transformation on the right-moving coordinate (see eq. (\ref{eq2j}) and the comments in the corresponding paragraph). This coordinate on the T-dualized circle $\hat{S}^{1}_{R_{D}}$ with $R_{D} = \alpha'/R$ now takes the form

\beqn
\hat{X}^{25}(\sigma,\tau) = X^{25}_{L}(\sigma,\tau) - X^{25}_{R}(\sigma,\tau) & = & c + 2 \alpha' p^{25} \sigma + \mbox{osc}  \non
 & = & c + 2 \alpha' \frac{n}{R} \sigma + \mbox{osc}  \non 
 & = & c + 2 n R_{D} \sigma + \mbox{osc}.   \label{eq2q2}
\eeqn
Thus the boundary conditions have changed for the dual coordinate from Neumann
to Dirichlet ones, i.e. the end points of the string are fixed to the values
$\hat{X}^{25}(\sigma = 0) = c$ respectively $\hat{X}^{25}(\sigma = \pi) = c
\ \mbox{mod} \ 2\pi R_{D}$. Another way of saying
this is that the open string end points are constrained to move within
D$24$-branes which are seperated from each other by a multiple of $2\pi R_{D}$
and are therefore identified as we have compactified the dual coordinate on a
circle of radius $R_{D}$ (see fig. \ref{figdb2}). 
\begin{figure}[h]
\begin{center}
\input db2.pstex_t
\end{center}
\caption{$c=0 \ \mbox{mod}\ 2\pi R_{D}$}
\label{figdb2}
\end{figure}
Like for the closed string T-duality exchanges winding and momentum quantum numbers for the open string. Before T-dualizing the winding number has been zero. After the transformation the center of mass momentum is zero as can be seen from (\ref{eq2q2}), i.e. not only the end points but also the center of mass of the open string is constrained to move in a 24 (spatial) dimensional hyperplane. On the other hand, as the string end points are now fixed in the compactified dimension, it makes sense to talk about winding states. Actually these are states for which $n$ is nonzero in (\ref{eq2q2}) because $\hat{X}^{25}$ and $\hat{X}^{25} + 2 \pi n R_{D}$ are identified for any $n \in \mathbb{Z}$. Let us summarize this point. Before T-dualizing the open string ends move within a D$25$-brane wrapped around the compact dimension (which is an elegant way to say that they can move freely in space time). In the dual description however the D$25$-brane has changed to a D$24$-brane. This is a general feature: T-dualizing in a certain compact direction $X^{i}$ turns a Dp-brane which is wrapped around this circle (i.e. the open strings obey Neumann boundary conditions in the $X^{i}$ direction) into a D$(p-1)$-brane. The inverse is also true. If the Dp-brane is fixed in the $X^{i}$ direction before T-dualizing (i.e. the open strings obey Dirichlet boundary conditions along $X^{i}$) it is turned into a D$(p+1)$-brane which is wrapped around the compact $i^{th}$ dimension. \\

This fact is also crucial for the incorporation of D-branes and open strings in type II string theory. We will see in section \ref{charges} that Dp-branes with p even only appear in the type IIA theory and those with p odd exist in type IIB (see e.g. table \ref{tab1}). As was explained above T-duality in an odd number of directions switches from IIA to IIB theory and thus it is necessary for consistency that T-duality in one direction changes the value of p by plus or minus one.  \\
     
Next we consider the massless spectrum of the bosonic open string. In the dual picture these massless states are given by states without winding\footnote{This is true for generic values of $R_{D}$, but as in the closed string case extra massless states appear for the self dual radius $R_{D} = R = \sqrt{\alpha'}$.}. This can be seen from the mass formula 
\be
M^{2} = (p^{25})^{2} + \frac{1}{\alpha'} (N-1) = \left( \frac{n}{\alpha'} R_{D} \right)^{2} + \frac{1}{\alpha'} (N-1).   \label{eq2r2}
\ee
Thus the massless states are given by ($N=1$, $n=0$):
\be
\alpha^{\mu}_{-1}|0\rangle, \quad \alpha^{25}_{-1}|0\rangle ,   \label{eq2s2}
\ee
where the first state is a $U(1)$ gauge boson and the second one a scalar
whose VEV describes the $\hat{X}^{25}$ position of the D-brane in the dual
space. This easily generalizes to the case of a Dp-brane. Then
$\alpha^{\mu}_{-1}|0\rangle$ gives again a massless $U(1)$ gauge boson and the VEVs
of $\alpha^{i}_{-1}|0\rangle$ are the $(25-p)$ coordinates of the Dp-brane in the space directions transversal to it. As the open string has no momentum in these directions the same holds for the $U(1)$ gauge bosons. Thus it has become conventional to speak of the gauge theory living on the world volume of the Dp-brane (which amounts to a dimensionally reduced gauge theory). \\

So far we have only considered open strings charged under a $U(1)$ gauge
group. We now want to generalize the results to non abelian gauge groups. To do
so let us consider orientable open strings whose end points carry charges
under a non abelian gauge group. Consistency requirements restrict the choice
to $U(n)$ for orientable strings (and $SO(n)$ respectively $USp(n)$ for
non orientable ones) \cite{GSW}. To be more precise one end transforms under
the fundamental representation ${\bf n}$ of $U(n)$ and the other one under its
complex conjugate ${\bf \bar{n}}$. The ground state wavefunction is thus not only
specified by the center of mass momentum of the string but additionally by the
charges of the end points, giving rise to a basis $|k;ij\rangle$ ({\em Chan Paton
  basis}). Generally the open string
states can be characterized by their charges with respect to the generators of
the Cartan subalgebra $U(1)^{n}$ of $U(n)$, which can be taken as the $n$
different $n \times n$ matrices with just one entry $1$ on the diagonal. The
states $|k;ij\rangle$  of the Chan Paton basis are now those states which carry
charge $+1$ respectively $-1$ under the $i^{th}$ respectively $j^{th}$
$U(1)$ generator. Obviously the whole open string transforms as a ``bifundamental'' under the tensor product
$\bf{n} \otimes \bar{\bf{n}}$ which is just the adjoint of $U(n)$. From this point of
view it seems more appropriate to take as basis for the ground state
wavefunction the combinations $|k;a\rangle = \sum_{i,j=1}^{n} |k;ij\rangle
\lambda^{a}_{ij}$. The factors $\lambda^{a}_{ij}$ are the matrix elements of
the $U(n)$ generators and are called {\em Chan Paton
  factors}. The $|k;a\rangle$ transform under the adjoint of $U(n)$ if the $i$ index transforms with $\bf{n}$ and the $j$ index with $\bar{\bf{n}}$. The usual vector at the massless level $\alpha^{\mu}_{-1} |k,a\rangle$ is now a $U(n)$ gauge boson. \\

The new feature in toroidal compactification of the open string with gauge
group $U(n)$ is the possible appearance of Wilson lines (see footnote
\ref{fn1}). If we compactify the $25^{th}$ dimension on a circle with radius $R$ a possible background field with non trivial line integral along this circle is 
\be
A^{25} = \frac{1}{2\pi R} \mbox{diag} \left(\theta_{1}, \ldots , \theta_{n}
\right) = -i \Lambda^{-1} \partial_{25} \Lambda \, , \quad \Lambda =
\mbox{diag} \left( e^{i X^{25} \theta_{1} / 2 \pi R} , \ldots , e^{i X^{25} \theta_{n} / 2 \pi R} \right).   \label{eq2t2}
\ee
If $\theta_{i}=0$ (or all equal to another constant value) for $i=1,\ldots ,m$ and $\theta_{j} \neq 0$ (and all pairwise different) for $j=m+1,\ldots ,n$ the gauge group is broken to $U(n) \rightarrow U(m) \times U(1)^{n-m}$ (compare our discussion of the Wilson lines of the heterotic string in section \ref{heterotic}). Thus the $\theta_{i}$ play the role of Higgs fields. Another important characteristic of the Wilson line is that it changes the value of momentum $p^{25}$, where the change depends on the Chan Paton quantum numbers of the state. Therefore we use the Chan Paton basis in the following. In particular we have for a ground state $| l/R;ij\rangle$ 
\be
p^{25}_{(ij)} = \frac{l}{R} + \frac{\theta_{j} - \theta_{i}}{2 \pi R}.   \label{eq2u2}
\ee
To see this consider first the case of a $U(1)$ gauge theory and a point
particle with charge $q$. Under a gauge transformation with $\Lambda^{-1}$ the
background gauge field vanishes and one knows from ordinary quantum mechanics
that the wavefunction of the particle picks up a phase 
\beqn
\exp \left( -iq \int_{x^{25}_{0}} dx^{25} A^{25}  \right) ,
\eeqn
where $x^{25}_0$ is a reference
point. Thus it is no longer periodic under $X^{25} \rightarrow X^{25} + 2 \pi R$
but gets a phase $\exp(-iq\theta)$. As the wavefunction in the momentum
representation is a plane wave, this non periodicity is equivalent to a shift in the canonical momentum 
\be
p^{25} \rightarrow \frac{l}{R} - \frac{q \theta}{2 \pi R}.   \label{eq2v2}
\ee
Let us now turn to string theory. The background gauge field (\ref{eq2t2}) is an element of the Cartan subalgebra, i.e. of the subgroup $U(1)^{n}$ of $U(n)$. The string states with Chan Paton quantum numbers $|ij\rangle$ have charges $+1$ $(-1)$ under the $i^{th}$ $(j^{th})$ $U(1)$ factor and are neutral with respect to the others. Thus (\ref{eq2u2}) follows immediately from (\ref{eq2v2}). If we insert (\ref{eq2u2}) into (\ref{eq2r2}) we get 
\be
M^{2} = \frac{(2 \pi l + \theta_{j} - \theta_{i})^{2}}{4 \pi^{2} R^{2}} + \frac{1}{\alpha'} (N-1),   \label{eq2w2}
\ee
from which we see that for generic $\theta$ values the only massless states are the diagonal ones ($i=j$) giving a gauge group $U(1)^{n}$ (for $l=0$ and $N=1$). If some of the $\theta$'s are equal we get additional massless states enhancing the gauge symmetry and confirming our discussion above. \\

\begin{figure}[h]
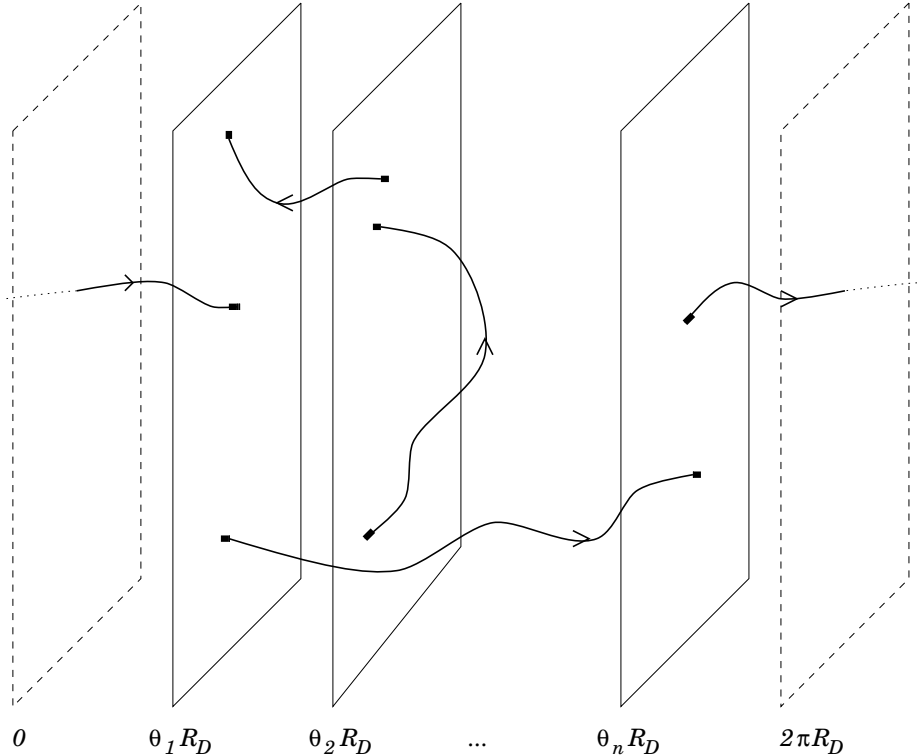

\begin{center}
\input dbcp.pstex_t
\end{center}
\caption{D-brane configuration in the presence of a Wilson line}
\label{figdbcp}
\end{figure}

Now we want to interpret the situation in the dual picture. From (\ref{eq2q2}) and (\ref{eq2u2}) we see that the dual coordinate for a string with Chan Paton labels $|ij\rangle$ is given by
\be
\hat{X}^{25}_{(ij)} (\sigma, \tau) = c + \left( 2l + \frac{\theta_{j} - \theta_{i}}{\pi} \right) R_{D} \sigma + \mbox{osc}.   \label{eq2x2}
\ee
If we set $c = \theta_{i} R_{D}$ we get 
\beqn
\hat{X}^{25}_{(ij)}(\sigma=0, \tau) & = & \theta_{i} R_{D}  , \non
\hat{X}^{25}_{(ij)}(\sigma=\pi, \tau) & = & 2 \pi l R_{D} + \theta_{j} R_{D}.   \label{eq2y2}
\eeqn
Thus we have $n$ different D-branes, whose positions (modulo $2 \pi R_{D}$) are given by $\theta_{j} R_{D}$ (see fig. \ref{figdbcp}). From (\ref{eq2w2}) we see that generically only open strings with both end points on one and the same D-brane yield massless gauge bosons (gauge group $U(1)^{n}$) and strings which are stretched between different D-branes give massive states with masses $M \sim (\theta_{i} - \theta_{j}) R_{D}$. Obviously the masses decrease with smaller distances and vanish if the two D-branes take up the same position. In this picture the coincidence of D-brane positions leads to the encountered gauge symmetry enhancement. In particular if all D-branes are stuck on top of each other the gauge group is $U(n)$. \\

A feature of D-branes which we just mention for later use but without
proof is that they break (part of the) supersymmetry (see e.g. \cite{POL3,BA}). In the special situation above all the $n$ D-branes were parallel
to each other. In this case half of the supersymmetries are broken (leading to
$N=4$ in $D=4$ for type II string theory). To break even more supersymmetries one needs more general
configurations of intersecting D-branes. Two orthogonal D-branes for example
break altogether $3/4$ of supersymmetry (giving $N=2$ in $D=4$) if the
number $d_{\bot}$ of dimensions in which just one of the two branes is
extended (but not both) is $4$ or $8$ (examples are a D$4$-brane together
with a D$0$-brane or two completely transverse D$2$-branes). If $d_{\bot}$
is however $2$ or $6$ (it is always even) all of the supersymmetry is
broken. In situations with rotated D-branes it is also possible to get $N=1$
supersymmetry in $D=4$ (i.e. $1/8$ of the SUSY generators are
preserved). For some further details the reader is referred to \cite{POL3}.

\section{Non perturbative phenomena}
\label{nonperturbative}

In the previous chapters interactions in string perturbation theory were defined by a summation
over all conformally inequivalent world sheets, each  weighted with a factor
of the string coupling constant according to the topology of the respective
Riemann surface. In this
chapter we are going to look for explicit non perturbative states in string
theory to get a more complete description of the spectrum and possible
interactions. One strategy to find states of non perturbative nature is to
look for non trivial solutions of the classical equations of motion of the low
energy approximation to string theory. We will find these equations to have
brane solutions which depend only on a subset of the coordinates of
space-time, such that the sources of the fields are multidimensional
membranes. They are solitonic in the sense that their masses are proportional
to inverse powers of the string coupling $M^2_{\mbox{\scriptsize sol}} \sim
1/g^2_{\mbox{\scriptsize S}}$, while the states of the perturbative spectrum
have masses proportional to the string coupling constant:
$M^2_{\mbox{\scriptsize per}} \sim g^2_{\mbox{\scriptsize S}}$. They are then called \emph{p-branes}, if they extend into $p$
spatial dimensions, thus have $(p+1)$-dimensional worldvolume. Further we shall demonstrate that also the D-branes discussed earlier as hyperplanes in
space-time, on which open strings might end with Dirichlet boundary
conditions, display characteristic features of solitonic states. By an
indirect way of reasoning we shall argue that they are in fact dynamical
objects, i.e. they interact with open strings, thereby couple to gravity and gauge fields, and
fluctuate in shape and position. Their perturbative degrees of freedom are
open strings ending on the brane, which describe the fluctuations of the
D-brane by their perturbative spectrum. Also we shall find their mass spectrum
to be of intermediate range, $M^2_{\mbox{\scriptsize D}} \sim
1/g_{\mbox{\scriptsize S}}$, which indicates that they are in between elementary perturbative states and solitons and therefore caused them being adressed as half-solitonic. The three types of states, perturbative, solitonic and half-solitonic, are mixed by various (conjectured) S, T and U dualities from the previous chapters. They leave the spectrum invariant but transform the moduli, thereby in some cases interchanging the perturbative and non perturbative regimes of the theory. \\

\subsection{Solitons in field theory}
\label{solfield}
 
To motivate later discussions and interpretations, we first give an introduction into field theoretical solitons, in particular the black holes of Einstein gravity and the t`Hooft-Polyakov monopole of non abelian gauge theories. Such solitons are defined to be non trivial solutions to the field equations with finite action. Regarding the obvious symmetries of their spectrum, we then point out the idea of S-duality and how it is supposed to be realized in supersymmetric gauge theories. The bridge to the quantum theory will be built by realizing the BPS nature of some of the classical solutions after embedding these into supersymmetric theories, which is assumed to guarantee their existence in the spectrum of the quantum theory, too. \\

\subsubsection{Black holes}
\label{bh}

We start with discussing classical solutions to the Einstein equations that are not of perturbative nature as they involve large deviations from the Minkowski flat space-time. The classical theory of general relativity originates from the vacuum Einstein action
\beqn
S_{\mbox{\sz E}} = \int{d^4x\ \sqrt{-g} R}
\eeqn
of pure gravity, where $g$ is the determinant of the metric $g_{\mu \nu}$ and $R$ the curvature scalar. By the usual methods one then finds the Einstein equation, the equation of motion of the metric field without matter, which (in $d > 2$) simply demands the space-time to be Ricci flat:
\beqn
R_{\mu \nu} = 0.
\eeqn
Perturbative solutions to these equations are for instance given by gravitational waves, freely propagating, small deviations from the flat metric. The most prominent among the non-per\-tur\-ba\-tive solutions are black holes \cite{T1} and the prototype of such is the Schwarzschild metric
\beqn
ds^2_{\mbox{\sz S}} = - \left( 1- \frac{2m^\prime}{r} \right) dt^2 + \left( 1- \frac{2m^\prime}{r} \right)^{-1} dr^2 + r^2 \left( d\theta^2 + \sin^2 \left( \theta \right) d\phi^2 \right) .
\eeqn
It is the unique stationary solution of the vacuum Einstein equations outside
a star that has collapsed to a pressureless fluid of matter. The parameter
$m^\prime$ is related to the fourdimensional Newton constant $G_{\mbox{\sz
    N}}$ by $m^\prime = G_{\mbox{\sz N}} m/c^2$, $m$ being the mass of the
black hole, which is derived by comparing the asymptotic large $r$ behaviour
of the $g_{00}$ component to the classical non-relativistic Newton law. The
Schwarzschild metric suffers from two obvious divergencies. First the event
horizon at $r=2m^\prime$ which is only a coordinate divergence and involves
the change of the metric signature $(-+++) \rightarrow (+-++)$, in a way the
interchange of radial and time directions, while the curvature and even
$R_{\mu \nu \rho \sigma}R^{\mu \nu \rho \sigma}$ stay finite. When observed
from the radial infinity, where the metric tends to flat Minkowski space-time,
no geodesic can reach the horizon at finite values of the affine parameter
which can in case of timelike geodesics be taken to be its proper time. 
There is also the physically more dramatic divergence at $r=0$, where the
square of the curvature tensor $R_{\mu \nu \rho \sigma} R^{\mu \nu \rho
  \sigma}$ diverges as an inverse power of the radial coordinate. Such
divergencies cannot be removed by conformal transformations, they imply that
geodesics, particle trajectories, cannot be completed into these points, which
is said to be a generic feature of any gravitational collapse. Concerning the
global causal structure, the horizon signals the notion, that all timelike or
null geodesics from inside $r<2m^\prime$ are leading into the spacelike
singularity, therefore no matter can escape its destiny of falling into the
singularity at the core, which occurs even in finite proper time. Generalizations of the Schwarzschild solution are given by the Reissner-Nordstr\"om solution for charged black holes, obtained by adding a Maxwell term $-\frac{1}{4} F^2$ to the vacuum action, and the Kerr solution that incorporates also rotating black holes and is the most general form of any stationary solution to the Einstein equations of the vacuum plus only electromagnetism. These metrics are related to more complicated global structures of space time and display different types of curvature and coordinate singularities. As we shall later on refer to the charged black holes, we here give the solutions for the metric and the electromagnetic potential of the Reissner-Nordstr\"om case:
\beqn
ds^2_{\mbox{\sz RN}} &=& -\left( 1- \frac{2m^\prime}{r} + \frac{q^2}{r^2} \right) dt^2 + \left( 1- \frac{2m^\prime}{r} + \frac{q^2}{r^2} \right)^{-1} dr^2 + r^2 \left( d\theta^2 + \sin^2 \left( \theta \right) d\phi^2 \right) ,\non 
A^\mu &=& \delta^{0 \mu} \frac{q}{r}.
\eeqn
Obviously $q$ can be interpreted as the total electric charge of the black
hole. Further analysis shows, that only in the case $m^\prime \geq q$ the
curvature singularity at the radial origin is screened by a horizon. If $m^\prime > q$ the metric actually has two horizons
\beqn
r_\pm = m^{\prime} \pm \sqrt{m^{\prime 2} -q^2}
\eeqn
which can both be removed by analytic extension. Different from the Schwarzschild case, the singularity itself is timelike and the causal structure more involved. The case of an extremal charged black hole is reached if $m^\prime =q$ which demands the horizons to coincide and the metric simplifies to 
\beqn
ds^2_{\mbox{\sz ex}} = - \left( 1- \frac{2m^\prime}{r} \right)^2 dt^2 + \left( 1- \frac{2m^\prime}{r} \right)^{-2} dr^2 + r^2 \left( d\theta^2 + \sin^2 \left( \theta \right) d\phi^2 \right) .
\eeqn  
An important property of this situation is its vanishing surface gravity 
\beqn
\kappa_\pm = \frac{r_\pm - r_\mp}{2r^2_\pm} =0,
\eeqn
which is a measure of the local accelaration that appears to affect a particle
near the horizon, as it is being watched from infinity. Note that the causal
structure of space-time of this solution still is differing from the
Schwarzschild solution, though the metric might look somewhat similar. Besides
the differences mentioned all black hole solutions have got the feature in
common, that (as long as the positive energy hypothesis holds), any curvature
singularity is shielded by a horizon which, viewed from spatial infinity
(expressed in asymtotically flat coordinates), cannot be reached by any
particle (any space- or timelike geodesic) in finite time. \\

We shall now introduce concepts of particle physics and thermodynamics into this classical picture to explain the related problems of the loss of information and Hawking radiation. As there is no complete quantum theory available that allows a consistent treatment of the gravitational interactions together with the strong and electroweak interactions, we have to rely on semiclassical methods and sometimes heuristic arguments. Let us first look for the state equation of black hole thermodynamics. Even the most general black hole metric, the Kerr black hole, does not depend on the time and azimutal angle coordinate $\phi$, such that these define vector fields along which the action is constant, i.e. Killing vector fields:
\beqn
\eta \equiv \frac{\partial}{\partial t}, \quad \zeta \equiv \frac{\partial}{\partial \phi} .
\eeqn
While in a curved background one cannot simply integrate the field equations over some space-like region to get conserved quantities, as usually done in field theory on Minkowski flat space-time, the projections of the energy momentum tensor onto such Killing vector fields are conserved. Thus to any Killing vector field $\xi$ one can associate a conserved charge by integrating its covariant derivative over the boundary of some spacelike region $V$ 
\beqn
Q_\xi (V) \equiv \frac{1}{16\pi G} \oint_{\partial V}{d\sigma_{\mu \nu}\ D^\mu \xi^\nu} = \frac{1}{8\pi G} \int_V{d\sigma_\mu\ D_\nu D^\mu \xi^\nu} = \frac{1}{8\pi G} \int_V{d\sigma_\mu\ R^{\mu \nu} \xi_\nu} .
\eeqn
In the presence of a black hole the volume integral is conveniently split off into a surface integral over the horizon $H$ and a volume integral over the space time $ST$ outside:
\beqn
Q_\xi (V) = \frac{1}{8\pi G} \int_{ST}{d\sigma_\nu \ D_\mu D^\nu \xi^\mu} + \frac{1}{16\pi G} \oint_{H}{d\sigma_{\mu \nu}\ D^\mu \xi^\nu}  .
\eeqn
The two Killing vector fields $\zeta$ and $\eta$ in particular produce conserved charges that contain besides other terms the total energy (mass) $m^\prime$ and the angular momentum $J$ of the Kerr black hole. Using the explicit expressions one can derive the mass formula of Smarr:
\beqn
m^\prime = \frac{\kappa}{4\pi} A + 2\Omega_{\mbox{\sz H}} J + \Phi_{\mbox{\sz H}} q ,
\eeqn
where $A$ is the area of the
horizon, $\Omega_H$ the constant angular velocity at the horizon, $J$ the
total angular momentum, $\Phi_H$ the electric potential at the horizon. We
notice that the extremal Reissner-Nordstr\"om solution with vanishing $\kappa$
and $\Omega_H$ has mass equal to its electric potential energy: $m^\prime =
\Phi_{\mbox{\sz H}} q$. This will be identified to be the BPS mass bound later
on. By analyzing the dimensionality of the various quantities (c.f. Euler theorem on homogeneous functions), one further finds the differential state equation to be: 
\beqn 
dm^\prime = \frac{\kappa}{8\pi} dA + \Omega_{\mbox{\sz H}} dJ + \Phi_{\mbox{\sz H}} dq ,
\eeqn
which is the so called first law of black hole mechanics. Interpreting this as a thermodynamical relation, suggests to define the black hole (Hawking) temperature $T_{\mbox{\sz H}} \equiv \kappa /2\pi$ and its (Bekenstein-Hawking) entropy $S_{\mbox{\sz BH}} \equiv A/4$. It is in fact clear that the black hole must carry entropy anyway, as throwing matter into it would otherwise destroy such. (On the other hand no such process could be watched from the asymptotically flat region.) The Hawking temperature we have recovered here is also known from other arguments, which explains the seemingly arbitrary choice of numerical coefficients. The second law of thermodynamics now translates into the statement that the horizon of an isolated black hole has non-decreasing area. One should suspect that this black hole entropy can be computed by a state counting procedure as usual in quantum statistics. This task has never been solved in the context of QM or QFT, as the quantum degrees of freedom of the black hole are still unknown. We shall later address this question from the point of view of string theory. \\

We now briefly turn to Hawking radiation \cite{HAW}, which refers to the fact that black holes can radiate particles exhibiting a thermal spectrum. The formalism of quantum field theory on curved space-time allows under certain circumstances, as are met when black holes are formed, non unitary changes of the Fock space basis. From this it follows that the vacuum before the collapse may look like a many particle state afterwards. Along these lines one can deduce the claimed statement that black holes in fact radiate. In a more heuristic manner it can be figured that after pair-creation of a particle and an antiparticle by vacuum fluctuations for instance of the photon field near the horizon, one of the two particles is drawn inside the horizon, while the other one appears to be radiated away from the black hole. By a semiclassical analysis of a field propagating along a trajectory close to the future event horizon Hawking could demonstrate that the spectrum of this radiation is approximately that of a black body thermal radiation at the Hawking temperature $\kappa /2\pi$:
\beqn
n(\omega ) = \frac{1}{e^{2\pi \omega /\kappa} -1} .
\eeqn
The black hole obviously can thermalize its entropy and energy by this mechanism. A very special case is the extremal Reissner-Nordstr\"om black hole with $\kappa=0$, so that no Hawking radiation is emitted and it appears to be stable. Although this analysis does not take into account the effect of decreasing the mass of the black hole in the process (back reaction) it is convenient to cite the riddle of loss of information when throwing matter into the hole that later on radiates away thermalized \cite{GID}. The formerly pure quantum state of such matter has lost its coherence. A proper unitary description of the whole phenomenon is expected to clarify the question but it still remains to be found. Apparently it should have to include a consistent quantum treatment of the gravitational field, an outstanding challenge so far. Later on we shall reveal some of the attempts that have been undertaken in string theory recently. \\

\subsubsection{Magnetic monopoles}
\label{magneticmonopoles}

The second type of solitonic objects we shall discuss are the magnetic monopoles of gauge theories. We will proceed very much like in the previous section by looking for classical solutions to the field equations, which have finite energy densities and thus have possibly relevant contributions to the quantum theoretical path integral. The magnitude of these contributions will depend on the coupling and is in general non perturbative. While we could not give a rigorous answer to the question, if black holes are stable states of the (unknown) quantum theory, this issue can be addressed in gauge theory, at least in some supersymmetric extension, and leads to the notion of so called BPS saturated states. We now start with the most simple setting, classical electrodynamics and its Maxwell equations. \\
 
The idea of magnetic monopoles traces back to Dirac who figured out what happened when symmetrizing the Maxwell equations 
\beqn
\vec \nabla \vec{E} = \rho_{\mbox{\sz e}}, & & \quad \vec \nabla \times \vec{B} - \frac{\partial}{\partial t} {\vec{E}} = \vec{j}_{\mbox{\sz e}} ,\non
\vec \nabla \vec{B} = 0, & & \quad \vec \nabla \times \vec{E} + \frac{\partial}{\partial t} {\vec{B}} = 0 
\eeqn
by generalizing the electromagnetic duality
\beqn
\vec{E} \rightarrow \vec{B}, \quad \vec{B} \rightarrow -\vec{E}  \label{eqelmagdual}
\eeqn
of the vacuum equations to a symmetry of the full classical electromagnetic theory. This is possible by introducing magnetic currents $j^\mu_{\mbox{\sz m}}$ into the equations. The corresponding particles are called magnetic monopoles. Such a single stationary and point-like monopole of magnetic charge $g$ at the origin generates a singular magnetic field 
\beqn
\vec{B} = g \frac{\vec{r}}{4\pi r^3}
\eeqn
which itself is created by a vector potential, that cannot be defined globally. In fact it is singular at least on a half line, the so called Dirac string. More mathematically speaking, the Bianchi identity of the field strength tensor is no longer satisfied and thus it cannot be exact globally anymore. The presence of such a Dirac string, although classically meaningless, can be probed by a Bohm-Aharonov experiment. Moving an electron around this string and demanding its wave function to be single valued imposes the famous Dirac quantization condition on the product of the two charges:
\beqn \label{dirac}
ge = 2\pi n ,
\eeqn
where $n$ is some arbitrary integer. It offers an explanation for charge quantization, even if only by postulating objects yet unobserved. Another way of deriving it \cite{WY} is finding that the vector potentials defined on different parts of a sphere around the magnetic charge can only be matched together on the whole sphere up to gauge transformations 
\beqn
A_\mu \rightarrow A_\mu + \partial_\mu \chi
\eeqn
in the overlapping regions, where $\chi$ is only defined modulo $2 \pi g$,
while the transition function $\exp (ie\chi )$ should be defined uniquely. For
$n \geq 1$ the Dirac condition tells us that one of the two charges is
obviously larger than $1$ and therefore naiv perturbation theory as an
expansion in this parameter impossible. The apparent symmetries of this
generalized setting of electromagnetism allow an electromagnetic duality
interchanging the electric and magnetic fields like in (\ref{eqelmagdual})  and simultaneously the two currents $j^\mu_{\mbox{\sz e}} \leftrightarrow j^\mu_{\mbox{\sz m}}$ and thereby linking the perturbative and non perturbative regimes. This might be thought of as the most simple case of S-duality.\\

We shall pursue this idea further by inspecting classical Yang-Mills gauge theory which will then be embedded in its supersymmetric extension. The challenge to find an exact solution to a phenomenologically interesting classical field theory, that exhibits properties of a particle, i.e. has local support in space and is constant in time, was solved by the t`Hooft-Polyakov \cite{THO,POL} solution of the $SU(2)$ Yang-Mills-Higgs (or Georgi-Glashow) model. Such solutions are called solitons and they carry properties of monopoles. The basic idea is to embed the electromagnetic $U(1)$ gauge group in a larger simple group which is spontaneously broken to single out the electromagnetic $U(1)$ as unbroken symmetry, but now with charge quantization necessarily implicit. The model is constructed from the Lagrangian
\beqn
{\cal L} = -\frac{1}{4} \tr\ F^2 + \frac{1}{2} D^\mu \Phi^a D_\mu \Phi^a - V(\Phi) ,
\eeqn
where the gauge field $A^a_\mu$ and the scalar Higgs field $\Phi^a$ take their values in the adjoint representation of the $SU(2)$ gauge group. The potential of the Higgs field is assumed to create a non-vanishing expectation value alike the double well potential. The derivation of the equations of motion and the energy momentum tensor is straightforward, the former are
\beqn
D_\mu F^{a \mu \nu} &=& e \eps^{abc} \Phi^b D^\nu \Phi^c ,\non
\left( D_\mu D^\mu \Phi \right) ^a &=& - \frac{\partial V(\Phi)}{\partial \Phi^a}
\eeqn
and the latter comes out in the symmetrized version:
\beqn
\Theta^{\mu \nu} = - F^{a\mu \rho} F^{a\nu}_{\rho} + D^\mu \Phi^a D^\nu \Phi^a - \eta^{\mu \nu} {\cal L}.
\eeqn
From its time component one can read off the energy density and realizes that the classical vacuum has a constant Higgs field with vacuum expectation value given by the minimum of the potential. Choosing 
\beqn
V(\Phi) = \frac{\lambda}{4} \left( \Phi^a \Phi^a -v^2 \right) ^2 ,
\eeqn
this is $\langle \Phi^a \Phi^a \rangle = v^2$. The constant Higgs field leaves only a smaller gauge group $U(1)$ of rotations around the direction of its expectation value intact and the manifold of degenerate vacua (the coset space) is topologically equivalent to a sphere:
\beqn
{\cal M} _{\mbox{\sz vac}} = SU(2)/U(1) \simeq S^2 .
\eeqn 
Any solution to the field equations that wants to have finite energy must
locally look like a vacuum field configuration at spatial infinity. Therefore
at large $r$ any Higgs field has to obtain the vacuum value $\Phi^a \Phi^a
=v^2$. This relation defines the sphere ${\cal M}_{\mbox{\sz vac}}$ in the
isospin space, whose points are connected by gauge transformations of the
coset group, rotations in that space. Thereby the asymptotic value of the
Higgs field induces a mapping of the spatial infinity $S^2$ onto the set of
degenerate vacua $S^2$, which can be characterized by its integer winding
number $n$ and is an element of the second homotopy class of $S^2$:
\beqn
\pi_2 \left( S^2 \right) = \mathbb{Z}.
\eeqn
The constant Higgs field belongs to $n=0$, but we shall also find solutions to
the field equations with finite energy that have non-vanishing winding
number. We construct the concrete form of such a vector potential leading to a
monopole solution by demanding the energy density to be decreasing faster than
$1/r^2$ when $r\rightarrow \infty$ to obtain a finite total energy. Consequently the covariant derivative of the Higgs field has to decrease faster than $1/r$ and from 
\beqn
\partial_\mu \Phi^a + e \eps^{abc} A_\mu^b \Phi^c \rightarrow 0 \quad \mbox{ faster than 1/r} 
\eeqn
one can find the most general expression for the gauge potential
\beqn
A_\mu^a = -\frac{1}{ev^2}\eps^{abc} \Phi^b \partial_\mu \Phi^c +\frac{1}{v} \Phi^a A_\mu
\eeqn
with some arbitrary smooth $A_\mu$. The corresponding field strength points into the direction of the Higgs field:
\beqn
F^{a\mu \nu} = \frac{\Phi^a}{v} \left( \partial^\mu A^\nu - \partial^\nu A^\mu - \frac{1}{ev^3} \eps^{def} \Phi^d \partial^\mu \Phi^e \partial^\nu \Phi^f \right) .
\eeqn
Integrating the field equations over the transverse space we then find the magnetic charge of the monopole being proportional to the winding number $n$ of the Higgs field around the sphere at radial infinity 
\beqn
g = -\frac{1}{2} \int_{S^2}{\eps^{ijk} F_{jk} dx_i} = \frac{1}{2ev^3} \int_{S^2}{\eps^{ijk}\eps^{abc} \Phi^a \partial_j \Phi^b \partial_k \Phi^c \ dx_i } = \frac{4\pi n}{e} ,
\eeqn
which explains its being called a topological charge in contrast to the electric Noether charge. The Dirac quantization condition is thus reproduced up to a factor $1/2$, which is due to the fact, that the fermions we could possibly add into the model were half integer charged, as they were to lie in the fundamental representation of the $SU(2)$. That two seemingly very different arguments reproduced the same quantization condition can be traced back to the topological statement that 
\beqn
\pi_2 \left( SU(2)/U(1) \right) = \pi _1 \left( U(1) \right) .
\eeqn
The lefthandside of the equation gives the range of winding numbers for the Higgs, whereas the righthandside is the possible number of windings of the electromagnetic gauge field $\chi$ around the Dirac string. In the latter case we had to plug in the source term by hand to spoil the Bianchi identity, in the former case this is achieved by the winding of the Higgs field automatically. \\

The general solution for the electromagnetic potential allows a very 
special and symmetric choice \cite{PSO}. On the one hand the field configurations can neither be gauge invariant under the full $SU(2)$ nor be invariant under the $SO(3)$ of spatial rotations because of the spontaneous symmetry breaking and the non trivial behaviour of the Higgs field at radial infinity. On the other hand any spatial rotation can be accompanied by an appropriate gauge transformation to cancel the variation of the fields, as the angular dependence of the Higgs field at radial infinity is locally pure gauge. Such we can have solutions which are invariant under simultaneous rotations and gauge transformations. The arbitrariness that remains can be summarized in the according ansatz 
\beqn
\Phi^a &=& \frac{\bar{r}^a}{er} H(ver) ,\non
A^a_i &=& -\eps^a_{ij} \frac{\bar{r}^j}{er} \left( 1- K(ver) \right), \non
A^a_0 &=& 0, 
\eeqn
where $\bar{r}^a$ is the radial unit vector in the isospin space and $\bar{r}^i$ in the Minkowski space. To render charge and mass finite one has to impose boundary conditions on $H(r)$ and $K(r)$ that allow finite results for the integrated energy-momentum tensor as well as for the integration of the field equation of the electromagnetic field. Defining mass and charge by such integrals over spatial regions in the usual manner one can derive the BPS (Bogomolny-Prasad-Sommerfield) bound on the lowest possible mass of a monopole:
\beqn
M_{\mbox{\sz m}} \geq vg 
\eeqn  
with equality if and only if $V(\Phi) =0$. 
(Note as an aside that extremal Reissner-Nordstr\"om black holes 
are indeed BPS saturated.) We shall later see that from the point of 
view of the supersymmetric extension of this model the states that 
satisfy the equality are very special. In this case the equations of 
motion can be translated into the differential (Bogomolny) equations \cite{BOG}
\beqn
x \frac{dK(x)}{dx} = -K(x) H(x) , \quad x \frac{dH(x)}{dx} = H(x)-\left( K(x)^2 -1 \right) 
\eeqn
which are solved according to the relevant boundary conditions by 
\beqn
H(x) = x \coth(x) -1 , \quad K(x) = \frac{x}{\sinh(x)} .
\eeqn
Plotting the two functions, one immediately realizes how they generalize the well known onedimensional (topologically charged) ``kink'' solution. Substituting back one finds that these mo\-no\-poles satisfy the BPS mass bound with equality holding as well as the Dirac charge quantization condition with minimal winding number (topological charge) $4\pi =eg$, so that we are tempted to think of them as being the elementary magnetic charges, stable per definition, if charge conservation holds. It was also noticed that the same Yang-Mills-Higgs model can have solutions that carry both electric and magnetic charges \cite{JZ}. If one also allows for the topological term 
\beqn
{\cal L} \rightarrow {\cal L} - \frac{\theta e^2}{32\pi^2} \ ^* \hspace{-0.07cm} F_{\mu \nu} F^{\mu \nu} 
\eeqn
to be present, where the star marks the Hodge dual, their magnetic charge can be related to the $\theta$-angle, the imaginary part of the gauge coupling \cite{WI1}. This term is proportional to the instanton number, it violates parity and locally it can be written as a total derivative. Globally it might lead to an additional violation of the Bianchi identity and an extra boundary term in the field equation, modifying the electric charge of the state. Together we have then got two topological charges, the winding $n_{\mbox{\sz e}}$ of the gauge field and the winding $n_{\mbox{\sz m}}$ of the Higgs field, enough to have dyonic states carrying both types of charges. The Dirac condition and the BPS bound are modified to be
\beqn
\frac{q}{e} &=& n_{\mbox{\sz e}} - \frac{n_{\mbox{\sz m}} \theta}{2\pi} ,\non
M_{\mbox{dyon}} &\geq & v \sqrt{q^2 + g^2} ,
\eeqn
where $n_{\mbox{\sz m}}$ is related to the magnetic charge $g$ of the dyon by
the former Dirac condition, while $q$ is the electric charge of the dyon, now
no longer being quantized in integer units of $e$. The above discussion was a
little sketchy and the situation appears to have become more complicated than
before introducing dyons. We shall return to it from another point of view
after looking at supersymmetric extensions of the model. But now already we
can address the issue of duality that had been conjectured by Montonen and Olive \cite{MONOLI}. They observed that in the model we considered not only
the dyonic charges saturate the BPS bound but also the perturbative degrees of
freedom, i.e. the massive gauge bosons, the remaining massless photon and also
the Higgs field. Also they found that monopoles do not excert any force onto
each other by a cancellation of attraction and repulsion, mediated by the
Higgs and gauge boson exchange respectively. This lead to the assumption that
there might be some kind of mapping from the perturbatively light fields onto
the non perturbative dyon spectrum acting in some unknown way on the moduli
space spanned by the vacuum expectation value of the Higgs and the couplings, which could be figured to be a symmetry of the theory, a duality. Montonen and Olive conjectured that there could exist an entire charge lattice of states which were to be permuted by duality transformations. For the sake of charge conservation these have to map the lattice onto itself and they therefore found the most natural form of the duality to be an $SL(2,\mathbb{Z})$ group acting on the point lattice in the complex plane. Thus the duality has been called $S$-duality, as well as its generalization in string theory, already mentioned in previous chapters. A couple of unanswered questions remained, among them the problems that quantum corrections should be expected to modify the classical potential and spectrum considerably and further that it was not possible to match the dyons and the perturbative fields together into multiplets of the Lorentz group. The most important progress in these issues was achieved in supersymmetric models. \\

\subsubsection{BPS states}
\label{bpsstates}

The success in identifying dualities in supersymmetric gauge theories 
\cite {WI2} has been very impressive over the last couple of years. 
In gauge theories with one supersymmetry (${\cal N} =1$) Seiberg discovered dualities between the large distance (IR) behaviour of electric and magnetic versions of the same theory \cite{SEI1}. The most prominent example of all is the treatment of ${\cal N}=2$ extended supersymmetric models by Seiberg and Witten that allowed to compute the spectrum of the theory with gauge group $SU(2)$ exactly and identified the condensation of monopoles as a mechanism of quark confinement \cite{SEIWIT1,SEIWIT2}. The solution of the theory also involved a duality, when the gauge symmetry had been broken spontaneously to leave only a single gauge boson massless, i.e. there exists a version of S-duality on the Coulomb branch of the moduli space. Much earlier it had already been found that the amount of supersymmetry necessary to implement the S-duality in its full conjectured extent is even larger and that only ${\cal N}=4$ supersymmetric Yang-Mills (SYM) theory would allow the dyons and the gauge bosons to sit in the same representations of the Lorentz group \cite{OSB}. So let us briefly discuss the way topological charges are introduced in supersymmetric gauge theories and stress the particular role that is being played by states that saturate the BPS bound. \\

The most general algebra of ${\cal N}$ supercharges that can be constructed in $d=4$ dimensions is given by (two component Weyl formalism):
\beqn
\{ Q^i_\alpha , \bar{Q}^j_{\dot{\beta}} \} &=& 2 \delta^{ij} \sigma_{\alpha \dot{\beta}}^\mu P_\mu ,\non
\{ Q^i_\alpha , Q^j_\beta \} &=& \eps_{\alpha \beta} U^{ij}, \non
\{ \bar{Q}^i_{\dot{\alpha}} , \bar{Q}^j_{\dot{\beta}} \} &=& \eps_{\dot{\alpha} \dot{\beta}} V^{ij} .
\eeqn
where $i,j$ label the different supercharges, $\alpha, \beta, \dot{\alpha}, \dot{\beta}=1,2$ the spinor
components and $C=\gamma_0$ is the charge conjugation matrix, $P^\mu$ the
momentum operator, $U^{ij}= -U^{ji}$ and $V^{ij}=-V^{ji}$ being called central
charges. The $Q^i_\alpha$ are the generators of the
supersymmetry and their conjugation is defined by 
\beqn
Q^i_\alpha \equiv \left( C_{\alpha \beta} \bar{Q}^{i \beta} \right)^\dag .
\eeqn 
Applied to the fields of a given theory they generate the
minimal field content of the supersymmetrized version of that theory. They
carry a representation of the Lorentz group $SO(3,1)$ as well
as of some internal symmetry that acts on the supercharges. In the absence of central charges the algebra simplifies and we can easily find its representation in terms of a physical, i.e. supersymmetric, multiplet of states. In the case of a massless multiplet $(P^2=0)$ we can choose the Lorentz frame in which $P^\mu = (E,0,0,E)$ and look for a representation of the little group $SO(2)$ that leaves $P^\mu$ invariant, which can afterwards be promoted to a representation of the full Lorentz algebra by application of the boost operators. (Supersymmetry transformations and boosts commute.) Thus we obtain the simplified algebra
\beqn
\{ Q^i_\alpha , \bar{Q}^j_{\dot{\beta}} \} = 4E \delta^{ij} \left( \begin{array}{cc} 0 & 0 \\ 0 & 1 \end{array} \right) _{\alpha \dot{\beta}} , 
\quad \{ Q^i_\alpha , Q^j_\beta \} = 0, 
\quad \{ \bar{Q}^i_{\dot{\alpha}} , \bar{Q}^j_{\dot{\beta}} \} = 0 .
\eeqn
Obviously the upper components anticommute and by the usual positive norm
argument they have to annihilate the physical Hilbert space, thus they are
trivially represented. In other words: Massless states leave half of the
supersymmetry unbroken. The lower components form a Clifford algebra and act
as raising and lowering operators of the helicity operator, the generator of
rotations around the direction of $P^\mu$. They create an antisymmetric tensor
representation of the internal $U({\cal N})$ symmetry from a given state of lowest helicity $s$:
\beqn
\vert P^\mu, s \rangle, \quad \frac{1}{\sqrt{4E}} ( Q^i_2 )^* \vert P^\mu, s \rangle, \quad \frac{1}{4E} ( Q^i_2 )^* ( Q^j_2 )^* \vert P^\mu, s \rangle, ...
\eeqn
The series terminates after application of ${\cal N}$ raising operators, which
in the case of ${\cal N}=1$ gives a spectrum consisting only of two kinds of
fields. We shall only be interested in multiplets that contain no higher spin
than $s=1$. The possibilities we are then left with are two complex scalars
and a single fermion (chiral multiplet) or a fermion with a vector (vector
multiplet), while in ${\cal N}=2$ one could have a vector with a complex
scalar and two fermions (vector multiplet) or two fermions with two complex scalars (hypermultiplet), counting always on-shell degrees of freedom after eliminating auxiliary fields. Multiplets with $s\leq 1$ that involve more massless fields can always be reduced to tensor products of these irreducible representations. Such multiplets can be thought of as the minimal field content a supersymmetric theory has to contain.\\

We do not intend to look at the case of massive fields explicitly as the situation gets more complicated and we shall not be interested in the massive fields with vanishing central charges in the later discussion. One gets aware at once that the matrix of the supercharge anticommutator has no vanishing eigenvalues anymore and all supercharges have to be represented non trivially. The shortest possible multiplets get very much longer than the massless multiplets above, their internal symmetry is found to be $USp(2{\cal N})$ . A general classification can be found in the standard literature \cite{WBA}, a very recent review of this and some of the following is \cite{PW}. \\

We now also allow central charges in the superalgebra, which implies ${\cal N}>1$, then look for the short multiplets as before and will find them to be BPS saturated. Therefore we take massive states with rest frame momentum $P^\mu=(m,0,0,0)$ whose superalgebra is
\beqn
\{ Q^i_\alpha , \bar{Q}^j_{\dot{\beta}} \} &=& 2m \delta^{ij} \delta_{\alpha \dot{\beta}}  ,\non
\{ Q^i_\alpha , Q^j_\beta \} &=& \eps_{\alpha \beta} U^{ij} ,\non
\{ \bar{Q}^i_{\dot{\alpha}} , \bar{Q}^j_{\dot{\beta}} \} &=& \eps_{\dot{\alpha} \dot{\beta}} V^{ij} .
\eeqn
By using the mentioned internal symmetry one can redefine the supercharges in a way that combines the former ones into chiral components and allows to diagonalize the right hand side of the anticommutator with eigenvalues $Z_i$ (no summation):
\beqn
\{ \tilde{Q}^{i+}_\alpha , ( \tilde{Q}^{j+}_\beta )^\dag \} &=& 2 \left( m-Z_i \right) \delta^{ij} \delta_{\alpha \beta}  ,\non
\{ \tilde{Q}^{i-}_\alpha , ( \tilde{Q}^{j-}_\beta )^\dag \} &=& 2 \left( m+Z_i \right) \delta^{ij} \delta_{\alpha \beta} ,
\eeqn
By the positive definiteness of the operator $\tilde{Q}^{i\pm}_\alpha ( \tilde{Q}^{i\pm}_\alpha )^\dag$ (again no summation) 
we immediately get the BPS bound on the eigenvalues of the central charge matrix: $Z_i \leq m$. 
Also we can identify the short or BPS multiplets that correspond to the massless multiplets in the case of vanishing central charges. They are created when all the eigenvalues saturate the bound: $Z_i=m$. 
Generally the multiplets contain less and less states when more and more
eigenvalues satisfy this relation, as more supercharges get represented
trivially then. These states are of great importance as they are believed to
be not affected (at least not too much) by renormalization, when the classical
supersymmetric theory is taken to be the bare Lagrangian of a quantum field
theory. One then generally expects that the procedure of renormalization might
change some or all parameters of the theory, floating into some fixed point of
the renormalization group and thus spoiling all the results of classical
analysis which we dealt with so far. On the other hand if nothing very
dramatic happens and all parameters vary sufficiently smoothly, the number of
physical fields should remain unchanged. More precisely stated, if charge and
mass are coefficients of relevant or marginal operators in a Lagrangian that
is at the foundation of a quantum theory, for BPS states their values should
after renormalization still coincide if supersymmetry is present and the number of fields unchanged. In contrast to the coefficients of irrelevant operators the renormalized values of such parameters are not determined by renormalization alone but have to be fixed by an experiment. A natural value is assumed to be of the order of the quantum corrections, but in principle any value can be obtained by adjusting the bare parameters order by order in perturbation theory as desired. In supersymmetric field theories there often occurs a cancellation of perturbative or even all quantum corrections, which then renders the protected parameter to be a free modulus of the theory, exactly determined by its bare value. (At least the potential of any supersymmetric field theory is not affected by perturbative quantum corections.) This is also the reason why we can imagine consistently to vary the string coupling from strong to weak by changing its bare value, which is the only free modulus of string theory. Because of such reasoning one hopes that once the BPS states are identified in the classical approximation of some supersymmetric theory, i.e. the lowest order of its perturbative expansion, they should exist also in the non perturbative, large coupling sector of the moduli space. Establishing duality relations between perturbative and non perturbative sectors of theories very much relies on comparing the BPS spectra of both theories in the respective domain. There is in fact a case in which the duality argument can be made rigorous. This is the maximal ${\cal N}=4$ supersymmetric extension of Yang-Mills theory which we already mentioned earlier to be the only SYM theory that provides the correct multiplets to have full Montonen-Olive S-duality implemented \cite{OSB}. This theory is also conformally invariant and has all its $\beta$-functions vanishing exactly, so that BPS states indeed remain unaffected by renormalization even if non perturbative effects are taken into account. \\

We now return to our previous subject of monopoles and dyons in the Georgi-Glashow model and relate central charges of the superalgebra of its ${\cal N}=2$ supersymmetric extension to topological charges of solitonic solutions of the field equations \cite{WI2}. The Lagrangian of the SYM theory that includes the former model 
\beqn
{\cal L} &=& -\frac{1}{4} F_{\mu \nu}^a F^{\mu \nu}_a +\frac{1}{2} \sum_{i=1}^2{\left( \bar{\Psi}^a_i i\gamma_\mu D^\mu \Psi^a_i +  D_\mu \Phi_i^a D^\mu \Phi^a_i \right)} \non
& & +\frac{1}{2} g^2 \Tr \left[ \Phi_1,\Phi_2 \right] ^2 +\frac{1}{2} ig\eps_{ij} \Tr \left( \left[ \bar{\Psi}_i,\Psi_j \right] \Phi_1 + \left[ \bar{\Psi}_i,\gamma_5 \Psi_j \right] \Phi_2 \right)
\eeqn
contains a single ${\cal N}=2$ vector multiplet (a vector gauge boson with field strength $F^a_{\mu \nu}$, two fermions $\Psi^a_i$, a scalar $\Phi^a_1$ and a pseudo-scalar $\Phi^a_2$) in the adjoint representation of the gauge group. If one computes the variations of the supersymmetry transformations in terms of the fields and keeps track of all boundary terms, one finds that the supersymmetry only holds up to the boundary terms
\beqn
U &=& \int{d^3x\ \partial_i \left( \Phi_1^a F_a^{0i} +\Phi^a_2 \frac{1}{2} \eps^{ijk} F_{ajk} \right) } ,\non
V &=& \int{d^3x\ \partial_i \left( \Phi_1^a \frac{1}{2} \eps^{ijk} F_{ajk} +\Phi^a_2 F_a^{0i} \right) } .
\eeqn    
These are the generalizations of the topological charges given by the winding of Higgs and gauge field of the former model. If non vanishing they demand to implement central charges $U$ and $V$ into the superalgebra:
\beqn
\{ Q^i_\alpha ,Q^j_{\beta} \} = \eps^{ij} \eps_{\alpha \beta} U , \quad \{ \bar{Q}^i_{\dot{\alpha}} , \bar{Q}^j_{\dot{\beta}} \} = \eps^{ij} \eps_{\dot{\alpha}  \dot{\beta}} V .
\eeqn
Thus the numerical values of the central charges are equal to the topological charges in the SYM theory, which in the non supersymmetric case we identified with the electric and magnetic charges of the classical monopole solutions. BPS monopoles are then special solutions that also saturate the BPS bound and which are annihilated by one half of the supercharges. In other words the presence of a BPS monopole somewhere in the universe can be thought of as a topological non trivial modification of the vacuum that imposes certain boundary conditions on the fields at infinity and breaks half of the supersymmetry by these conditions. The other half of the superalgebra acts on the monopole state by creation of fermionic solutions to the field equations (Dirac equations) in this background. The space of these fermionic zero-modes is parametrized by the moduli of the monopole solution, its position and charges, i.e. the number of fermionic zero-modes in the monopole background is related to the dimension of the moduli space of the monopole. All the states saturating the BPS mass bound we have been discussing so far, Reissner-Nordstr\"om extremal black holes and Prasad-Sommerfield monopoles, can be embedded in supersymmetric theories to yield BPS states. While we only deduced their properties and existence from classical analysis, the supersymmetry is assumed to protect them against renormalization by quantum effects, so that they remain BPS multiplets. In this sense classical arguments are extrapolated towards quantum exactness. Reviews of the whole material can also be found in \cite{OLI,H}.

\subsection{Solitons in string theory}
\label{stringsolitons}

In analogy to solutions to classical field equations of the previous chapter we shall now discuss solitonic BPS objects in string theory. In the following sections we shall find two similar but not completely identical types of candidates for such states, solitonic p-branes and D(irichlet)-branes. We would like to point out some of their differences and similarities. We will start with an introduction into the way how the calculus of differential forms allows to deduce the possible spatial dimension of charged objects in string theory on the grounds of the ranks of the tensor fields occurring in the low energy effective action, the supergravity theories in ten dimensions. These objects will then first be established classically as p-branes, solutions of the supergravity field equations, that only depend on a subset of the coordinates. They are interpreted as higher dimensional monopoles and black holes, as they are sources for the generalized electromagnetic tensor fields as well as for the gravitational metric field, that are extended in spatial directions also. Integrating the dual electromagnetic field strengths over the space transverse to the worldvolume of the branes then allows to introduce the notion of non perturbative, topological charges into the theory. The second type of states, the D-branes, have been reviewed as defects of space-time which have (perturbative) open string world sheets ending on them. 
By indirect arguing many indications have been found, that they are intermediate states between perturbative excitations and solitonic p-branes, that share many properties of the latter. In addition to their being classical solutions of the effective supergravity theory they further provide us with a prescription of how to handle their perturbative degrees of freedom via the open strings ending on them, which by the continuity of BPS states can be traced back into the strongly coupled non perturbative regime. This is the key to a perturbative quantum description of strongly coupled string phenomena. We shall find an example of such methods in the treatment of black holes.

\subsubsection{Extended charges as sources of tensor fields}
\label{charges}

In this introductory section we first explain the necessary mathematics to generalize classical electromagnetism to higher dimensions, i.e. the calculus of differential forms \cite{DU,NE,TE}. Taking the entries of the field strength tensor $F^{\mu \nu}$ of the usual Maxwell theory as coefficients of the 2-form 
\beqn
F = 2 dA = \partial_{[\mu} A_{\nu ]} dx^\mu \wedge dx^\nu = F_{\mu \nu} dx^\mu \wedge dx^\nu
\eeqn
and the current $j^{\mu}$ correspondingly as a 1-form $j_{\mbox{\sz e}} \equiv j^\mu dx_\mu$ we can rewrite the inhomogeneous part of Maxwell`s equations 
\beqn
d \host F = \host j_{\mbox{\sz e}}, \quad \host F _{\mu \nu} &=& \frac{1}{2} \eps_{\mu \nu \rho \sigma} F^{\rho \sigma}
\eeqn
and represent the electric charge inside some spatial region by the integral of the field equation over the transverse space ${\cal M}$
\beqn
e = \int_{\partial {\cal M}}{^*F} = \int_{\cal M}{\host j_{\mbox{\sz e}}} ,
\eeqn  
using Stoke's theorem. The homogeneous part of Maxwell's equations becomes the Bianchi identity $dF=0$. The dualized set of equations including the magnetic current is then given by adding ad hoc a term to the field strength, that is not closed, but its Hodge dual is, thus spoiling the Bianchi identity, but not modifying the electric charge definition: 
\beqn
F & \rightarrow & F = 2 dA +\omega , \non
d \host F &=& \host j_{\mbox{\sz e}}, \quad dF = d\omega = j_{\mbox{\sz g}} .
\eeqn
The magnetic charge becomes
\beqn
g = \int_{\partial \tilde{\cal M}}{F} = \int_{\tilde{\cal M}}{j_{\mbox{\sz g}}}.
\eeqn  
The gauge freedom corresponds to adding an exact form to the 1-form potential:
\beqn
A \rightarrow A + d\Lambda.
\eeqn
All these statements can easily be generalized to differential forms of higher degree:
\beqn
\mbox{potential:}\ & & A^{(p+1)} =  A_{\mu_1 ... \mu_{p+1}} dx^{\mu_1} \wedge ... \wedge dx^{\mu_{p+1}} , \non
\mbox{gauge freedom:}\ & & A^{(p+1)} \rightarrow A^{(p+1)} + d\Lambda^{(p)}, \non
\mbox{field strength:}\ & & F^{(p+2)} = (p+2)\ dA^{(p+1)}, \non
\mbox{electric charge:}\ & & e = \int_{\cal M}{^*j^{(d-p-1)}_{\mbox{\sz e}}}, \non
\mbox{magnetic charge:}\ & & g = \int_{\tilde{\cal M}}{ j^{(p+3)}_{\mbox{\sz g}}}.
\eeqn
One finds that in general $j_{\mbox{\sz g}}$ is a $(p+3)$-form, if the
electric charge is an object extending into $p_{\mbox{\sz e}} \equiv p$ space
dimensions, while the dual magnetic charge lives in $p_{\mbox{\sz m}} \equiv
d-4-p_{\mbox{\sz e}}$ dimensions. In the classical $d=4$ Maxwell theory we had
of course $p{\mbox{\sz e}}= p_{\mbox{\sz m}} =0$. From this generalized
setting the Dirac quantization condition for the product of the two charges
can be retained unchanged as in the dualized Maxwell theory. In a
Bohm-Aharonov experiment one would have to move extended objects around
extended singularities and the magnetic potential integrated over non trivial
cycles around its singularities leads to a magnetic charge quantized by exactly the same formula (\ref{dirac}) in integer units of the inverse electric charge times $2\pi$. \\

Thus we have found a rule that allows to deduce from the rank of some antisymmetric tensor field, rewritten in the language of a differential form of same degree, the spatial dimension of the corresponding charge that is the source of the generalized electromagnetic field corresponding to that tensor field. As well this charge is quantized in analogy to Dirac's condition. We are now in the position to discuss the dimensionality of the charges we do expect in the various string models by inspecting their bosonic field contents, summarized in table \ref{tab1}.
\begin{table}[h]
\caption{Forms and charges in various string models}
\begin{center}
\begin{tabular}{lcccc}
\vspace{0.2cm}
Model & Potential & Field Strength & $p_{\mbox{\sz e}}$ & $p_{\mbox{\sz m}}$ \\
\vspace{0.2cm}
Universal NSNS sector: & & & & \\
\vspace{0.2cm}
Heterotic, IIA, IIB & $B^{(2)}$ & $F^{(3)}$ & 1 (fundamental string) & 5 (NS-5-brane) \\
\vspace{0.2cm}
RR sector: & & & & \\
\vspace{0.2cm}
IIA & $A^{(1)}$ & $F^{(2)}$ & 0 (D-Particle) & 6 \\
\vspace{0.2cm}
    & $A^{(3)}$ & $F^{(4)}$ & 2 & 4 \\
\vspace{0.2cm}
IIB & $A^{(0)}$ & $F^{(1)}$ & -1 (D-Instanton) & 7 \\
\vspace{0.2cm}
    & $A^{(2)}$ & $F^{(3)}$ & 1 (D-String) & 5 \\
\vspace{0.2cm}
$\hspace{2cm}$ (self-dual) & $A^{(4)}$ & $F^{(5)}$ & 3 & 3 \\
\vspace{0.2cm}
M-theory ($d=11$) & $A^{(3)}$ & $F^{(4)}$ & 2 (Membrane) & 5 \\
\end{tabular}
\end{center}
\label{tab1}
\end{table}
Some comments are necessary which will become clear only in the next sections
to follow. The universal NSNS sector is common to both heterotic, the IIA and
the IIB model. Its 2-form potential couples to the respective fundamental
string itself, i.e. the world sheet coordinates of the string couple to the
space time antisymmetric tensor, as usual in the $\sigma$-model approach. The
solitonic object of this sector is the NS-5-brane, solitonic in the sense of a
p-brane as found in the universal bosonic sector of $d=10$ supergravity. It is
the magnetic dual of the string and couples magnetically to the NSNS 3-form field strength. All other charged states listed in the table come from the RR sector, they are assumed to be realized as D-branes. These are string solitons which can perturbatively be described by open strings with Dirichlet boundary conditions and have an interpretation as p-branes in the low energy approximation by supergravity. The meaning of these statements will be explored in the following sections. All these states are related by several kinds of (conjectured) duality transformations that also connect the perturbative and non perturbative degrees of freedom. Particularly the fundamental string and the NS-5-brane of the universal IIB sector are related to the D1-brane, the so called D-String, and the D5-brane of the RR sector of the IIB model via S-duality. The 3-branes of the IIB are selfdual, referring to the selfdual field strength tensor $F^{(5)} = \host F^{(5)}$ of that theory and the (-1)-brane is apparently an object local in space and time and therefore usually adressed as an instanton. All these states can be identified as states arising in some particular compactification of states known from $d=11$ M-theory, which we shall turn to in the final chapter. \\

\subsubsection{Solutions of the supergravity field equations: p-branes}
\label{pbranes}

The next task will be to derive solutions to the low energy field equations of
string theory \cite{DU,ST,CHS}. These describe the various fields whose
sources are multidimensional objects that carry mass and charges corresponding to the
various field strengths. We use the field equation of the effective theory,
supergravity in $d=10$ dimensions, to find solitonic states that display
finite action. Their support in space has to be localized and all 
fields must decrease fast enough at infinity but possibly with non 
trivial winding. Thereby we always focus on the bosonic degrees of freedom, taking the fermionic fields then given automatically by supersymmetry. The massless bosonic fields of the string spectra are the graviton $G_{\mu \nu}$, the antisymmetric tensor field $B_{\mu \nu}$ with 3-form field strength and the dilaton $\Phi$ from the universal sector, as well as the various $A^{(p+1)}$ potentials or their field strengths $F^{(p+2)}$ from the IIA and IIB RR sectors. Their effective action is dictated by the according supergravity theory:
\beqn \label{actioneinstein}
S_{\mbox{\sz eff}}^{\mbox{\sz E}} = \frac{1}{2\kappa^2} \int{d^{10}x\ \sqrt{-G^{\mbox{\sz E}}} \left( R \left( G^{\mbox{\sz E}} \right) -\frac{1}{2} \nabla_\mu \Phi \nabla^\mu \Phi - \sum_{n} \frac{1}{2n!} e^{a\Phi} F^{\mu_1 \cdots \mu_n} F_{\mu_1 \cdots \mu_n} \right) } ,
\eeqn 
here written in the so called Einstein frame and all 
topological Chern-Simons terms being dropped. In the 
universal sector we have $a=-1, n=3$ while in the Ramond sectors $a=(5-n)/2$ and $n$ runs over the relevant form degrees. When commenting on the $n=5$ case we shall eventually ignore the problem that no consistent action for the self-dual 5-form is known so far. The action (\ref{actioneinstein}) can be rephrased into the $\sigma$-model (or string) frame by a Weyl rescaling
\beqn
G^{\mbox{\sz E}}_{\mu \nu} 
\rightarrow G^{\mbox{\sz E}}_{\mu \nu} e^{\Phi/2} 
\equiv G^\sigma_{\mu \nu} ,
\eeqn
which results in:
\beqn \label{sugrasigma}
S_{\mbox{\sz eff}}^\sigma &=& \frac{1}{2\kappa^2} \int{d^{10}x\ \sqrt{-G^\sigma} \left( e^{-2\Phi} \left( R \left( G^{\sigma} \right) +4 \nabla_\mu \Phi \nabla^\mu \Phi - \frac{1}{12} F^{\mu \nu \rho} F_{\mu \nu \rho} \right)  \right. } \non
& & - \left. \sum_{n}{ \frac{1}{2n!} F^{\mu_1 \cdots \mu_n} F_{\mu_1 \cdots \mu_n} } \right)  ,
\eeqn 
the sum now only running over the Ramond fields, of course. The
latter form can also directly be derived as a low energy effective action of
string theory using the formalism of world sheet $\sigma$-models. An important
thing to notice is, that the dilaton has obtained a universal coupling to all
fields of the universal sector but decouples from the fields of the RR
sector. Thus we expect to get different dependences of masses and charges on
the string coupling $\langle \exp \left( \Phi \right) \rangle$ for branes
originating from tensor fields of the different sectors. The masses of the
branes coupling to NSNS fields should be expected to behave like $g_{\mbox{\sz
    S}}^{-2}$, while those from the RR sector will
interpolate between $g_{\mbox{\sz S}}^{-2}$ and $g_{\mbox{\sz S}}^0$. To avoid certain difficulties and for the sake of brevity we now take the somewhat simpler type of Einstein action
\beqn
S^{\mbox{\sz E}}_{\mbox{\sz eff}} = \frac{1}{2\kappa^2} \int{d^dx\ \sqrt{-G^{\mbox{\sz E}}} \left( R^{\mbox{\sz E}}- \frac{1}{2} \nabla^\mu \Phi \nabla_\mu \Phi - \frac{1}{2n!} e^{a\Phi} F^{\mu_1 \cdots \mu_n} F_{\mu_1 \cdots \mu_n} \right) } 
\eeqn
keeping only a single tensor field and derive the various equations of motion:
\beqn \label{sugraequ}
R^{\mbox{\sz E}}_{\mu \nu} &=& \frac{1}{2} \partial_\mu \Phi \partial_\nu \Phi + S_{\mu \nu}, \non
S_{\mu \nu} &\equiv & \frac{1}{2(n-1)!} e^{a\Phi} \left( F_{\mu \rho_1...\rho_{n-1}} F_{\nu}^{\ \rho_1...\rho_{n-1}} - \frac{n-1}{n(d-2)} F^2 G^{\mbox{\sz E}}_{\mu \nu} \right) , \non
\nabla_\mu \left( e^{a\Phi} F^{\mu \mu_2...\mu_n} \right) &=& 0 ,\non
\partial_\mu \partial^\mu \Phi &=& \frac{a}{2n!} e^{a\Phi} F^2 .
\eeqn
The first one is a generalized Einstein equation, the third one the vacuum Maxwell equation and the last one describes a Klein-Gordon field coupling somehow to electromagnetism. Later on we shall have to add a source term into the action to get non trivial solutions. This will lead to additional delta function like, singular sources on the right hand side of the equations of motion. To make an ansatz, we now split the coordinates into the $x_\mu,\ \mu=0,...,p$, on a $(p+1=n-1)$-dimensional hyperplane which is taken to contain the charged object and the transverse spatial directions $y_M,\ M=p+1,...,d-1$. We then demand usual Poincare $P(1,p)$ invariance in the first $p+1$ coordinates, the worldvolume of the brane, and isotropy, $SO(d-p-1)$ invariance, in the rest of space to be satisfied by the solution. All the fields necessarily have to be independent of the internal $x_\mu$ coordinates because of translation invariance. We also need to determine the amount of supersymmetry that is left unbroken by the ansatz. Therefore we have to look for the decomposition of the tendimensional Poincare invariant supercharges into $(p+1)$ and $(d-p-1)$-dimensional spinors, themselves invariant under the demanded space-time symmetries. For $d=10$ it is found that the tendimensional chirality condition $(1- \Gamma_{11}) \eps_{10} =0$ implies that also
\beqn
(1-\Gamma_{(p+2)}) \eps_{(p+1)}=0, \quad (1-\Gamma_{(10-p)}) \eps_{(9-p)}=0,
\eeqn
where $\Gamma_{(D+1)} \equiv \Gamma_0 \cdot \ldots \cdot \Gamma_{(D-1)}$ is the respective chirality operator and $\eps_D$ an arbitrary invariant spinor in the $D$ dimensional subspaces. By inspecting the eigenvalues of these $\Gamma$ matrices one finds that for the cases we consider one half of the supersymmetry is broken by the brane ansatz, which indicates that we might be dealing with a BPS state. In fact it is possible to go the other way round and construct the solutions we shall uncover by just their property of breaking exactly one half of the supersymmetry \cite{CHS}. Splitting off the metric according to the ansatz we can always write it      
\beqn
ds_{\mbox{\sz E}}^2 = e^{2A(r)} dx_\mu dx^\mu + e^{2B(r)} dy_M dy^M ,
\eeqn
where both contractions of indices only involve flat metrics and $r \equiv \sqrt{y_M y^M}$ is the radial distance in the space orthogonal to the brane. The ansatz for the metric obviously respects Lorentz and translation invariance on the brane and also rotation invariance in the transverse directions. For the field strength tensor there are two different choices one can think of, as we have the two options to construct the electric charge from the field strength tensor itself or the magnetic charge from its dual, such that one has the two options:
\beqn \label{feldstaerke}
F^{\mbox{\sz e}}_{M \mu_2...\mu_n} &=& \eps_{\mu_2...\mu_n} \partial_M e^{C(r)} , \non
F^{\mbox{\sz m}}_{M_1...M_{\tilde{n}}} &=& g\ \eps_{M_1...M_{\tilde{n}} M}
\frac{y^{M}}{r^{\tilde{n}+1}} .
\eeqn
The form degrees are related by $\tilde{n} = d-n$, as both field strengths are Hodge dual in the d-dimensional space-time. This completes the ansatz. It of course remains to be verified that charge and mass of these states take finite values. We now omit lots of technical details which can be found for instance in \cite{DU,ST}. Finally one can reduce the three undefined functions to a single one and a couple of parameters, related by:
\beqn \label{allglsg}
e^{2A(r)} &=& H(r) ^{-\frac{4 \tilde{d}}{\Delta (d-2)}} ,\non
e^{2B(r)} &=& H(r) ^{\frac{4d}{\Delta (d-2)}} ,\non
e^{C(r)} &=& \frac{2}{\sqrt{\Delta}} H(r)^{-1} ,\non
e^{\Phi(r)} &=& H(r) ^{\frac{2a}{\zeta \Delta}}.
\eeqn
The remaining function $H(r)$ is harmonic in the transverse space, i.e. obeys the Laplace equation:
\beqn
\delta_{NM} \partial^N \partial^M H(r) = 0.
\eeqn
Only if we add a source on the right hand side of 
this equation, we get non trivial solutions for $H(r)$, 
as integrable, globally harmonic functions necessarily vanish. 
Adding a source that only has support on the volume of the brane, 
a $(p+1)$-dimensional charged current, now corresponds to turning 
the Laplace equation into a Poisson equation \cite{DGHR}. 
An explicit form for such a current will be discussed later. The solution for $H(r)$ with a brane at the transverse origin can then be written 
\beqn \label{harmonfu}
H(r) = 1+ \alpha/ r^{\tilde{d}}, \quad \alpha >0, 
\eeqn
the parameters being given by:
\beqn
\Delta &=& a^2 + \frac{2 (p+1) \tilde{d}}{d-2} ,\non
\zeta &=& \left\{ \begin{array}{cl} +1 & \mbox{for electric solutions}  \\ -1 & \mbox{for magnetic solutions} \end{array} \right. ,\non
\tilde{d} &=& d-p-3.
\eeqn
In the magnetic case the integration constant $\alpha$ is related to the magnetic charge parameter:
\beqn
\alpha = \frac{g \sqrt{\Delta}}{2\tilde{d}} ,
\eeqn
while in the electric case it is fixed by the charge parameter of the source terms in the action. In analogy to the definition of charges in electromagnetism we can integrate the field strength $(p+2)$-form over a surface that encapsulates the p-brane and obtain the charge of an object that has a $(p+1)$-dimensional worldvolume. This charge is the source of the metric field in the transverse directions, as well as of the tensor field. Only in the case of a $1$-brane we know the proper quantum theory of the desired object, which is simply the fundamental string. Taking this example one can add a source term into the action (\ref{sugrasigma}) which has only support on a two dimensional submanifold of space-time and whose action there is given by the $\sigma$-model action of string theory:
\beqn \label{actionsigma}
S_{\sigma} = -\frac{T_2}{2} \int{d^2\xi\ \left(\sqrt{h} h^{\alpha \beta}
    \partial_\alpha X^\mu \partial_\beta X^\nu G_{\mu \nu} + \eps^{\alpha
      \beta} \partial_\alpha X^\mu \partial_\beta X^\nu B_{\mu \nu} - 2\pi \al
    \sqrt{h} R \Phi  \right) } . 
\eeqn
All the analysis can be carried out with the only exception that delta
function source terms appear on the right hand side of the field
equations. This is interpreted in the following way: The fundamental, microscopic string is the source of a macroscopic field configuration, the $1$-brane solution of the low energy approximation to string theory. The fields that appear in the world sheet theory as the massless modes of the string are now (in the worldvolume theory) organized in supergravity multiplets and their source is the twodimensional string again. Its electric charge under the tensor field is given by integrating the field equation 
\beqn
e_{\mbox{\sz ST}} = \frac{1}{\sqrt{2\kappa^2}} \int_{\partial {\cal M}^8}{e^{-\Phi(r)} \host F(r)} = \sqrt{2\kappa^2} T_2.
\eeqn
The mass of this $1$-brane is defined by the integral over the time component of the energy momentum tensor, which can also be calculated from the $\sigma$-model source and for the given solution is found to saturate the BPS bound, it is equal to the charge of the brane. Despite from this case it is an open question what ``material'' branes are made of in general. We can interpret the 1-brane as the solution coming from the string and the NS-5-brane as its $d-4-1=5$ dimensional magnetic dual. For the rest of the p-brane solutions one has no elementary particles at hand. Only by the discovery of Polchinski it became plausible that the desired objects are related to D-branes.\\

Let us now point out an aspect of duality in the brane picture. The electric
and magnetic p-brane solutions we have found correspond to a particle like
state, which is the source of the electric field strength tensor $F^{(n)}$, as
well as a solitonic object, the source for the magnetic dual field $\host
F^{(d-n)}$. This setting allows a notion of duality in the sense that starting
from the dual tensor in the original action and splitting coordinates
accordingly would have interchanged the role of the two charges. In other
words, calling the $n-2$ dimensional state the elmentary and the $d-2-n$
dimensional the solitonic one is a matter of convention as long as we do not
decide which field strength is to be called fundamental. Thus the NS-$5$-brane might be as fundamental as the string itself, though we cannot say very much concerning its quantum theory, which is supposed to be given by a quantization of its coordinates in the spirit of the Polyakov action approach to string theory. From computing its (topological) charge by using the magnetic dual tensor field one can deduce the generalized Dirac quantization condition from a Bohm-Aharonov experiment
\beqn
2\kappa^2 T_2 T_6=2\pi n,
\eeqn
where $T_6$ is the tension of the $5$-brane. \\

We can summarize that we found solutions to classical low energy effective string theory, that are extended and carry mass and charge under the various tensor fields. Their particular values saturate the BPS bound and this allows us to strongly believe in the existence of these states also in the unknown quantum theory and further expect them not to be renormalized in a way that would spoil their being BPS saturated. The severe and in general unsolved problem remains: how to give a description of the quantum theory involving branes of higher dimension. \\

We next apply the solutions for the metric and the dilaton to various values for $d$ and $p$ to get some examples which we of course choose from the string spectra. The expressions for the field strengths are obtained as easy. For the fundamental string in $d=10$, having a worldvolume of dimension $2$ and correspondingly $n=3$, the metric and the dilaton field read
\beqn
ds_{\mbox{\sz E}}^2 &=& \left( 1+ \frac{\alpha}{r^6} \right)^{-3/4} dx_\mu^2 + \left( 1+ \frac{\alpha}{r^6} \right)^{1/4} dy_M^2 ,\non
e^\Phi &=& \left( 1+ \frac{\alpha}{r^6} \right)^{-1/2},
\eeqn
an expression that appears to display singularities not completely unfamiliar from those found for the black hole solutions in general relativity. In fact there are ``black brane'' solutions that have horizons shielding their singularities. They are in many respects similar to higher dimensional black holes \cite{HOS} and we shall eventually return to these examples when discussing black holes in string theory. For the NS-5-brane with $n=7$ we get instead 
\beqn
ds_{\mbox{\sz E}}^2 &=& \left( 1+ \frac{\alpha}{r^2} \right)^{-1/4} dx_\mu^2 + \left( 1+ \frac{\alpha}{r^2} \right)^{3/4} dy_M^2 ,\non
e^\Phi &=& \left( 1+ \frac{\alpha}{r^2} \right)^{1/2}.
\eeqn
Comparing the solutions for the string and the NS-5-brane, the two regions of space-time, the brane and the
transverse space appear exchanged. Also the coupling has been inverted, as supposed for the electric-magnetic duality of fundamental perturbative states and monopoles. Particularly for the self-dual 3-brane of the Ramond sector with $n=5$ we get:
\beqn
ds_{\mbox{\sz E}}^2 &=& \left( 1+ \frac{\alpha}{r^4} \right)^{-1/2} dx_\mu^2 + \left( 1+ \frac{\alpha}{r^4} \right)^{1/2} dy_M^2 ,\non
e^\Phi &=& 1,
\eeqn
where there is no distinction between the two types of states, electric or magnetic. Finally the D5-brane metric differs only in the dependence of the dilaton from its S-dual, the NS-5-brane:
\beqn
e^\Phi = \left( 1+ \frac{\alpha}{r^2} \right)^{-1/2} ,
\eeqn
as the duality transformation exchanges perturbative and non perturbative states. In this sense the dualities of string theory are manifest in the supergravity brane solutions. S-duality is an involution of the IIB supergravity that can map the exponential of the dilaton to its inverse and leave the metric in the Einstein frame invariant. Thus the D5-brane gets mapped to the above NS-5-brane solution, while the fundamental string gets mapped to an object with
\beqn
e^\Phi = \left( 1+ \frac{\alpha}{r^6} \right)^{1/2} ,
\eeqn  
and otherwise unchanged metric, which is just the D1-brane solution. In a more general analysis also T-duality can be recovered. After performing a dimensional reduction of IIA and IIB supergravity on a circle one finds a unique supergravity in $d=9$ that again allows an involution of its superalgebra and field content, in general exchanging the fields and charges that originate from IIA and IIB. Decompactifying afterwards, one recognizes that this entire transformation realizes the T-duality transformation which is known from the perturbative string theory and which maps odd dimensional IIB branes into even dimensional IIA branes and vice versa. This has to be discussed in the string frame, where any Dp-brane has 
\beqn
ds_\sigma^2 &=& \left( 1+ \frac{\alpha}{r^{\tilde{d}}} \right)^{-1/2} dx_\mu^2 + \left( 1+ \frac{\alpha}{r^{\tilde{d}}} \right)^{1/2} dy_M^2 ,\non
e^\Phi &=& \left( 1+\frac{\alpha}{r^{\tilde{d}}} \right) ^{-(p-3)/4}.
\eeqn
Compactifying along the brane worldvolume now leads to a brane wrapped around the circle and the rank $p+1$ of the tensor field coupling to the brane is effectively reduced by one. The involution of the ninedimensional supergravity next is of such nature that a tensor field of this kind is mapped to another one that already in the tendimensional theory had the lower rank $p$, so that after decompactifying we end up with a brane of one dimension less. On the other hand starting with a compactification in a direction transvers to the brane, the rank of the tensor field is unchanged at first and it is then being mapped to a tensor field that had originally higher rank, but lost an index during the compactification. By decompactifying the additional dimension opens up and the brane gains an extra dimension in the end. (This discussion is so far limited to the case of D-branes and has to be modified for NS-branes.) We shall be discussing the role of the (elevendimensional) superalgebra in M-theory later on, which reveals the particular importance of central charges for the existence of corresponding branes and should thereby make the above explanations also more transparent.

\subsubsection{D-branes as p-branes}

Another type of extended geometrical objects in string theory are the D-branes we discussed earlier. These have formerly been introduced as fixed hyperplanes in space-time, where open strings can end on. The necessity of their existence had already been demonstrated by T-duality \cite{POL2} when Polchinski found an interpretation \cite{POL1} that allowed to view them as dynamical, charged objects that fluctuate in shape and position and couple to the RR fields of the string world volume theory. This induced a tremendous amount of work on D-branes and related subjects which left very little doubt about the statement that D-branes are some sort of non perturbative states of string theory, relatives of the solitonic p-branes from the previous section. They have a perturbative description by open strings ending on them and their BPS nature conserves generic features of this in the non perturbative regime. There are numerous reviews of D-brane physics, only to mention \cite{POL3,BA,PCJ}, so that we restrict ourselves to illustrate Polchinski's original computation of the tension and charge by regarding a scattering process of strings emitted from and absorbed by D-branes. \\

We first show why the p-branes of the NSNS sector cannot be the sources of the RR fields. Let us recall the world sheet origin of the various fields that occur in the low energy effective string actions. In general the states are created by mode operators of the coordinate and spin fields subject to the constraints of superconformal invariance imposed by the super Virasoro operators. The space time fields are then given by the polarizations of such states, for a massless state in the universal NSNS sector e.g.
\beqn
\vert G_{\mu \nu}, B_{\mu \nu}, \Phi, k^\mu \rangle_{NSNS} = \left( G_{\mu \nu} \alpha_{-1}^{(\mu} \bar{\alpha}_{-1}^{\nu)} + B_{\mu \nu} \alpha_{-1}^{[\mu} \bar{\alpha}_{-1}^{\nu]} + \Phi \alpha_{-1}^\mu \bar{\alpha}_{-1}^\nu \eta_{\mu \nu} \right) \vert 0, k^\mu \rangle_{NSNS} ,
\eeqn
and their equations of motion express the restriction imposed by the
constraints. For these NSNS fields the equations of motion can similarly be determined by the
$\sigma$-model approach, which consists in taking the action
(\ref{actionsigma}) with the coordinates coupling to the space-time fields as
a quantum field theory of the coordinates in the usual sense. The space-time
fields are coupling constants of this theory and conformal invariance implies
the vanishing of all the $\beta$-functions. These can be computed by standard
methods in terms of the bare values of the space-time fields. Demanding them
to vanish yields the equations of motion order by order in the $\sigma$-model loop expansion
parameter $\al$. The one-loop result reproduces the equations known from the bosonic sector of $d=10$ supergravity \cite{FMP}. \\

The fields of the RR sector on the other hand originate from the tensor
product of the space time spinor fields $s_a$ and $\bar{s}_a$ of the left and right moving
sectors. Thus they are polarizations $H_{ab}$ which after expanding
into gamma matrices look like:
\beqn \label{rrvertex}
\vert H_{\mu_1...\mu_n}, k^\mu \rangle_{RR} = \bar{s}_a^T \left( \sum_{n=1}^{10}{\frac{i^n}{n!} H_{\mu_1...\mu_n} \left( \Gamma_0 \Gamma^{\mu_1...\mu_n} \right)^{ab} } \right) s_b \vert 0,k^\mu \rangle_{RR}.
\eeqn
To be more precise, the $H$'s are the field strengths of the RR fields. Their
equations of motion are derived by exploring the Dirac equations that come
along with super Virasoro constraints as well as chirality conditions, while a
corresponding $\sigma$-model, necessarily involving a coupling to the spin
vertex, is not known. After rewriting the Lorentz tensors into differential
forms the equations can be summarized by simply demanding the forms to be harmonic
\beqn
dH = d\host H =0 ,
\eeqn  
which allows to introduce potentials. As we have seen
earlier, the degree of the potential determines the dimension of the brane
which it naturally couples to. Looking at the RR $3$-form with $2$-form
potential, everything appears to be quite similar to the NSNS $3$-form field
strength, except that the latter couples to the coordinates of the world
sheet, while the former couples to the spin field. But this crucial fact
prohibits to interpret the fundamental closed string as the source of the RR
fields. Assume a scattering process of an incoming and outgoing closed string
with a vertex operator (\ref{rrvertex}) of the RR field inserted. Because of the RR fields coupling only via their field strength to the world sheet, this
amplitude always carries a power of the external momentum. While the diagram itself is to be interpreted as the coupling of the world volume tensor field to the macroscopic string, its zero momentum limit is the charge of the string under that field, which is vanishing. Therefore we are forced to conclude that the
p-brane solutions of the supergravity field equations, which had fundamental strings as electric or magnetic
sources, cannot be the states that carry the RR charges. There must be different elementary objects in string theory as RR charged fields. These may then have a low energy description as p-branes that are charged with respect to the RR field strengths. D-branes are of course thought to be the right choice. \\

The observation of Polchinski now was the following. One computed the one-loop scattering amplitude of an open string in the vacuum but with Dirichlet boundary conditions at its ends (signaled by putting $L_0^D$ instead of $L_0$). This can alternatively be seen as two D-branes of any type II model exchanging a closed string at tree level \cite{GR1}:
\beqn
\langle D \vert e^{- 2\pi^2 \left( L_0+\bar{L}_0-2 \right) /t} \vert D \rangle_{\mbox{\sz tree}} \leftrightarrow  \langle 0 \vert e^{-2t \left( L^D_0-1 \right) } \vert 0 \rangle_{\mbox{\sz 1-loop}} .
\eeqn
The length of the closed string is parametrized by $2\pi^2 /t=l$, while the open string circling around the cylinder has the length $t$.  
\begin{figure}[h]
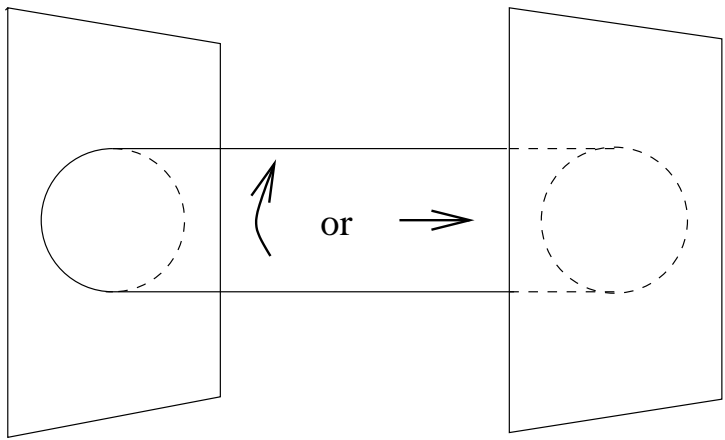

\begin{center}
\input figscatt.help
\end{center}
\caption{Open string one-loop or closed string tree-level diagram}
\label{figscatt}
\end{figure}
The result for this matrix element included a contribution of the RR fields
of the closed string to the scattering amplitude, indicating a coupling of the
RR fields to the brane. One then compares this to the same tree level amplitude in the
supergravity approximation to low energy type II string theory with a
Dirichlet brane effective action added as a source of the fields and a Wess-Zumino coupling of the RR field to the coordinates of the brane. Comparing the leading order contributions that come from the dilaton, graviton and RR fields separately allowed to deduce the charge density and tension of the brane. In fact all contributions cancel (``no force rule'') but apparently the D-branes feel a repulsion via RR fields. We first look at the string calculation of the open string one-loop vacuum diagram. In general one has to compute the path integral
\beqn
Z_{\mbox{\sz vac}} &=& \int{ \frac{{\cal D}h^{\alpha \beta} {\cal D}X^\mu
    {\cal D}\psi_{\mbox{\sz L}} {\cal D}\psi_{\mbox{\sz R}} }{\mbox{Vol}}\ \exp \Big( - \frac{1}{4\pi \al} \int_{\mbox{Cyl}}{d\tau d\sigma\ \sqrt{-h} \Big( h^{\alpha \beta} \partial_\alpha X^\mu \partial_\beta X_\mu  } } \non 
& & + \psi_{\mbox{\sz L}} \partial_- \psi_{\mbox{\sz L}} + \psi_{\mbox{\sz R}} \partial_+ \psi_{\mbox{\sz R}} \Big)  \Big)   .
\eeqn
The fields have to obey the appropriate boundary conditions, Dirichlet at the ends and periodicity or antiperiodicity around the cylinder. While we shall compute $Z_{\mbox{\sz vac}}$ as the one-loop zero-point function of the open Dirichlet string, we could have alternatively called it the tree-level two-point function of two boundary states in closed string theory. What causes all the problems in computing such integrals is in particular the integration measure which has to be divided by the volume Vol of the local symmetry group of superconformal transformations and super Weyl rescalings. A mathematically more rigorous treatment can be found in \cite{POL4}, we indicate the outcome here. First we split the (constant) zero-modes of the coordinate fields from the integration, which gives a prefactor equal to the (infinite) volume $V_{p+1}$ of the D-brane only, as constant shifts transverse to the brane are prohibited. (Remember that the centre of mass of the string also has to move on a hyperplane.) Then we can formally perform the Gaussian integrations 
\beqn
Z_{\mbox{\sz vac}} \sim \Dt^{-1/2} \left( \frac{1}{4\pi \al} \frac{1}{\sqrt{-h}} \partial_\alpha \left( \sqrt{-h} h^{\alpha \beta} \partial_\beta \right) \right) \Dt^{1/2} \left( \frac{\partial_-}{4\pi \al} \right) \Dt^{1/2} \left( \frac{\partial_+}{4\pi \al} \right)  
\eeqn 
apply the old ``$\ln \Dt^{1/2} = \frac{1}{2} \Tr \ln$'' trick as well as
\beqn \label{heatkernel}
\ln \left( {\cal O} \right) = - \int_0^\infty{\frac{dt}{t}\ \left( e^{-t{\cal O}} - e^{-t} \right) } ,
\eeqn
which holds for any operator ${\cal O}$ with spectrum in the right half plane of positive real part, finally getting
\beqn
\ln Z_{\mbox{\sz vac}} = V_{p+1} \Tr \left( \int_0^\infty{\frac{dt}{t} e^{-t\pi \al (k^2 +M^2)}} \right) .
\eeqn
Here the information about the oscillator spectrum has been translated into
the degeneracy of the mass levels which are being summed over. We have
included a factor of $2$ for the possible orientations of the string and
omitted the second term in (\ref{heatkernel}) demanding a different kind of
regularization for the integral later on. (In unoriented type I theory the
factor $2$ is missing, but on the other hand one is forced by the orientifold
projection only to consider pairs of branes at mirror positions, which in the end brings back the
appropriate factor.) All the functional determinants and traces involved have to be performed on the right spaces and superspaces, such that all the difficulties in obeying constraints and boundary conditions have only been hidden away so far. The functional trace over the string spectrum consists of an integral over momenta $k$ on the brane as well as a sum over all oscillator levels of spin and coordinate fields:
\beqn
\ln Z_{\mbox{\sz vac}} &=& V_{p+1} \int{\frac{d^{p+1}k}{(2\pi)^{p+1}} \Tr_{\mbox{\sz osc}} \left( \int_0^\infty{\frac{dt}{t} e^{-t\pi \al (k^2 +M^2)} } \right) } \non
&=& V_{p+1} \int_0^\infty{\frac{dt}{t} \left( 4\pi^2 \al t \right)^{-(p+1)/2} \Tr_{\mbox{\sz osc}} \left( e^{-t\pi \al M^2/2} \right) } .
\eeqn
The mass $M$ of a state is given by the formula (\ref{eq2w2}) and especially depends on the separation $r$ of the branes. The knowledge of the discrete mass spectrum of the theory, that is gained from the oscillator expansion of the fields subject to the super Virasoro constraints, now allows to actually perform the sum over oscillators. The procedure for the spin fields is a little tedious as all allowed periodicity conditions for going around the cylinder (NS and R combinations or spin structures) have to be regarded. This can be found in the standard literature, e.g. chapter $9.4$ of \cite{GSW}, and has been explicitly went through in \cite{GR1}. The result can be written most easily using Jacobi $\theta$ and Dedekind $\eta$ functions:
\beqn
\ln Z_{\mbox{\sz vac}} = \frac{V_{p+1}}{2} \int_0^\infty{\frac{dt}{t} \left( 4\pi^2 \al t \right)^{-(p+1)/2} e^{-r^2t \pi \al} \sum_{s=2,3,4}{(-1)^s \frac{\theta_s^4 \left( 0 \vert \frac{it}{2} \right)}{\eta^{12} \left( \frac{it}{2} \right) }}} .
\eeqn
The sum in the integrand is actually vanishing by some identity of $\theta$
functions but we can split off the two contributions that cancel each other
and by their respective periodicity condition identify the RR ($s=4$) and
NSNS ($s=2,3$) contributions separately. This implies that we reinterpret the
open string 1-loop amplitude in terms of the tree-level closed string
contributions. The poles of the amplitude arise from the UV region, i.e. small
$t$, and such we expand the integrand around $t=0$ getting the leading contribution:
\beqn
\sum_{s=2,3,4}{(-1)^s \frac{\theta_s^4 \left( 0 \vert \frac{it}{2} \right)}{\eta^{12} \left( \frac{it}{2} \right) }} = (1-1) t^4 + o\left( e^{-1/t} \right) .
\eeqn
We define the propagator 
\beqn
\Delta_{(d)} (x^2) = \frac{\pi}{2} \int_0^\infty{dt\ \left( 2\pi^2 t \right)^{-d/2} e^{-x^2/2\pi t}} = \int{\frac{d^d p}{(2\pi )^d} \frac{e^{ipx}}{p^2}} 
\eeqn
and write the final result
\beqn
\ln Z_{\mbox{\sz vac}} = V_{p+1} (1-1) 2\pi \left( 4\pi^2 \al \right)^{3-p} \Delta_{(9-p)}(r^2) + o\left( e^{-r/\sqrt{\al}} \right) .
\eeqn
We can summarize the discussion in the following way: The force between
D-branes due to the exchange of dilaton, graviton and RR field vanishes by a
cancellation of the attractive gravitational force and the repulsive electric
force. This is completely analogous to the behaviour of magnetic monopoles in
field theory, which do not excert any force on each other by a cancellation of
Higgs and vector boson exchange. In particular the contribution of the RR
field to the amplitude does not vanish. Therefore in the effective field theory the D-brane somehow couples to the RR fields and thus has to carry a corresponding charge. \\
 
The leading contribution of the above string amplitude can also be computed in the low energy approximation to string theory, which corresponds to widely separated D-branes, large $r$ or effectively small $\al$. We now demonstrate how this is done in the field theory that is given by the effective type II action for the bulk, an appropriate world volume action for the brane and a coupling term. This will allow to deduce the macroscopic charge density $\rho_p$ and brane tension $T_p$. To decouple the dilaton and graviton propagators one can rewrite the effective action of type II supergravity in the Einstein frame: 
\beqn
S^{\mbox{\sz II}}_{\mbox{\sz eff}} &=& \frac{1}{2\kappa^2} \int{d^{10}x\ \left( \sqrt{-G} \left( R^{\mbox{\sz E}} +\frac{1}{2} \left( d\Phi \right)^2 + \frac{1}{12} e^{-\Phi} \left( dB \right)^2 \right) \right. } \non
& & + \left. \sum_{p}{\frac{1}{2(p+2)!} e^{(3-p)\Phi /2} \left( dC^{(p+1)} \right)^2 } \right) .
\eeqn
Note the abuse of notation that mixes forms and functions in the integrand and leads to some changes of signs compared to earlier notations but helps very much to keep notations short. In a similar manner as was sketched above an effective action for open strings in the presence of a D-brane can be extracted from a proper $\sigma$-model \cite{TSE,LEI}:
\beqn
S_{\mbox{\sz eff}}^{\mbox{\sz D}} = T_p \int_{{\cal M}^{(p+1)}}{d^{p+1}\xi\ e^{(3-p)\Phi/4} \sqrt{- \dt \left( G +B+2\pi \al F \right) } } ,
\eeqn
where $G_{ij}$ and $B_{ij}$ are the pullbacks of the metric and the
antisymmetric NSNS tensor to the D-brane world volume and
$F^{(2)}=(p+2)dA^{(1)}$ is the field strength of the $U(1)$ gauge potential
$A^{(1)}$ that couples to the boundary of the open string. The world volume
coupling constant $T_p$ defines the string tension. The above action is taken
as the effective action of the D-brane itself by noticing that the
(perturbative) open strings ending on the brane effectively carry its degrees
of freedom as they are the only perturbative manifestation of D-branes in
string theory. Thus the low energy theory of D-branes is the perturbation
theory of open strings with Dirichlet boundary conditions. Note that the
dependence of the D-brane action on the dilaton differs from the type II
effective action by a factor $\exp \left( (p-3)\Phi/4 \right)$, which in the $\sigma$-model frame
boils down to the difference between the dilaton dependence $\exp \left( -2\Phi \right)$ and $\exp
\left( -\Phi \right)$. Thus the effective string tensions of a D-brane and a
p-brane differ by $\langle \exp \left( \Phi \right) \rangle = g_{\mbox{\sz S}}$. This fact leads to the name half-solitons for D-branes. Further we add the natural electric coupling of the RR field to the brane volume (the pullback of the RR field onto the worldvolume of the brane):
\beqn
S_{\mbox{\sz eff}}^{\mbox{\sz WZ}} = \rho_p \int_{{\cal M}^{(p+1)}}{d^{p+1}\xi\ C_{\mu_0...\mu_p} \partial_0 \xi^{\mu_0} \cdots \partial_p \xi^{\mu_p} } ,
\eeqn
where the coupling constant $\rho_p$ is the charge density. Finally one substitutes the identity
\beqn
1 = \int{d^{10}x\ \delta^{(p+1)}(x_{\|}-\xi) \delta^{(9-p)}(x_{\perp}-a) } 
\eeqn
into the D-brane action, where $a$ is the coordinate of the D-brane in the transverse space, in order to integrate out the world sheet coordinates (static gauge). Altogether we have got a field theory with an extended D-brane source for the NSNS fields and the RR tensor, whose propagation in the bulk of space-time is governed by type II supergravity. This theory might be ill defined as a quantum field theory but one can still extract the classical approximation by computing the tree level contributions to the two-point function of dilaton, graviton and RR field. Therefore we naively expand the functions of the fields to get all terms that are linear (sources) and quadratic (propagators) in the fields:
\beqn
& & S_{\mbox{\sz eff}}^{\mbox{\sz II}} + S_{\mbox{\sz eff}}^{\mbox{\sz D}} + S_{\mbox{\sz eff}}^{\mbox{\sz WZ}} =  \int{d^{10}x\ \left( -\frac{1}{2\kappa^2} \left( \sqrt{-G} R + \frac{1}{2} \left( d\Phi \right)^2 +\frac{1}{2(p+2)!} \left( dC^{(p+1)} \right)^2 \right) \right. } \\
& & + \sum_{i=1,2}{\left( T_p \delta^{(9-p)} (x_\perp -a_i) \left( -\frac{p-3}{4} \Phi + \sqrt{-G} \right) + \rho_p C^{(p+1)} \delta^{(9-p)} (x_\perp -a_i) + \cdots \right) }  +\cdots . \nonumber
\eeqn 
By dropping all coupling and higher terms we restrict ourselves to those, which are relevant for the tree-level two-point functions. We only did not explicitly expand the metric field of the pure gravity Einstein-Hilbert part around a flat background and avoid a discussion of difficulties concerning gauge fixing of the gravitational action etc. In this form one can immediately read off the propagators and source terms and quite easily compute the three contributions to the total tree level amplitude. The steps are displayed in \cite{BA} and the result reads 
\beqn
\ln Z_{\mbox{\sz vac}} = 2V_{p+1} \kappa^2 \left( \rho_p^2 - T_p^2 \right) \Delta_{(9-p)}\left( r^2\right) ,
\eeqn
which is compatible with the string result if 
\beqn
\rho_p^2 = T_p^2 = \frac{\pi}{\kappa^2} \left( 4\pi^2 \al \right)^{3-p}
\eeqn
holds. The RR charge density of the D-brane is equal to its tension, a version of the BPS mass bound. The modified Dirac quantization condition can be deduced from a Bohm-Aharonov experiment as usual. The product of the charge of the brane of $p$ spatial dimensions with the charge of its $d-4-p=6-p$ dimensional dual has to be an invisible phase factor:
\beqn
T_pT_{(6-p)}= \rho_p \rho_{(6-p)} = \frac{\pi n}{\kappa^2}.
\eeqn
This relation is satisfied by the above D-brane charge densities with $n=1$, thus D-branes are states with minimal RR charge and can be called elementary from this point of view. \\

We have only recalled the first of very many considerations that all lead to the conclusions that D-branes are the RR charged BPS states that were missing in the non perturbative part of the string spectrum before Polchinski's discovery. Let us collect the ``facts'' again: D-branes break one half of the supersymmetry by the same arguments as for p-branes. They carry charge density equal to their tension, and they satisfy the minimal version of the Dirac charge quantization condition. Their low energy limit is a supergravity p-brane solution and their perturbative degrees of freedom are the light modes of open strings ending on them.\\

\subsubsection{Black holes in string theory} 

In this chapter we shall review the application of the ideas concerning
D-branes, their interactions and duality relations to the entropy computation
and information loss dilemma of black holes. There have been several models in
various dimensions suggested after the first treatment in $d=5$. We shall
refer to \cite{SV,MS} and $d=4$, a more complete discussion of the whole
material can for instance be found in \cite{MAL}. The basic method consists in
taking configurations of D-branes and NS-5-branes, which have a low energy description in
terms of supergravity solutions, p-branes, whose fields, especially their
metric, can be written explicitly as in chapter \ref{stringsolitons}. These are then being compactified by a Kaluza-Klein procedure down
to say $d=4$ dimensions, which corresponds to choosing a space-time background
vacuum of appropriate topology, most simply the direct product of a sixdimensional torus $T^6$ with fourdimensional flat Minkowski space. 
While this background preserves ${\cal N}=8$ supersymmetry in $d=4$, which is further reduced to ${\cal N}=4$ by the presence of the branes, there have also been ${\cal N}=2$ compactifications on Calabi-Yau 3-folds been considered, which introduce a dependence of the black hole entropy on the topological properties of the Calabi-Yau manifold \cite{LU2}. If
this is done in a skilful manner the resulting fourdimensional metric can be
tuned to exactly resemble one of the metrics of black holes we know from general relativity. Particularly generalizations of Reissner-Nordstr\"om solutions can
be obtained this way. Of course we get a lot more fields by the supersymmetry,
which are in general supposed not to modify the conclusions severely and are
omitted in the following. We then intend to count the degeneracy of the
resulting state of a couple of D-branes, NS-5-branes and strings with fixed values of macroscopic energy and charges, whose
logarithm simply is the entropy of the resultant fourdimensional black
hole. But as we do not know the non perturbative degrees of freedom of a
D-brane, we have to make sure that we can change the coupling towards the
perturbative regime of the string moduli space without loosing control about
the states we are looking at. The first thing to notice is then that we shall
have to take BPS states, which means extremal Reissner-Nordstr\"om black hole
solutions. They can be assumed to exist in the non perturbative as well as in
the perturbative spectra. The second point is that the classical p-brane
solutions also include a dependence of the dilaton field on the radial
coordinate of the transverse space, which is of the kind that it generically
tends to blow up at the horizon $r=0$, thus preventing any small coupling
treatment at the position of the brane, no matter what the bare coupling is
chosen. (The classical value of the dilaton will coincide with its quantum expectation value, as the potential goes unrenormalized.) Thus the brane configuration we choose also has to take care to have regular dilaton at the horizon.\\

We now sketch how the model is constructed in detail and how the state
counting proceeds, following the concrete steps of \cite{MS,MAL}. One uses
$N_6$ D6-branes, $N_2$ parallel D2-branes and $N_5$ parallel NS-5-branes of the IIA string
theory, which are living in a space-time of topology $R^4 \times T^6$, all
being wrapped around the torus. Strictly speaking we have not shown that such many brane states with several intersecting D-branes and NS-branes exist and are stable. In fact, the ``no force'' condition allows to consider rather arbitrary configurations which can also be managed to be BPS saturated. The metric in the non compact four dimensions
then has to be tuned to be the Reissner-Nordstr\"om classical solution of
general relativity. Also one adds open strings carrying (quantized) purely left
moving momentum $N/R$ in one of the compact directions. The D6-branes have to
wrap around all the directions of the torus, while the D2-branes are taken to
intersect the NS-5-branes only in a onedimensional subspace of their
respective worldvolumes in a way that all branes have one compact direction in
common. The string momentum is supposed to flow exactly in that particular
direction. The low energy solution for the  field
strengths is then given similar to (\ref{feldstaerke}), the RR 4-form field
strength originates from the D2-brane sources, the 2-form field strength from
the D6-branes and the NSNS 3-form field strength from the NS-5-branes. These we shall not need again, but we have to make sure that the dilaton is regular. Defining the respective harmonic functions according to (\ref{harmonfu}) by  
\beqn
h_p \equiv 1 +\frac{N_p q_p^{(4)}}{r} ,
\eeqn
we can write the solution for the dilaton
\beqn
e^{-2\Phi} = h_2^{-1/2} h_5^{-1} h_6^{3/2} ,
\eeqn
which proves the regularity of the dilaton at the position $r=0$ of the branes
as well as in the asymptotically flat region $r \rightarrow \infty$. Thus we
can take the coupling to be small everywhere by choosing its finite value at the horizon extremely small. The unique dependence of the
different harmonic functions on the radial coordinate through $1/r$ results
from the fact that they all have a threedimensional uncompactified transverse
space and $1/r$ is the appropriate Green's function of the Laplacian. The
values for the charge parameters $q_p^{(4)}$ in $d=4$ are in fact given by
applying the dimensional reduction prescription to the charges in ten
uncompactified dimensions. The tendimensional fields are decomposed according
to the lower dimensional Lorentz group and then taken to be independent of the
higher compact dimensions. The higher dimensions can thus in the case of a torus compactification
trivially be integrated out, which for the mass of a p-dimensional D-brane
wrapped around the torus gives 
\beqn
m_{\mbox{\sz D}}^{(p)} = \frac{R_9}{g_{\mbox{\sz S}} \al} \frac{R_8}{\sqrt{\al}} \cdots \frac{R_{10-p}}{\sqrt{\al}} .
\eeqn  
This mass is then inserted in Newtons formula for the gravitational potential
and compared to the (classical) large radius behaviour of the $g_{00}$ component of the metric deduced in (\ref{allglsg}) 
\beqn
g_{00} \sim \frac{1}{2} \frac{q^{(4)}_p}{r} .
\eeqn
Thus we can express the parameters in the harmonic
functions in geometric quantities and Newton's constant. A similar charge parameter has to be defined for the discrete momentum of the open strings:
\beqn 
k= \frac{N q^{(4)}}{r}.
\eeqn
The solution for the
metric from (\ref{allglsg}) is the low energy approximation for a single brane. The rules for superposing several such branes (``harmonic function rule'') lead to the string frame line element
\beqn
ds^2_{\mbox{\sz RN}} &=& - \frac{1}{\sqrt{H_2 H_6}} dt^2 +  H_5 \sqrt{H_2 H_6} \left( dx_1^2 + dx_2^2 + dx^2_3\right) + \frac{K}{\sqrt{H_2 H_6}} \left( dt -dx_9 \right)^2 \non
& & + \frac{H_5}{\sqrt{H_2 H_6}} dx_4^2 +\sqrt{\frac{H_2}{H_6}} \left( dx_5^2 + \cdots + dx_8^2 \right) + \frac{1}{\sqrt{H_2H_6}} dx_9^2 ,
\eeqn
where the $H_p$ and $K$ denote the tendimensional ancestors of the $h_p$ and $k$ before dimensional reduction. After reducing to $d=4$ and switching to the Einstein frame we can get back to the desired Reissner-Nordstr\"om metric by a proper choice of the numbers of branes and the radii of the compactification. The (thermodynamic) Bekenstein-Hawking entropy can then be found by computing the area of the horizon
\beqn
S_{\mbox{\sz BH}} = \frac{A}{4G^4_{\mbox{\sz N}}} = 2\pi\sqrt{N_2 N_6 N_5 N}
\eeqn
and thermodynamical properties can be discussed \cite{KLO}. This is the point of view of the low energy approximation through supergravity and its p-brane solutions. \\

We now turn to string theory by regarding the quantum degrees of freedom of
the state that consits of the branes and strings we have put together to match
the Reissner-Nordstr\"om metric. As we have managed to get a solution with a
regular dilaton at the horizon, we feel free to change the coupling constant
from the non perturbative to the perturbative regime and discuss perturbations
of the brane state, which are small fluctuations of the D-branes' positions and
shape that can be described by weakly coupled open strings ending on the
D-branes. Counting the degeneracy of the black hole state in four dimensions
now means counting all possible configurations of strings attached to a given
set of intersecting branes that leave the macroscopic value of energy, mass
and charges invariant, i.e. one has to count the number of states of the open strings
stretching between the D2-branes and the D6-branes that carry the momentum
$N/R$ along the compact direction common to all the branes. Thus the
degeneracy of the string spectrum will be
responsible for the (statistical) entropy of the black hole. Again in other
words: The quantum degrees of freedom of a macroscopic fourdimensional black
hole are open strings travelling in the internal compact space carrying some given amount
of energy and momentum. The details of the degeneracy calculated in a proper
way are given in \cite{SV,MS}, we shall be content to use only a brief and
heuristic treatment. First one has to notice that all the D-branes are cut into half infinite branes at the intersections with the NS-5-branes, thus the
number of D-branes is multiplied by the number of NS-5-branes present: $N_2N_6N_5$. 
\begin{figure}[h]
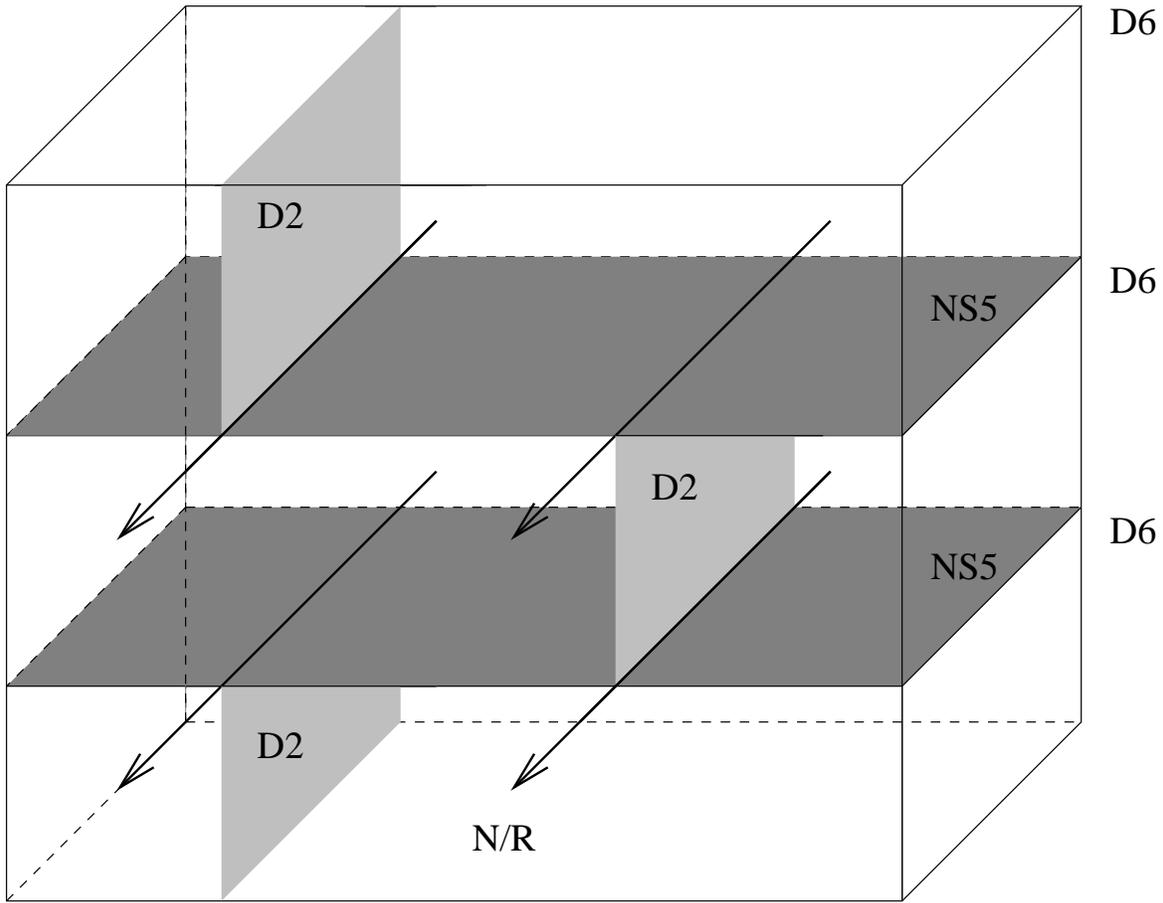

\begin{center}
\input figbh.help
\end{center}
\caption{The internal brane geometry of a fourdimensional black hole}
\label{figbh}
\end{figure}
This configuration is depicted in figure \ref{figbh} with arrows indicating the momentum flow. (For a
general deduction how branes can end on branes see \cite{T6,S1}.) Take next
into account the different possibilities of boundary conditions an open string
can have here (ND, NN, DN, DD) or equivalently the combinations of branes
which it can end on (2-2,2-6,6-2,6-6). Counting the two orientations this
together gives $2N_2N_5N_6$ different ways to attach an open string to the D-branes. Further one has to observe that the massless excitation level of these
strings contains two fermionic and two bosonic on-shell degrees of
freedom. Thus we got a gas of $N_F=4N_2N_5N_6$ fermions and the same number $N_B$ of bosons moving in a twodimensional space-time and carrying momentum $N/R$ and corresponding energy. The state density $d(N_B,N_F,N)$ of such systems is known from conformal field theory to grow exponentially with energy for high excitation levels, according to \cite{CA}
\beqn
d(N_B, N_F,N) \sim \exp \left( 2\pi \sqrt{\frac{(2N_B +N_F)N}{12}} \right) .
\eeqn
The logarithm of this is the macroscopic entropy 
\beqn
S = \ln \left( d(N_B, N_F,N ) \right) \sim 2\pi \sqrt{N_2N_5N_6N} ,
\eeqn
which perfectly matches the Bekenstein-Hawking result. The key ingredient to this remarkable result can be seen in the special property of string theory (or twodimensional conformal field theory) to have exponentially growing state density. Thus we can say that the specific input of string theory into this derivation appears crucial for the correct magnitude of the quantum degeneracy of a macroscopic black hole state of general relativity. \\

The methods we have sketched above have also been applied to calculate the
entropy of non extremal black holes \cite{HST}. These can be constructed most
easily from the extremal model by adding right moving momentum $N_R /R$
carried by additional open strings (not to be confused with the left and right
moving sectors of a single closed string) in the compact direction common to all the branes. Naively the above state counting procedure can be repeated and the left and right moving strings at small coupling contribute independently to the degeneracy:
\beqn
S = \ln \left( d_R (N_B,N_F,N_R) d_L(N_B,N_F,N_L) \right) .
\eeqn
The result is again in perfect agreement with the
Bekenstein-Hawking area law. In this picture Hawking radiation is figured to
arise from a recombination of left and right moving open strings forming a
closed string that leaves the brane and moves freely in the bulk. This enables
to compute the spectrum of the radiation by an evaluation of the tree-level partition function, which allows by standard methods of statistical physics to obtain the expectation values of the occupation numbers of string oscillator levels. This is in fact the spectral density of the closed string radiation from the black hole. Indeed one finds a black body spectrum (not surprisingly for a set of free harmonic oscillators) and one can identify the Hawking temperature in terms of the charges and momentum quantum numbers. As this is a completely unitary computation in a fully quantum theory, there is a priori no way left for any information (coherence) to dissipate away. Any pure state will stay to be one forever. On the other hand, as has been pointed out already by the authors of \cite{HST}, the success of the naive application of the procedure that was employed to compute the entropy of the extremal black hole is not clear. It very essentially relies on the opportunity to change from the strong to the weak coupling regime. Starting from a situation with many D-branes present we would assume to have relevant (if not dominating) corrections due to interactions near the horizon, as the large number of branes leads to an increase of the dilaton there. These corrections are suppressed by going to the very small coupling regime. The continuity of the former D-brane configuration and in particular the existence of all of the degenerate quantum states in both regimes was guaranteed by their BPS nature in the extremal case. This argument does no longer hold for non extremal black holes that are clearly non BPS. They should be affected by renormalization in presumably very different manners for different values of the string coupling constant. For instance in \cite{MAL}, there have been arguments given, why open string loop corrections might not change the results of the small coupling computation, but a generally excepted answer is not available.

\section{M-theory}
\label{mtheory}

In this final chapter we shall hardly be able to give anything more than a
little taste of what M-theory is meant to be, while its final version surely is
still under construction anyway. The large amount of symmetry that is found in
the spectrum and the interactions of string theory has lend a lot of heuristic
evidence to the very tempting conjecture there might be a unique theory of
gravity and supersymmetric gauge theory at the heart of it. This is assumed to
incorporate all degrees of freedom encountered in string theory in a single
type of theoretical model, the perturbative string excitations as well as all
the branes we have found so far. Also it has to reproduce the low energy
theory of string theory, the two types of $d=10$ supergravity, in a consistent
manner. An astonishing fact is that these requirements appear to be met quite
naturally if one takes this M-theory to be living in an elevendimensional
space-time, that after compactification of a single spatial direction yields
in different limits the string theories we like. This explains them all to be connected by duality transformations, as they originate from the same mother theory. The only free parameter in the procedure appears to be the radius of the compact additional dimension, while string theory in the critical dimension also has a single free parameter, its coupling constant. Let us start by briefly indicating some motivating evidence for the eleventh dimension. \\

The maximal dimension, in which any supergravity theory could be defined is
$d=11$. Higher dimensions necessarily lead to spin $5/2$ fields in the theory,
which one does not know how to deal with consistently \cite{NA}. This theory
is thus naturally assumed to be the low energy effective theory that
approximates M-theory. There is a unique supergravity in $d=11$, being scale
invariant, i.e. it does not have any free parameters. This makes us believe,
that it itself does not come from a compactified even higher dimensional and
even more unknown theory, as the compactification should include some scale
parameters according to the geometry. As it is also non chiral, in a
dimensional (Kaluza-Klein type) reduction by a compactification in one direction it leads to type IIA supergravity. The field content of its bosonic sector contains only a 3-form potential $A^{(3)}$ plus the metric field. The effective action follows:
\beqn
S_{\mbox{\sz eff}}^{(d=11)} = \frac{1}{2\kappa^2_{(11)}} \int{d^{11}x\ \sqrt{-g} \left( R - \frac{1}{48} \left( dA^{(3)} \right)^2 \right) } .
\eeqn
As earlier we did not write fermionic fields and also left out topological Chern-Simon terms. A Kaluza-Klein reduction to $d=10$ dimensions is performed by putting the eleventh coordinate $x_{10}$ on a circle of radius $R_{11}$. The fields will then be decomposed with respect to the tendimensional Lorentz group and reveal the field content of $d=10$ type IIA according to
\beqn
A_{\mu \nu \rho} \rightarrow B_{ij}\sim A_{ij10},\ A_{ijk} ,\non
g_{\mu \nu} \rightarrow \Phi \sim g_{1010},\ A_{i}\sim g_{i10},\ g_{ij},
\eeqn
where indices $i,j$ only run from $0$ to $9$. To make this more precise we can write the elevendimensional line element
and potential:
\beqn
ds_{11}^2 &=& e^{4\Phi /3}  \left( dx_{10} + A_i^{(1)} dx^i \right)^2 + e^{-2\Phi /3} ds_{10}^2, \non
A_{\mu \nu \rho} & \rightarrow & \{A_{ijk}, B_{ij} \delta_{\rho 10} \}.
\eeqn 
The tendimensional fields then are taken to be independent of the eleventh,
internal coordinate for consistency reasons. This assures that any solution
derived in the lower dimensional theory is also a solution of the original
one. The eleventh dimension can now be integrated over trivially and yields a
prefactor in front of the action, relating the tendimensional gravitational coupling constant $\kappa$ to the elvendimensional $\kappa_{(11)}$ by 
\beqn \label{couplrel}
\kappa^2 = \frac{\kappa_{(11)}^2}{2\pi R_{11}} .
\eeqn
The string coupling constant $g_{\mbox{\sz S}} = \langle \exp \left( \Phi \right) \rangle$ in the tendimensional IIA theory is also related to the
radius $R_{11}$ of the compactification by 
\beqn
R_{11} = g_{\mbox{\sz S}}^{2/3},
\eeqn 
thus the string coupling constant is
revealed to be given by the radius of the additional dimension of
M-theory. This implies that the perturbative regime of string theory does not
know anything about the eleventh dimension as the small coupling sector
corresponds to the small radius region. Vice versa the non perturbative regime
of string theory should imply a decompactification of the eleventh
dimension. M-theory does by this mechanism automatically include the
non perturbative effects of string theory \cite{WIT4}. Also the other degrees
of freedom of the $d=10$ supergravity theory match together with their
ancestors in eleven dimensions. For the fermionic fields this check is as easy
as the one we made above for the bosonic fields, while it also appears to hold
(as it better has to) in the non perturbative sector of the spectrum
\cite{SCH}. \\

In the non perturbative part of its spectrum the elevendimensional
supergravity has by the sake of its $4$-form field strength an electric
membrane (M2-brane) as well as a solitonic M5-brane of the same nature as the
p-branes we discussed earlier. From these the states of the type IIA string
spectrum can be constructed by a proper choice of the coordinates which are
being compactified. The string itself can be seen to be a membrane wrapped
around the compact circle, the string D$2$-brane is an uncompactified
M2-brane, and similarly the D4-brane and the NS-5-brane descend from the
M5-brane. A less obvious case is the D$6$-brane, which can be traced back to a Kaluza-Klein monopole of M-theory \cite{T5}. The respective string tensions can be computed on both sides of this correspondence as a test. They are related by the minimal Dirac quantization conditions and one pair of brane tensions can be fixed by hand. But afterwards the relation (\ref{couplrel}) between the two coupling constants has to hold, when comparing the rest of the tensions, and in fact, it does. \\

A more systematic way to see, how the fields and non perturbative states of
low energy string theory emerge from $d=11$ is to analyze the ${\cal N}=1$
supersymmetry algebra \cite{T2,T3,T4}. We shall indicate, how a couple of the
features of string theory, including the different string models, their
spectra and the duality transformations that relate them, are understood to be
descendants of the supersymmetry in M-theory. This reasoning gives a very comprehensive and systematic treatment of separate issues in
string theory. So let us look at the most general
superalgebra in eleven dimensions. A given (Majorana spinor) super charge
$Q_\alpha$ has $32$ real components, such that $\{ Q_\alpha , Q_\beta \}$ is a
symplectic $Sp(32)$ matrix with $32 \cdot 33/2=528$ independent entries. Under
the Lorentz subgroup $SO(1,10)$ of $Sp(32)$ it can be decomposed into
irreducible representations, which are a vector (the momentum), a second rank
tensor and a fifth rank tensor. The last two are the possible central terms in the algebra, which thus reads \cite{AGIT}:
\beqn
\{ Q_\alpha , Q_\beta \} =  \left( \Gamma^\mu C \right)_{\alpha \beta} P_\mu + \frac{1}{2} \left( \Gamma^{\mu \nu} C\right)_{\alpha \beta} Z_{\mu \nu} + \frac{1}{5!} \left( \Gamma^{\mu_1 \cdots \mu_5} C\right)_{\alpha \beta} Z_{\mu_1 \cdots \mu_5} .
\eeqn
The terms are not central in the sense of central charges, which we introduced
in chapter \ref{bpsstates}. Those were only allowed for extended
supersymmetry, while the central terms here break the Lorentz invariance. If
we take the tensor fields in the Lagrangian to have support only in a particular subspace, the
invariance is restored on the transverse coordinates, which is a hyperplane, worldvolume of a p-brane. The presence of the charges then modifies the field equations of the RR tensor fields by adding sources which are topological in the sense that they are locally exact forms. They are vanishing unless the branes wrap around non trivial cycles in space-time, for instance
\beqn
Z^{\mu \nu} = q \int_{{\cal M}^{(2)}}{dx^\mu \wedge dx^\nu} ,
\eeqn
which is zero, if the brane volume ${\cal M}^{(2)}$, a 2-cycle, is contractible. By similar arguments as we used in chapter \ref{bpsstates} one can now deduce that there can be M2-branes and M5-branes as presumably elementary and solitonic BPS states. In fact one also suspects dual M9-branes and M6-branes to exist. The M2-brane was the first to be established as a solution to the field equations \cite{BST} in a similar manner as we followed in chapter \ref{stringsolitons}. There have also been extensive researches for its quantum theory, the quantization of its coordinates \cite{WLN}. Also the M5-brane is thought to be rather well understood in terms of its worldvolume action \cite{BKN}. One can, for instance, give explicit formulas for the metrics and field strengths and discuss the singularity structures, the masses and the charges, which verifies them to saturate the BPS mass bound. The latter dual branes are still somewhat mysterious and are being under observation. \\

The easiest way to get an impression what might be happening when these M-theory states, or better say $d=11$ supergravity states, are being compactified down to the critical string dimension, is again to look at the dimensional reduction of the superalgebra
\beqn
\{ Q_\alpha , Q_\beta \} &=& \left( \Gamma^\mu C\right)_{\alpha \beta} P_\mu
+ \left( \Gamma^{10} C\right)_{\alpha \beta} P_{10} + \left( \Gamma^\mu
  \Gamma^{10} C\right)_{\alpha \beta} Z_{\mu 10} + \frac{1}{2} \left( \Gamma^{\mu \nu} C\right)_{\alpha \beta} Z_{\mu \nu}  \non
& & + \frac{1}{4!} \left( \Gamma^{\mu \nu \rho \sigma} \Gamma^{10} C
\right)_{\alpha \beta} Z_{\mu \nu \rho \sigma 10} + \frac{1}{5!} \left( \Gamma^{\mu \nu \rho \sigma \lambda} C\right)_{\alpha \beta} Z_{\mu \nu \rho \sigma \lambda} ,
\eeqn
where the indices now run from $0$ to $9$ only. We notice that the central terms in $d=10$ originate from the central terms of $d=11$ as well as from the eleventh entry of the momentum. All the central terms we need as charges of the tensor fields in the $d=10$ IIA supergravity are present, such that all the brane solutions we have constructed and conjectured in the previous chapters could have been foreseen from this simple analysis. In particular we recognize our earlier statement that the states carrying Kaluza-Klein momentum $P_{10}$ in the compact direction from the tendimensional point of view look like D0-branes, charged under the scalar central term of the IIA super algebra. The relation to IIB  is a little more subtle. It was however found that a torus compactification of M-theory leads to IIB compacified on a circle, where the $SL(2,\mathbb{Z})$ acting on the complex modulus of the torus is exactly mapped to the IIB selfduality group that acts on the complex combinations of the NSNS and RR scalars and 2-forms \cite{SCH}. Along similar lines one can proceed further to recover more dualities of string theory. By combining chiral components of the supercharges and after performing a chirality flip on one half of the components, one gets from IIA to a chiral type IIB superalgebra of only say left handed supercharges. This can be related to the T-duality of the two type II string theories in $d=10$. Also one can truncate the IIA theory down to an ${\cal N}=1$ theory by a one sided parity operation keeping only invariant supercharges. This then yields the superalgebra of the heterotic theory with all its central terms. It also reveals that these central terms involve only the sum $P_\mu +Z_\mu$ of momentum and topological winding charge, therefore it is unaffected by exchanging the two. In this way the heterotic T-duality arises very naturally from symmetries of the truncated $d=11$ superalgebra, which are invisible from ten dimensions. A large part of the web of string dualities has thus been observed emerging from the superalgebra of the elvendimensional ancestor. \\

After having returned to the perturbative T-duality we started with, we like to stop our journey at this point, leaving all the more advanced topics to more specialized reviews, a couple of which we have cited above. The most prominent omissions we have left out surely include the developments initiated by \cite{HANWI} which allowed the study of gauge theories via the D-brane world volume effective field theories or via M-branes \cite{WITT} alternatively. Neither did we explore the ideas concerning the more realistic non extremal and non supersymmetric black holes in detail and completely omitted a discussion of the Maldacena conjecture of the duality between IIB string theory on anti de Sitter space and an ordinary conformal field theory on its boundary \cite{MALC,WIC,KPC}.

\clearpage

\begin{appendix}

\section{Compactification on $T^2$ and T-duality}

\begin{figure}[h]
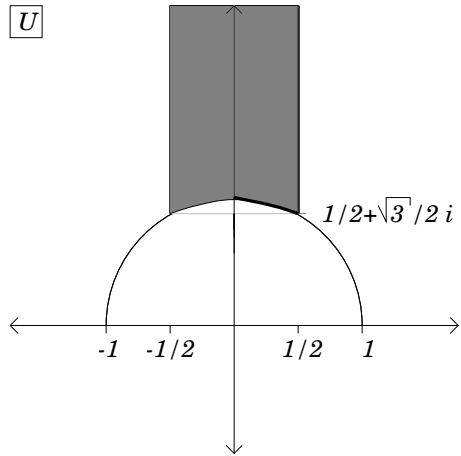

\begin{center}
\input fund.pstex_t
\end{center}
\caption{Moduli space of the complex structure modulus of a torus.}
\label{figfund}
\end{figure}

In order to illustrate the statements of section \ref{torus} we will now focus on the compactification of a bosonic string on a twodimensional torus. We have then four background fields, three coming from the metric ($G_{ij}$) and one from the antisymmetric tensor ($\tilde{B}_{ij} = \epsilon_{ij} B$) spanning the classical moduli space 
\be
{\cal M}_{\mbox{\sz class}} = \frac{SO(2,2,\mathbb{R})}{SO(2)_{L} \times SO(2)_{R}} \simeq \left. \frac{SL(2,\mathbb{R})}{U(1)}\right|_{T} \times \left. \frac{SL(2,\mathbb{R})}{U(1)}\right|_{U} \simeq \mathbb{H}|_{T} \times \mathbb{H}|_{U}.   \label{eq2z}
\ee
Here we have introduced the two complex moduli
\beqn
U & = & U_{1} + i U_{2} = \frac{G_{12}}{G_{22}} + i \frac{\sqrt{\mbox{det}G}}{G_{22}} \in \mathbb{H} , \non 
T & = & T_{1} + i T_{2} = \frac{1}{2} \left( B + i \sqrt{\mbox{det}G} \right) \in \mathbb{H}  , \label{eq2a2}
\eeqn
where $U$ is the complex structure modulus describing the form of the torus
and $T$ is the K\"ahler modulus ($\sqrt{\mbox{det}G}$ gives the volume of the
torus). That means we represent the two dimensional lattice in the complex
plane. It is possible to express the metric in terms of $U$ and $T$ as:
\be
G = \frac{2 T_{2}}{U_{2}} \left( \begin{array}{cc}
                               U_{1}^{2} + U_{2}^{2} & U_{1} \\
                               U_{1} & 1 
                               \end{array}
                        \right).
\label{eq2b2}
\ee
One can also write $p_{L}^{2}$ and $p_{R}^{2}$ in terms of the moduli:
\beqn
\left( \vec{p}_{L}\right) ^{2} & = & \frac{1}{2 T_{2} U_{2}} \left| (n_{1} - U n_{2}) - T(m_{2}+Um_{1}) \right|^{2}  , \non
\left( \vec{p}_{R}\right)^{2} & = & \frac{1}{2 T_{2} U_{2}} \left| (n_{1} - U n_{2}) - \bar{T}(m_{2}+Um_{1}) \right|^{2},   \label{eq2c2}
\eeqn
where $(n_{1},n_{2})$ are the momentum numbers and $(m_{1},m_{2})$ the winding numbers\footnote{This can be checked after a straightforward but tedious calculation with the help of (\ref{eq2t}).}. The spectrum is given in (\ref{eq2u}). It can be shown, as pointed out at the end of section \ref{torus}, that its symmetry group is 
\be
\Gamma_{\mbox{\sz T-duality}} = SO(2,2,\mathbb{Z}) \simeq SL(2,\mathbb{Z})_{U} \times SL(2,\mathbb{Z})_{T} \times \mathbb{Z}_{2}^{I} \times \mathbb{Z}_{2}^{II}.   \label{eq2d2}
\ee
We will demonstrate that this is indeed a group of transformations under which the
spectrum is invariant. To be honest we will just focus on the $\left( \vec{p}_{L}\right) ^{2}
+ \left( \vec{p}_{R}\right) ^{2} $-part of eq. (\ref{eq2u}). The number operators 
\beqn
N_{L/R} = \sum_{n>0} \alpha^{i}_{L/R, -n} (G,B) \, G_{ij} \, \alpha^{j}_{L/R, n} (G,B)
\eeqn
can be shown \cite{GPR} to be manifestly invariant under T-duality because of
the non trivial transformation of the oscillators which compensates the
transformation of the metric. That (\ref{eq2d2}) is precisely the entire T-duality group relies on the general result stated at the end of section \ref{torus}, which can also be found in \cite{GPR}. (In fact the symmetry group contains one further element, namely symmetry under the worldsheet parity transformation $\sigma \rightarrow -\sigma$, implying $B \rightarrow -B$.)
\begin{figure}[h]
\begin{center}
\begin{picture}(240,90)(0,0)
  \thicklines
  \put(0,0){\circle*{3}}
  \put(90,0){\circle*{3}}
  \put(180,0){\circle*{3}}
  \put(30,45){\circle*{3}}
  \put(120,45){\circle*{3}}
  \put(210,45){\circle*{3}}
  \put(60,90){\circle*{3}}
  \put(150,90){\circle*{3}}
  \put(240,90){\circle*{3}}
  \LongArrow(0,0)(90,0)
  \LongArrow(0,0)(30,45)
  \DashLine(0,0)(120,45)3
  \DashLine(0,0)(210,45)3
  \LongArrow(112,42)(120,45)
  \LongArrow(196,42)(210,45)
\end{picture}
\end{center}
\caption{Different basis vectors can define the same lattice.}
\label{figlat}
\end{figure}
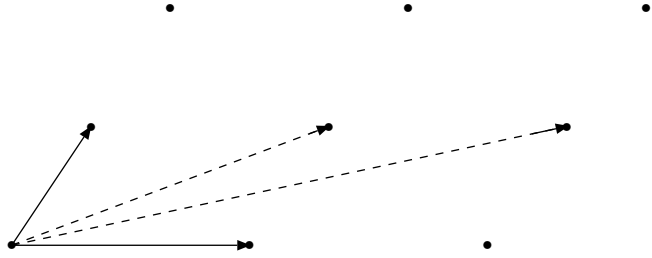
The first $SL(2,\mathbb{Z})_{U}$ in (\ref{eq2d2}) reflects the fact that the target space is a two dimensional torus, whose complex structure modulus always has an $SL(2,\mathbb{Z})$ symmetry. Not all values of $U$ lead to different complex structures but only those in the fundamental region ${\cal M}=\mathbb{H}/SL(2,\mathbb{Z})$ (see fig. \ref{figfund}, the thick lines of the boundary belong to the moduli space, the thin ones not). The transformation 
\beqn
U \rightarrow \frac{a U + b}{c U + d} \quad \mbox{with} \quad \left(\begin{array}{cc} a & b \\ c & d \end{array} \right) \in SL(2,\mathbb{Z}) ,
\eeqn 
does not change the complex structure of the torus and just amounts to another choice of basis vectors for the same lattice (see fig. \ref{figlat}). This symmetry of the spectrum is classical in the sense that it does not need any special features of the string but just depends on a symmetry of the target space. The second $SL(2,\mathbb{Z})_{T}$ is stringy from nature. Its generators are
as usual $T \rightarrow T+1$, which is a shift in $B$, and $T \rightarrow
-1/T$, which for $B=0$ amounts to an inversion of the torus
volume. Invariance under the first transformation can be understood from
(\ref{eq2q}). If $B_{ij}$ is constant, the second term is a total
derivative (namely $\, B_{ij} \partial_{\alpha} (\epsilon^{\alpha \beta} X^{i} \partial_{\beta} X^{j})$) and thus its contribution topological. An integer shift in $B$ (i.e. in general a shift by an antisymmetric matrix with integer entries) amounts to a shift of the action by an integer multiple of $2 \pi$ and therefore does not change the path integral. \\

If one considers the special case of a background with $B=G_{12}=0$, i.e. a lattice with basis $\{R_{1}, i R_{2} \}$, leading to $G_{11}=R_{1}^{2}$, $G_{22}=R_{2}^{2}$ and $\sqrt{\mbox{det}G}=R_{1} R_{2}$, we have $U= i R_{1}/R_{2}$ and $T= i R_{1} R_{2}/2$. The element $T \rightarrow -1/T$, $U \rightarrow U$ acts on the lattice according to $R_{1}/\sqrt{2} \rightarrow \sqrt{2}/ R_{2}$ and $R_{2}/\sqrt{2} \rightarrow \sqrt{2}/ R_{1}$. \\

The first $\mathbb{Z}^{I}_{2}$ exchanges the complex structure and K\"ahler
moduli, $U \leftrightarrow T$, and is a two dimensional example of mirror symmetry. It is related to the T-duality of one of the circles making up the torus. This becomes clear if one looks again at the special case of $U= i R_{1}/ R_{2}$ and $T= i R_{1} R_{2}/2$. It translates to $R_{1} \rightarrow R_{1}$ and $R_{2}/\sqrt{2} \rightarrow \sqrt{2}/ R_{2}$. The corresponding T-duality for the second circle is achieved by a composition of this $\mathbb{Z}_{2}^{I}$ transformation and twice the $SL(2,\mathbb{Z})_{U}$ transformation $U \rightarrow -1/U$, $T \rightarrow T$, i.e. $R_{1} \leftrightarrow R_{2}$ (altogether this amounts to $T \rightarrow -1/U$ and $U \rightarrow -1/T$). The second $\mathbb{Z}_{2}^{II}$ is given by $(T, U) \rightarrow (-\bar{T}, -\bar{U})$, which translates into $B \rightarrow -B$ and $G_{12} \rightarrow -G_{12}$. 
It is an easy exercise to verify explicitly that the ensembles of all
different values $\left( \vec{p}_{L}\right)^{2}$ respectively $\left( \vec{p}_{R}\right)^{2}$ from
(\ref{eq2c2}) are seperately unchanged under the above transformations and
therefore the (target space) perturbative spectrum is invariant. Of course,
like in the one dimensional case, the single states are in general not
invariant and winding and momentum numbers mix under the transformations. To be more precise, the winding and
momentum numbers of the transformed states (tilded quantities) can be
expressed via the old ones according to table \ref{tab3}. Obviously if
$(n_{1}, m_{1}, n_{2}, m_{2})$ take all values of $\mathbb{Z}^{4}$, the same
is true for $(\tilde{n}_{1}, \tilde{m}_{1}, \tilde{n}_{2}, \tilde{m}_{2})$.
\begin{table}[h]
\begin{center}
\begin{tabular}{l c c c c}
Transformation & $\tilde{n}_{1}$ & $\tilde{n}_{2}$ & $\tilde{m}_{1}$ & $\tilde{m}_{2}$ \\ [2mm]
$U \rightarrow -\frac{1}{U}$ & $n_{2}$ & $-n_{1}$ & $m_{2}$ & $-m_{1}$  \\ [2mm]
$U \rightarrow U+1$ & $n_{1} - n_{2}$ & $n_{2}$ & $m_{1}$ & $m_{1} + m_{2}$ \\ [2mm]
$T \rightarrow -\frac{1}{T}$ & $m_{2}$ & $-m_{1}$ & $n_{2}$ & $-n_{1}$ \\ [2mm]
$T \rightarrow T+1$ & $n_{1} - m_{2}$ & $m_{1}+n_{2}$ & $m_{1}$ & $m_{2}$ \\ [2mm]
$T \leftrightarrow U$ & $n_{1}$ & $m_{2}$ & $m_{1}$ & $n_{2}§$ \\ [2mm]
$U \rightarrow -\bar{U}$, $T \rightarrow -\bar{T}$ & $n_{1}$ & $-n_{2}$ & $m_{1}$ & $-m_{2}$ \\
\end{tabular}
\end{center}
\caption{Transformation of winding and momentum numbers}
\label{tab3}
\end{table}
The invariance of the mass spectrum 
is of course only a necessary condition for the whole theory to be invariant under the T-duality group. It is possible to show that the partition function is also unchanged to all orders in perturbation theory. \\

T-duality is the remainder of a spontaneously broken gauge symmetry. Only at special points in the moduli space it is partially or
completely restored. These points correspond to fixed points or higher
dimensional fixed manifolds of some of the symmetry transformations of the
T-duality group \cite{GPR,DHS}. To illustrate this fact take the example of the circle compactification at the self dual radius $R_{\mbox{\sz fix}}$, where the gauge group is $SU(2) \times SU(2)$ (see below). It can be shown in this case that there are nine massless scalars besides the dilaton, which transform as $(\bf{3},\bf{3})$ under $SU(2) \times SU(2)$. The 3-3 component can furthermore be identified as the modulus $\delta R$ for the radius of the compactification circle (or to be more precise for its difference from the self dual radius). Moving away from the self dual radius amounts to giving an expectation value to $\delta R$ and thereby breaking the gauge symmetry to the generic group $U(1) \times U(1)$. However rotating by $\pi$ around the 1-axis in one of the $SU(2)$s changes the sign of the 3-3 component. This shows, that decreasing the value of the radius from the self dual value is gauge equivalent to increasing it. This fact survives the breaking of the gauge group in the form of T-duality. \\

The gauge symmetry enhancement at special loci in the moduli space happens of course also in our two dimensional example. The $\mathbb{Z}_{2}^{I}$ transformation (i.e. $T \leftrightarrow U$) has the fixed line $T=U$. In the special case of $B=G_{12}=0$ this amounts to choosing the self dual radius for $R_{2}$. From the experience with the circle compactification one therefore expects a symmetry enhancement according to $U(1)^{4} \rightarrow U(1)^{2} \times SU(2)^{2}$. This actually happens for $T=U$, which can be seen from the formulas for the left and right momenta (viewed as complex numbers), namely
\beqn
p_{L} & = & \frac{1}{\sqrt{2}} \frac{1}{T_{2}} \left( (n_{1} - T n_{2}) - T (m_{2} + T m_{1}) \right) , \non
p_{R} & = & \frac{1}{\sqrt{2}} \frac{1}{T_{2}} \left( (n_{1} - T n_{2}) - \bar{T} (m_{2} + T m_{1}) \right). \label{eq2e2}
\eeqn
For $B=G_{12}=0$ we have $\bar{T} = -T = -iT_{2}$ and thus four additional massless vectors (c.f. (\ref{eq2u}) and see table \ref{tab6}).
\begin{table}[h]
\begin{center}
\begin{tabular}{c c c c c c c c}
$m_{1}$ & $m_{2}$ & $n_{1}$ & $n_{2}$ & $p_{L}$ & $p_{R}$ & $N_{L}$ & $N_{R}$ \\ [2mm]
0 & $\pm 1$ & 0 & $\mp 1$ & 0 & $\pm i \sqrt{2}$ & 1 & 0 \\ [2mm]
0 & $\pm 1$ & 0 & $\pm 1$ & $\mp i \sqrt{2}$ & 0 & 0 & 1 \\ [2mm] 
\end{tabular}
\end{center}  
\caption{Additional $SU(2)$ gauge bosons.}
\label{tab6}
\end{table}
\noindent The last two columns are determined by the level matching condition $\frac{1}{2} |p_{L}|^{2} + N_{L} = \frac{1}{2} |p_{R}|^{2} +N_{R}$. If the oscillators carry indices of non compact dimensions, the corresponding vertex operators $:\bar{\partial}X^{\mu} \exp \left( ik \cdot X \right) \exp\left( \pm i \sqrt{2} X^{25}_{L} \right):$ and $:\partial X^{\mu} \exp \left( ik \cdot X\right) \exp\left( \pm i \sqrt{2} X^{25}_{R}\right) :$ represent four new gauge bosons, which together with the generically (for all values of the radius) existing gauge bosons $:\partial X^{25}_{L} \bar{\partial} X^{\mu} \exp \left( ik \cdot X\right) :$ and $:\partial X^{\mu} \bar{\partial} X^{25}_{R} \exp \left( ik \cdot X \right) :$ combine to the gauge fields of $SU(2)_{L} \times SU(2)_{R}$ (in the Cartan-Weyl basis). This can be checked directly by considering the currents 
\beqn
j^{1}(z) & = & \sqrt{2} :\cos \left( \sqrt{2} X^{25}_{L/R} (z)\right): \hspace{.4cm} = \hspace{.4cm} :\frac{1}{\sqrt{2}} \left( \exp \left( i \sqrt{2} X^{25}_{L/R}(z) \right) + \exp \left( - i \sqrt{2} X^{25}_{L/R}(z) \right) \right):\ ,   \non
j^{2}(z) & = & \sqrt{2} :\sin \left( \sqrt{2} X^{25}_{L/R} (z)\right) : \hspace{.4cm} = \hspace{.4cm} :\frac{1}{\sqrt{2} i} \left( \exp \left(i \sqrt{2} X^{25}_{L/R}(z)\right) - \exp \left( - i \sqrt{2} X^{25}_{L/R}(z) \right)\right):\ ,   \non
j^{3}(z) & = & i \partial X^{25}_{L/R}(z).   \label{eq2f2}
\eeqn
One can verify that the algebra formed by their Laurent coefficients is given
(for the left resp. right moving currents seperately) by a so called {\em level one SU(2) Kac-Moody algebra}
\be
\left[ j^{k}_{m}, j^{l}_{n} \right] = m \delta_{m+n} \delta^{kl} + i \epsilon^{klq} j^{q}_{m+n},   \label{eq2g2}
\ee
which reduces for the $j^{k}_{0}$ elements to the usual $SU(2)$ Lie algebra
(the fact that we get a much bigger (infinite) algebra is of course due to the
$z$-dependence of the currents in (\ref{eq2f2})). For more details on this
point see e.g. \cite{LT}. \\
 
The $U(1)^{2}$ related to the vectors $:\partial X^{24}_{L} \bar{\partial}
X^{\mu} \exp \left( ik \cdot X \right) :$ and $:\partial X^{\mu} \bar{\partial} X^{24}_{R} \exp \left( ik
  \cdot X \right) :$ is enhanced at the self dual value for the radius $R_{1}$. Since
the T-duality transformation for this circle is given by $T \rightarrow
-1/U$ and $U \rightarrow -1/T$ (see above) the fixed point is
$T = -1/U$. Again we have for $B=G_{12}=0$ four additional gauge bosons which enlarge the symmetry to an $SU(2)_{L} \times SU(2)_{R}$, namely: 
\begin{table}[h]
\begin{center}
\begin{tabular}{c c c c c c c c}
$m_{1}$ & $m_{2}$ & $n_{1}$ & $n_{2}$ & $p_{L}$ & $p_{R}$ & $N_{L}$ & $N_{R}$ \\ [2mm]
$\pm 1$ & 0 & $\mp 1$ & 0 & 0 & $\mp \sqrt{2}$ & 1 & 0 \\ [2mm]
$\pm 1$ & 0 & $\pm 1$ & 0 & $\pm \sqrt{2}$ & 0 & 0 & 1 \\ [2mm] 
\end{tabular}
\end{center}
\end{table}

\noindent We have used 
\beqn
p_{L} & = & \frac{1}{\sqrt{2}} \left( n_{1} + \frac{1}{T} n_{2} -T m_{2} +m_{1} \right) \ , \non
p_{R} & = & \frac{1}{\sqrt{2}} \left( n_{1} + \frac{1}{T} n_{2} -\bar{T} \left(m_{2} - \frac{1}{T} m_{1}\right) \right)   \label{eq2h2}
\eeqn
and $\bar{T} = -T$. If we satisfy both conditions $U=T$ and $UT=-1$ at the same time, all the $U(1)$'s are enhanced to give the gauge group $SU(2)^{4}$. This is obviously the case for $T=U=i$, which is also a fixed point of the $SL(2,\mathbb{Z})_{T}$ and $SL(2,\mathbb{Z})_{U}$ transformations $U \rightarrow -1/U$ and $T \rightarrow -1/T$ and of the $\mathbb{Z}_{2}^{II}$ transformation $T \rightarrow -\bar{T}$, $U \rightarrow -\bar{U}$. This is a generalization of the circle compactification in the sense, that we have compactified on two orthogonal circles with self dual radii. On D-dimensional tori it is possible to get in a similar way the gauge group $SU(2)_{L}^{D} \times SU(2)_{R}^{D}$. \\

To get more general gauge groups we need a non vanishing $B$. Like before the right choice for $T$ and $U$ is a value that is invariant under a subgroup of (\ref{eq2d2}). It is easy to verify that $T=U= 1/2 + i \sqrt{3}/2$ is invariant under $T \leftrightarrow U$, $T \rightarrow -1/(T-1)$, $U \rightarrow -1/(U-1)$ and $T \rightarrow 1-\bar{T}$, $U \rightarrow 1-\bar{U}$. In this case the restored symmetry group is $SU(3)_{L} \times SU(3)_{R}$ as we get twelve additional massless gauge bosons, whose left respectively right momenta make up the root lattice of $SU(3)$ (i.e. we have chosen the compactification torus to be defined by the lattice dual to the root lattice of $SU(3)$). The new states are summarized in table \ref{tab4}. It is again possible to show that the internal parts of the corresponding vertex operators together with those of the generically present gauge bosons generate a level one $SU(3)_{L} \times SU(3)_{R}$ Kac-Moody algebra. 
\begin{table}[h]
\begin{center}
\begin{tabular}{c c c c c c c c}
$m_{1}$ & $m_{2}$ & $n_{1}$ & $n_{2}$ & $p_{L}$ & $p_{R}$ & $N_{L}$ & $N_{R}$ \\ [2mm]
0 & $\pm 1$ & 0 & $\mp 1$ & 0 & $\pm i \sqrt{2}$ & 1 & 0  \\ [2mm]
$\mp 1$ & $\pm 1$ & $\pm 1$ & 0 & 0 & $\pm \sqrt{2} \left( \frac{\sqrt{3}}{2} + \frac{i}{2} \right)$ & 1 & 0  \\ [2mm]
$\mp 1$ & 0 & $\pm 1$ & $\pm 1$ & 0 & $\pm \sqrt{2} \left( \frac{\sqrt{3}}{2} - \frac{i}{2} \right)$ & 1 & 0  \\ [2mm]
0 & $\mp 1$ & $\mp 1$ & $\mp 1$ & $\pm i \sqrt{2}$ & 0 & 0 & 1 \\ [2mm] 
$\pm 1$ & $\mp 1$ & 0 & $\mp 1$ & $\pm \sqrt{2} \left( \frac{\sqrt{3}}{2} + \frac{i}{2} \right)$ & 0 & 0 & 1 \\ [2mm]
$\pm 1$ & 0 & $\pm 1$ & 0 & $\pm \sqrt{2} \left( \frac{\sqrt{3}}{2} - \frac{i}{2} \right)$ & 0 & 0 & 1
\end{tabular}
\end{center}
\caption{Additional $SU(3)$ gauge bosons.}
\label{tab4}
\end{table}

\end{appendix}

\end{document}